\documentclass[fleqn,usenatbib]{mnras}

\usepackage{newtxtext,newtxmath}
\usepackage{makecell}
\usepackage{relsize}

\usepackage[T1]{fontenc}

\DeclareRobustCommand{\VAN}[3]{#2}
\let\VANthebibliography\thebibliography
\def\thebibliography{\DeclareRobustCommand{\VAN}[3]{##3}\VANthebibliography}

\usepackage{graphicx}	
\usepackage{amsmath}	

\title[Metal and dust evolution in REBELS]{Metal and dust evolution in ALMA REBELS galaxies: insights for future JWST observations}

\author[M. Palla et al.]{
Marco Palla,$^{1,2}$\thanks{E-mail: marco.palla@inaf.it}
Ilse De Looze,$^{1}$
Monica Relaño,$^{3,4}$,
Stefan van der Giessen$^{1,3}$,
Pratika Dayal$^{5}$,
\newauthor\, Andrea Ferrara$^{6}$,
Raffaella Schneider$^{7,8}$,
Luca Graziani$^{7,8}$,
Hiddo S. B. Algera$^{9,10}$,
Manuel Aravena$^{11}$,
\newauthor\, Rebecca A. A. Bowler$^{12}$,
Alexander P. S. Hygate$^{13}$,
Hanae Inami$^{9}$,
Ivana van Leeuwen$^{13}$,
\newauthor\, Rychard Bouwens$^{13}$,
Jacqueline Hodge$^{13}$,
Renske Smit$^{14}$,
Mauro Stefanon$^{15,16}$,
Paul van der Werf$^{13}$
\\
\\
$^{1}$Sterrenkundig Observatorium, Ghent University, Krĳgslaan 281 - S9, 9000 Gent, Belgium\\
$^{2}$INAF - OAS, Osservatorio di Astrofisica e Scienza dello Spazio di Bologna, via Gobetti 93/3, 40129 Bologna, Italy\\
$^{3}$Dept. Fisica Teorica y del Cosmos, Universidad de Granada, Spain\\
$^{4}$Instituto Universitario Carlos I de Fisica Teorica y Computacional, Universidad de Granada, E-18071, Granada, Spain\\
$^{5}$Kapteyn Astronomical Institute, University of Groningen, P.O. Box 800, 9700 AV Groningen, The Netherlands\\
$^{6}$Scuola Normale Superiore, Piazza dei Cavalieri 7, 50126 Pisa, Italy\\
$^{7}$Dipartimento di Fisica, Sapienza, Universita di Roma, Piazzale Aldo Moro 5, I-00185 Roma, Italy\\
$^{8}$INAF-Osservatorio Astronomico di Roma, via Frascati 33, I-00078 Monte Porzio Catone, Roma, Italy\\
$^{9}$Hiroshima Astrophysical Science Center, Hiroshima University, 1-3-1 Kagamiyama, Higashi-Hiroshima, Hiroshima 739-8526, Japan\\
$^{10}$National Astronomical Observatory of Japan, 2-21-1, Osawa, Mitaka, Tokyo, Japan\\
$^{11}$Departamento de Astronomia, Universidad de Chile, Camino del Observatorio 1515, Las Condes, Santiago 7591245, Chile\\
$^{12}$Jodrell Bank Centre for Astrophysics, Department of Physics and Astronomy, School of Natural Sciences, The University of Manchester, Manchester, M13 9PL, UK\\
$^{13}$Leiden Observatory, Leiden University, NL-2300 RA Leiden, Netherlands\\
$^{14}$ Astrophysics Research Institute, Liverpool John Moores University, 146 Brownlow Hill, Liverpool L3 5RF, UK\\
$^{15}$ Departament d’Astronomia i Astrofìsica, Universitat de València, C. Dr. Moliner 50, E-46100 Burjassot, València, Spain\\
$^{16}$ Unidad Asociada CSIC "Grupo de Astrofísica Extragaláctica y Cosmología" (Instituto de Física de Cantabria - Universitat de València)
}

\date{Accepted XXX. Received YYY; in original form ZZZ}

\pubyear{XXXX}

\begin{document}
\label{firstpage}
\pagerange{\pageref{firstpage}--\pageref{lastpage}}
\maketitle

\begin{abstract}

ALMA observations revealed the presence of significant amounts of dust in the first Gyr of Cosmic time. However, the metal and dust buildup  picture remains very uncertain due to the lack of constraints on metallicity. 
JWST has started to reveal the metal content of high-redshift targets, which may lead to firmer constraints on high-redshift dusty galaxies evolution. 
In this work, we use detailed chemical and dust evolution models to explore the evolution of galaxies within the ALMA REBELS survey, testing different metallicity scenarios that could be inferred from JWST observations.
In the models, we track the buildup of stellar mass by using non-parametric SFHs for REBELS galaxies. Different scenarios for metal and dust evolution are simulated by allowing different prescriptions for gas flows and dust processes.
The model outputs are compared with measured dust scaling relations, by employing metallicity-dependent calibrations for the gas mass based on the [C\,{\sc{ii}}]\,158\,$\mu$m line.
Independently of the galaxies metal content, we found no need for extreme dust prescriptions to explain the dust masses revealed by ALMA. However, different levels of metal enrichment will lead to different dominant dust production mechanisms, with stardust production dominant over other ISM dust processes only in the metal-poor case. This points out how metallicity measurements from JWST will significantly improve our understanding of the dust buildup in high-redshift galaxies.
We also show that models struggle to reproduce observables such as dust-to-gas and dust-to-stellar ratios simultaneously, possibly indicating an overestimation of the gas mass through current calibrations, especially at high metallicities.

\end{abstract}



\section{Introduction}
\label{s:intro}

The understanding of the evolution of galaxies across cosmic time is one of the fundamental goals of present-day astrophysics.
Over the last decades, a constantly increasing number of galaxies in the Epoch of Reionization (EoR, $z\gtrsim 6$) have been observed through their rest-frame UV window (e.g. \citealt{Bowens10,McLure13,Finkelstein15,Oesch16,Stefanon19}), which is now opening up to rest-frame near-IR (NIR) wavelengths thanks to the launch of the {\it James Webb Space Telescope} (JWST, e.g., \citealt{Harikane22,Naidu22,Atek22,Castellano22}).

However, observations of galaxies across redshifts have long demonstrated the severe impact of dust on galaxy detectability and information that can be extracted from UV-to-NIR emission (\citealt{Draine89,Draine03,Calzetti00,Casey14}).
Only in the last several years, the {\it Atacama Large Millimeter/submillimeter Array} (ALMA) has enabled the study of the dust content in distant galaxies at rest-frame far-IR (FIR) wavelengths (see \citealt{Hodge20} for a review). These observations have allowed, despite the uncertainties, fairly robust estimates of the dust mass budget in galaxies at $z\gtrsim 6$ (e.g. \citealt{Riechers13,Venemans18,Tamura18,Schouws22}). In particular, they reveal dust-to-stellar mass ratios ranging between $\simeq 0.01 -3\%$ for “normal” (i.e. with conditions representative of the galaxy population at these early epochs) star forming galaxies (e.g. \citealt{Watson15,Laporte17,Bakx20,Fudamoto21}).
At the same time, a range of theoretical models were tested to understand the role of different processes in the buildup of the observed dust masses, i.e. dust formation from supernovae (SNe) and asymptotic giant branch (AGB) stars, grain growth in the interstellar medium (ISM), and dust destruction from SN shocks. 
Either analytic (e.g. \citealt{Valiante09,Valiante17,Dayal10,deBenassutti14,Gioannini17}), semi-analytic (e.g. \citealt{Mancini15,Popping17,Ginolfi18,Triani20}) or hydrodynamical (e.g. \citealt{Mckinnon18,Graziani20}) schemes have been used to study the impact of these different processes on the dust content of high-redshift galaxies. However, these studies present quite different outcomes depending on their physical treatment of the ISM and dust physics (see, e.g., \citealt{Popping22}). Furthermore, the low number of sources with dust detections at $z\gtrsim6$ to compare with has strongly limited our understanding of dust evolution in the EoR.\\

The ALMA {\it Reionization Era Bright Emission Line Survey} (REBELS, \citealt{Bowens22}) program has significantly increased the number of star-forming galaxies with robust dust detections during the EoR. By searching for bright ISM cooling lines ([C\,{\sc{ii}}]158$\mu$m and [O\,{\sc{ii}}]88$\mu$m) in 40 UV-bright ($M_{\rm UV}<-21.5$) galaxies at $z\sim 6.5-9.5$, this survey has yielded a sample of 16 galaxies detected in the dust continuum, among which 15 galaxies are also detected in the [C\,{\sc{ii}}]158$\mu$m emission line (from now on referred to as [C\,{\sc{ii}}] line (\citealt{Inami22}, Schouws et al., in prep).
The fairly large number of sources detected in REBELS, together with the diversity probed in stellar masses and star formation rates (e.g. \citealt{Topping22}), represents a new benchmark for the study of galaxies with dust mass measurements at high-redshift (\citealt{Ferrara22,Sommovigo22a}, see also \citealt{Sommovigo22b}). 

\citet{Dayal22} provided the first comparison between this set of sources and galaxy evolution models. By using predictions from the \texttt{DELPHI} semi-analytical code, they showed that their model struggles to reproduce the observed dust masses in \citet{Sommovigo22a}, especially when dealing with galaxies in the low-mass end of the sample ($\log(M_*/{\rm M_\odot})\sim 9$, $M_*$ from Stefanon et al., in prep.). In particular, for such galaxies, dust scaling relations were only reproduced with their “maximal dust mass” model, which relies on extreme assumptions concerning the dust evolution (e.g., grain growth timescale below 1 Myr, no dust destruction by SN shocks).
Different conclusions are reached by \citet{DiCesare23}, who investigated the same galaxies using \texttt{dustygadget} simulations (\citealt{Graziani20}). In this case, predictions instead show an overestimation of the dust-to-stellar mass ratios relative to the data for galaxies at the high-mass end ($\log(M_*/{\rm M_\odot})>9.5$) of the sample. In turn, this suggests either an overproduction of dust in the simulated galaxies or an overestimation of the dust temperature (and hence an underestimation of the dust mass) in the data. However, the discrepancy in the results of \citet{Dayal22} and \citet{DiCesare23} can (at least) partly be attributed to the use of different sets of stellar and dust masses by \citet{DiCesare23} (from \citealt{Topping22} and \citealt{Sommovigo22b}, respectively).

Despite the progress in the characterization and understanding of dusty, $z\gtrsim6$ sources, there is still a large amount of information that is lacking for these objects.
Until now, our comprehension of the dust evolution in the EoR has been strictly limited to the analysis of the dust mass and dust-to-stellar mass ratio as a function of the stellar mass in different galaxies (e.g. \citealt{Dayal22}, see also \citealt{Algera23}). Fundamental diagnostic quantities for dust evolution, such as the dust-to-gas and dust-to-metal ratios, have been inaccessible and this prevented us from strictly pinning down the role of the different dust formation and destruction mechanisms in the high-redshift Universe (see \citealt{Schneider23} for a recent review).
However, the advent of JWST has opened up new and exciting possibilities in the investigation of these galaxies. 
Aside from the improvement in the photometric sampling towards the rest-frame optical and NIR, which can provide a much better characterization of the galactic SED and improve the stellar mass and SFR determination, the JWST/NIRSpec spectrograph (\citealt{Jakobsen22}) allows intermediate resolution spectroscopy ($\mathcal{R}\sim 1000 - 2700$) in a wavelength regime comprising several of the most prominent optical nebular emission lines (e.g., [OII]$\lambda\lambda 3726,3729$, [OIII]$\lambda\lambda 4960,5008$, H$\beta$, H$\gamma$).
In turn, this allows robust $\log({\rm O/H})+12$ metallicity determinations through strong-line diagnostics (see \citealt{Kewley08,Kewley19,Maiolino19}) and possibly also through the direct-${\rm T_e}$ method, in the case of detection of the auroral [OIII]$\lambda 4363$ line. 
Despite of the significant intrinsic uncertainty in these pioneering high-redshift metallicity measurements (especially in the case of strong-line diagnostics), these are fundamental for many reasons.
First, they allow to explore the upper mass end of the galactic scaling relations, such as the mass-metallicity relation (MZR), which is extremely poorly sampled above $10^{9-9.5} \, {\rm M_\odot}$ at $z\gtrsim6$ (\citealt{Langeroodi22,Nakajima23,Curti23}). 
Moreover, ISM abundances from JWST spectroscopic observations allow us to use dust diagnostics other than the dust-to-stellar mass ratio (hereafter, DtS) to study the dust evolution in high-redshift sources. In fact, metallicity-dependent calibrations of the galactic gas mass from the [C\,{\sc{ii}}] line luminosity (\citealt{Heintz21,Heintz22,Vizgan22}) open up the possibility to estimate the ISM gas and metal mass in galaxies and therefore probe other fundamental dust relations, such as the dust-to-gas ratio (DtG) and dust-to-metal ratio (DtM) as function of the galactic metallicity.
These dust-related diagrams have been crucial in assessing the role of mechanisms contributing to dust evolution in different galaxy types from the local universe up to redshift 4 ( \citealt{RemyRuyer14,Schneider16,DeVis17,Ginolfi18,Peroux20,Nanni20,DeLooze20,Galliano21}, see also \citealt{Galliano18} for a review).
Additionally, looking at the DtG and DtM in combination with the gas-independent DtS measurement may be crucial to directly test the goodness of the above-mentioned [C\,{\sc{ii}}]-gas mass calibrations. In fact, the latter will play a crucial role in the study of the high-redshift Universe since they are usually the only way to estimate the gas mass in objects at such redshifts (due to the difficulties in getting other emission lines and/or the Rayleigh-Jeans end of the FIR SED; see, e.g. \citealt{Heintz21} and references therein).\\

Therefore, the aim of this paper is to provide an overview of the possible scenarios of metal and dust evolution that may arise from the analysis of REBELS galaxies with [C\,{\sc{ii}}] and dust detections in the light of future JWST data.
Starting from galactic chemical evolution models including the most recent stellar yields and dust prescriptions (SN and AGB dust production, dust grain growth, dust destruction) from the literature, we implement the galactic star formation histories (SFHs) as determined by \citet{Topping22} for each specific REBELS source and test different setups of the gas flow history and dust evolution parameters to model various scenarios for the metal and dust evolution, respectively.  
In this way, we provide insights into the dominant dust production mechanisms and the efficiency of dust production of $z\gtrsim6$ galaxies, currently under the assumption of different chemical enrichment scenarios, which will be constrained in the future by means of rest-frame optical ISM line spectroscopy.

The results of this paper therefore stress the urgent need for JWST spectroscopic follow-up of galaxies observed in the FIR with ALMA. JWST has started to acquire low-resolution ($\mathcal{R}\sim 100$) data for several REBELS sources (GO program ID: 1626, P.I. Stefanon), but we need larger samples and higher spectral resolution to robustly determine chemical abundances.
Only by building statistically significant samples of JWST-observed galaxies, we can constrain the general metallic content of high-z dusty galaxies and therefore give tighter constraints on the dust regulation mechanisms during the EoR.\\

The paper is organised as follows: 
in Section \ref{s:data} we briefly present the data products from the ALMA REBELS program used throughout this paper, focusing on the [C\,{\sc{ii}}] luminosity-gas mass calibration adopted.
In Section \ref{s:models}, we illustrate the model framework adopted for our predictions, specifying the details about the different setups used to produce different metal and dust scenarios. 
In Section \ref{s:curr_constr}, we show what the shortcomings of current data constraints are in deciphering dust evolution at high-redshift, while 
in Section \ref{s:results1} we show our model predictions for different metal and dust enrichment scenarios that may arise in REBELS targets. 
Finally, in Section \ref{s:conclusion} we draw our conclusions.

\section{Data}
\label{s:data}

In this paper, we take advantage of a subsample of 13 galaxies for which both the [C\,{\sc{ii}}]$158 \mu$m line and the 1900 GHz continuum were detected (at $>$5$\sigma$ and $>$3$\sigma$, respectively). 
For a full analysis of the [C\,{\sc{ii}}]-detected sources, we refer the reader to Schouws et al. (in prep.), while the dust continuum detections are presented in \citet{Inami22}.\\

The stellar masses ($M_*$) adopted for these REBELS targets are presented in \citet{Topping22}. In particular, we make use of the estimations given by the SED-fitting code {\it Prospector} (\citealt{Johnson21}) under the assumption of a
non-parametric SFH. Such SFHs are particularly well-suited to model old stellar populations that may be present, avoiding the "outshining problem" caused by more recent bursts of star formation (see \citealt{Topping22} and references therein). Our selected subsample spans a stellar mass range of $\log({\rm M_*/M_\odot}) \simeq 9.0 - 10.3$, i.e. similar to the one encompassed in the full sample.

Among the different dust estimations available (e.g. \citealt{Ferrara22,Sommovigo22a}), we use the dust estimates presented in \citet{Sommovigo22b}, which are based on the methodology shown in \citet{Sommovigo21,Sommovigo22a}\footnote{the dust masses provided in \citet{Sommovigo22b} are the ones described in \citet{Sommovigo22a} but recomputed according to the stellar masses from \citet{Topping22} non-parametric SFHs}. This method allows us to get fairly robust estimates of the dust temperature and mass by means of only one FIR continuum data point with Bayesian methods, by assuming the well established SFR-[C\,{\sc{ii}}] (\citealt{DeLooze14}, see \citealt{Schaerer20,Ferrara22} for high-z probes) and Schmidt-Kennicutt relations \citealt{Kennicutt98}) and imposing a metallicity-dependent DtG relation (e.g. \citealt{RemyRuyer14}). We refer to \citet{Sommovigo22a} for a detailed explanation on the employed methodology.
It is worth mentioning that for two REBELS galaxies at the high-mass end of the sample (REBELS-25 and REBELS-38) additional FIR continuum measurements around $\sim 90\,\mu$m rest-frame are available (\citealt{Algera23}), allowing for more robust constraints on the dust SED and thus on dust temperature and mass. We comment on the effects of these new measurements throughout the paper.

To make predictions on dust scaling relations including DtG or DtM we need to have estimates of the amount of gas available in the analyzed sources. 
However, measurements through usual indicators (e.g. CO lines, FIR continuum tail) are extremely hard to get at $z\gtrsim5$, preventing analyses on high-z galaxy samples (see, e.g., \citealt{Heintz22}). Therefore, the most promising approach to
obtain a census of the gas content in EoR galaxies is to use a suitable tracer of the cold neutral gas phase, such as the [C\,{\sc{ii}}] line. 
In this paper, we adopt the calibration by \citet{Heintz21} to estimate the HI gas content relying on the [C\,{\sc{ii}}] line luminosity and the metallicity:
\begin{multline}
    \log\bigg(\frac{L_{\rm [C\,{\sc{II}}]}}{{\rm L_\odot}}\bigg)=\\
    =(0.87\pm 0.09) \log\bigg(\frac{Z}{Z_\odot}\bigg) - (1.48 \pm 0.12) + \log\bigg(\frac{M_{\rm HI}}{{\rm M_\odot}}\bigg),
    \label{eq:Heintz21}
\end{multline}
    
where $Z/Z_\odot$ is the relative solar metallicity with $12+\log({\rm O/H})=8.69$ for $\log(Z/Z_\odot)=0$ (\citealt{Asplund09}) and $M_{\rm HI}$ and $L_{\rm [C\,{\sc{II}}]}$ are in units of ${\rm M_\odot}$ and ${\rm L_\odot}$, respectively. 
It is worth reminding that the metallicity $Z$ is a free parameter of Eq. \eqref{eq:Heintz21}, as metallicity measurements of any of the REBELS galaxies has not been published yet.
For ${\rm H_2}$ instead, we make use of the average [C\,{\sc{ii}}]-to-${\rm H_2}$ ratio found in the EoR galaxy simulations by \citet{Vizgan22}:
\begin{equation}
    \frac{M{_{\rm H_2}}}{\rm M_\odot} = (18 \pm 10)\, \frac{L_{\rm [C\,{\sc{II}}]}}{\rm L_\odot}.
    \label{eq:Vizgan22}
\end{equation}
Other works in the literature indicate larger [C\,{\sc{ii}}]-to-${\rm H_2}$ conversion factors (e.g. \citealt{Zanella18}, 
see also \citealt{Aravena23}). However, such calibrations are based on $z\lesssim2$ galaxies. Moreover, the difference between the conversion factors is around a factor of 1.5 which we found to have no implications for the studied trends presented in this work.
It is worth noting that the adopted calibrations were already tested in the framework of REBELS galaxies by \citet{Heintz22} by adopting the fundamental $M_*$-SFR-metallicity relation by \citet{Curti20}. However, such a relation was established at $z<3$. In this work, we also explore the effects of metallicity on the \citet{Heintz21} calibration, testing different metal enrichment scenarios that could emerge from JWST observations of REBELS targets.

\begin{table*}
    \centering
    \caption{Summary of the derived properties of the REBELS galaxies within our sample. $L_{\rm [C\,{\sc{II}}]}$ are from Schouws et al. (in prep). $M_*$ are from non-parametric SED fitting presented in \citet{Topping22}. $M_{\rm dust}$ are from \citet{Sommovigo22b}. $M_{\rm gas,{\rm [C\,{\sc{II}}]}}$ ranges are computed by summing the sum of HI and H$_2$ masses assuming metallicities between 0.1 $Z_\odot$ and $Z_\odot$.} 
    \begin{tabular}{c | c  c  c  c  c  c}
         \hline\\[-1.95ex]
         Source name & Redshift & $\log(L_{\rm [C\,{\sc{II}}]}/{\rm L_\odot})$ & $\log(M_*/{\rm M_\odot})$ & $\log(M_{\rm dust}/{\rm M_\odot})$ &  $\log(M_{\rm gas,{\rm [C\,{\sc{II}}]}}/{\rm M_\odot})$ &  $\frac{M_{\rm gas,{\rm [C\,{\sc{II}}]}}}{M_*+M_{\rm gas,{\rm [C\,{\sc{II}}]}}}$\\[0.2cm]
         \hline\\[-1.95ex]
         REBELS-05 & 6.496 & 8.84$^{+0.06}_{-0.05}$ & 10.09$^{+0.27}_{-0.44}$ & 7.18$^{+0.37}_{-0.32}$ & [10.52, 11.22] & [0.73,0.93]\\[0.15cm]
         REBELS-08 & 6.749 & 8.87$^{+0.07}_{-0.06}$ & 9.56$^{+0.31}_{-0.55}$ & 7.22$^{+0.35}_{-0.29}$ & [10.55, 11.25] & [0.91,0.97]\\[0.15cm]
         REBELS-12 & 7.349 & 9.00$^{+0.12}_{-0.10}$ & 10.34$^{+0.27}_{-0.42}$ & 7.30$^{+0.37}_{-0.32}$* & [10.69, 11.39] & [0.69,0.92]\\[0.15cm]
         REBELS-14 & 7.084 & 8.57$^{+0.15}_{-0.11}$ & 9.22$^{+0.45}_{-0.78}$ & 7.02$^{+0.33}_{-0.27}$ & [10.25, 10.93] & [0.91,0.98]\\[0.15cm]
         REBELS-18 & 7.675 & 9.03$^{+0.04}_{-0.03}$ & 10.22$^{+0.34}_{-0.58}$ & 7.31$^{+0.37}_{-0.33}$ & [10.72, 11.42] & [0.76,0.94]\\[0.15cm]
         REBELS-19 & 7.369 & 8.94$^{+0.14}_{-0.10}$ & 9.38$^{+0.33}_{-0.58}$ & 7.23$^{+0.34}_{-0.31}$ & [10.62, 11.33] & [0.95,0.99]\\[0.15cm]
         REBELS-25 & 7.306 & 9.20$^{+0.03}_{-0.03}\dagger$ & 10.27$^{+0.22}_{-0.42}$ & 7.56$^{+0.24}_{-0.33}$* & [10.88, 11.58] & [0.8,0.95]\\[0.15cm]
         REBELS-27 & 7.090 & 8.79$^{+0.05}_{-0.05}$ & 10.16$^{+0.23}_{-0.37}$ & 7.14$^{+0.32}_{-0.37}$ & [10.47, 11.17] & [0.67,0.91]\\[0.15cm]
         REBELS-29 & 6.685 & 8.74$^{+0.08}_{-0.07}$ & 10.04$^{+0.33}_{-0.49}$ & 7.11$^{+0.37}_{-0.32}$ & [10.42, 11.12] & [0.71,0.92]\\[0.15cm]
         REBELS-32 & 6.729 & 8.90$^{+0.05}_{-0.04}$ & 10.18$^{+0.28}_{-0.45}$ & 7.22$^{+0.35}_{-0.33}$ & [10.58, 11.28] & [0.72,0.93]\\[0.15cm]
         REBELS-38 & 6.577 & 9.23$^{+0.04}_{-0.04}$ & 10.37$^{+0.20}_{-0.29}$ & 7.48$^{+0.28}_{-0.30}$* & [10.91, 11.61] & [0.78,0.95]\\[0.15cm]
         REBELS-39 & 6.847 & 8.90$^{+0.15}_{-0.11}$ & 9.24$^{+0.40}_{-0.67}$ & 7.20$^{+0.32}_{-0.30}$ & [10.58, 11.28] & [0.96,0.99]\\[0.15cm]
         REBELS-40 & 7.365 & 8.69$^{+0.12}_{-0.09}$ & 10.22$^{+0.32}_{-0.54}$ & 7.08$^{+0.38}_{-0.32}$ & [10.37, 11.07] & [0.59,0.88]\\[0.1cm]
         \hline
    \end{tabular}\\[0.1cm]
    \label{tab:data_summary}
    {\it * additional estimates based on multiple band observations are also available for these sources (\citealt{Algera23})\\[0.05cm]
    $\dagger$ additional estimate by reanalysis of [C\,{\sc{ii}}] line spectral features: $\log(L_{\rm [C\,{\sc{II}}]}/{\rm L_\odot})=9.23^{+0.05}_{-0.05}$ (\citealt{Hygate23})}
\end{table*}

A summary of the derived properties for the REBELS galaxies subject of this study is shown in Tab. \ref{tab:data_summary}. For the gas masses and the gas fractions, we use the sum of the HI and H$_2$ masses inferred from [C\,{\sc{ii}}] line luminosity calibrations, and we show the range of values that we obtain if we assume a metallicity between 0.1$Z_\odot$ and $Z_\odot$.

\section{Modelling dust evolution in REBELS galaxies}
\label{s:models}

To study the origin of trends in observed dust and metal scaling relations inferred from future JWST observations, we use detailed chemical evolution models for galaxies including interstellar dust evolution.
To this aim, we adopt a significantly revised version of the \texttt{ Chemevol}\footnote{\url{https://github.com/zemogle/chemevol}} source code (\citealt{DeVis21}), in which we implemented updated recipes for elemental nucleosynthesis in stars and dust processes in the ISM.

In the next subsections, we go over the details of the model, explaining the star formation histories (SFHs) implemented for the different targets and the adopted prescriptions to predict the metal and dust scaling relations in REBELS galaxies.

\subsection{Chemical and dust evolution model}
\label{ss:general_model}

In our galactic evolution models we assume that galaxies start forming at redshift $z=25$, by primordial gas accretion from the intergalactic medium (IGM) which progressively accumulates onto the system, and the galaxy evolves suffering galactic winds.
As stated above, we assume gas accreted from the IGM to be devoid of both metals and dust. However, we also made tests in which reaccretion of the gas ejected through galactic winds also fuels subsequent gas infall onto galaxies, finding no significant differences in the model results.
The models also relax the instantaneous recycling approximation (IRA), i.e. they take into account stellar lifetimes in the process of chemical and dust enrichment. Therefore, realistic predictions are possible for different chemical abundance ratios and dust enrichment.

\subsubsection{Chemical evolution prescriptions}
\label{sss:general_chemical}

The evolution of a chemical element ${\rm i}$ in the ISM can be written as (see also \citealt{Matteucci12}):
\begin{equation}
	\dot{M}_{\rm i}= - \psi(t)\,X_{\rm i}(t) + R_{\rm i}(t) + \dot{M}_{\rm i,inf}(t) - \dot{M}_{\rm i,out}(t),
	\label{eq:chem_evo}
\end{equation}
where $\psi(t)$ is the star formation rate (SFR) and $X_{\rm i}(t)$ is the fraction of the element ${\rm i}$ in the ISM at the galactic evolutionary time $t$, which in our framework starts from 0 at redshift $z=25$.

The first term on the right-hand side of Eq. \eqref{eq:chem_evo} corresponds to the rate at which an element ${\rm i}$ is removed from the ISM due to the star formation process. 

$R_{\rm i}(t)$ (see \citealt{Palla20a} for the complete expression) takes into account the nucleosynthesis from low-intermediate mass stars (LIMS, $m < 8{\rm M_\odot}$), core collapse (CC) SNe ($m > 8{\rm M_\odot}$) and Type Ia SNe. 
The IMF assumed is the one from \citet{Chabrier03}, in order to be consistent with the stellar masses and SFHs adopted from \citet{Topping22}.
The stellar yields are taken from the most recent prescriptions of \citet{Ventura13,Ventura18,Ventura20,Ventura21} for LIMS and \citet{Limongi18} (set R300) for CC-SNe. For Type Ia SNe instead, we adopt the yields by \citet{Iwamoto99}, as extensively tested in chemical evolution works (see \citealt{Palla21} and references therein).
For the latter, we assume a $t^{-1}$ delay time distribution (\citealt{Totani08,Maoz17}), which is extensively used in models of chemical evolution (e.g. \citealt{Yates13,Cote16,Vincenzo17}).

The third term in Eq. \eqref{eq:chem_evo} is the gas infall rate, which in the framework of chemical evolution models is generally parametrised with an exponential decaying law, even for high-redshift galaxies (e.g. \citealt{Romano17,Palla20a,Koba23}):
\begin{equation}
    \dot{M}_{\rm i,inf}= C_{\rm inf}\, X_{\rm i,inf} \, e^{-t/\tau_{\rm inf}},
    \label{eq:infall}
\end{equation}
where $C_{\rm inf}$ is the normalization constant constrained to reproduce the total gas mass accreting onto the system $M_{\rm gas,tot}$, $X_{\rm i,inf}$ is the fraction of the element ${\rm i}$ in the infalling gas (with primordial composition) and $\tau_{\rm inf}$ is the typical infall timescale, which is chosen according the shape of the galactic SFH adopted (see \ref{ss:setup_model}). However, during this work we also test multiple gas accretion episodes in case of SFHs showing strong peaks in SFR (see \ref{ss:setup_model}).

The last term in Eq. \ref{eq:chem_evo} accounts for galactic winds. Here, we assume that the outflow is proportional to the SFR, as in this form:
\begin{equation}
    \dot{M}_{\rm i,out}=\omega_{\rm i}\, \psi(t),
\end{equation}
where $\omega_{\rm i}$ is the wind mass loading factor for the element ${\rm i}$. Since our main focus is to follow the general scaling relations in the modeled galaxies, we do not use differential winds, i.e. we assume the same loading factor $\omega\simeq3$ for the different chemical elements in the gas. The choice of such a loading factor allows to explain the extended [C\,{\sc{ii}}] halo observed in several redshift $z\simeq6$ galaxies in terms of a star-formation driven outflow, with similar [C\,{\sc{ii}}] halo features relative to REBELS target stacks (see \citealt{Pizzati20,Fudamoto22}). \\

Despite of careful choice of model prescriptions, motivated either by consistency with the derivation of the physical quantities of REBELS galaxies or by other observational proofs, 
we want to highlight that a certain level of uncertainty persists in the modelling assumptions in high-redshift galaxies.
For example, the treatment of stellar generations in the early stages of galaxy formation, with its assumptions on the IMF and the choice of stellar yields, have a non-negligible impact on the modelling results, leading to different conclusions on the evolution of the baryonic components in high-redshift galaxies (e.g. \citealt{Naab17} and references therein).

However, our choice of prescriptions remains robust accross our main goal, which is giving a first exploration of the parameter space in the explanation of dust-scaling relations, rather than drawing definitive conclusions on the evolution of EoR, UV-bright galaxies as those of the REBELS sample. 
In fact, the level of uncertainty in the treatment of dust evolution is even higher than the one for the rest of the baryonic physics, with severe limitation persisting even in the local Universe (e.g. \citealt{DeLooze20,Galliano21,Calura23}).

\subsubsection{Dust evolution prescriptions}
\label{sss:general_dust}

Here we apply the same formalism used in previous works which included the chemical evolution of dust in galaxies of various morphological types and at different redshifts (e.g., \citealt{Calura08,Gioannini17,Gjergo20,DeLooze20,Galliano21}). The equation used to determine the amount of dust at time $t$ is:
\begin{equation}
    \dot{M}_{\rm dust}= - \psi(t)\,X_{\rm dust}(t) + R_{\rm dust}(t) + \frac{M_{\rm dust}}{\tau_{\rm grow}} - \frac{M_{\rm dust}}{\tau_{\rm destr}} - \dot{M}_{\rm dust,out}(t).
    \label{eq:dust_evo}
\end{equation}
Similar to Eq. \eqref{eq:chem_evo}, $X_{\rm dust}(t)$ represents the dust fraction in the ISM. 

The first term on the right-hand side of Eq. \eqref{eq:dust_evo} accounts for the amount of dust destroyed by astration, i.e. incorporated into new stars. 

$R_{\rm dust}(t)$ accounts for stellar dust production, which in our model is from AGB stars (i.e. the final stage of a LIMS before the final white dwarf fate) and CC-SNe\footnote{both theoretical and observational arguments support that Type Ia SNe do not/negligibly contribute to the dust production budget (see \citealt{Palla20a} and references therein)}. For AGB stars, we use the dust yields of \citet{DellAgli17,Ventura18,Ventura20,Ventura21}, while for CC-SNe we use the dust yields from \citet{Marassi19} (set ROT FE). It is worth noting that both yield prescriptions are metallicity dependent. As detailed in \ref{ss:setup_model}, we allow the SN dust yields to be reduced by a factor $K$ to simulate the uncertain effect of the SN reverse shock, which can
destroy much of the dust produced in the same SN explosion (e.g. \citealt{Bianchi07,Bocchio16,Kirchschlager19,Kirchschlager23,Slavin20}). In fact, different models of this phenomenon lead to very different survival rates, from more than 50\% to much less than 10\% (see \citealt{Micelotta19,Schneider23} for a review).

The third term on the right side of Eq. \eqref{eq:dust_evo} accounts for dust accretion (or growth) in the cold ISM component. In this process, pre-existing dust seeds grow in size due to accretion of refractory elements on their surface (e.g. \citealt{Savage96,Hirashita11,Konstantopoulou22}), enabling a significant increase of the global dust mass (e.g. \citealt{Dwek98,Asano13,Mancini15,DeVis17,DiCesare23}). As in Eq. \eqref{eq:dust_evo} the dust growth rate is regulated by the accretion timescale $\tau_{\rm grow}$, which can be expressed as:
\begin{equation}
    \tau_{\rm grow}=\frac{\tau_0}{(1-{\rm DtM}(t)/{\rm DtM}_{\rm max})},
\end{equation}
where DtM$(t)$ is the dust-to-metal ratio at a time $t$, DtM$_{\rm max}$ is the maximum dust-to-metal ratio (0.6, see Appendix \ref{app:DtMmax}) and $\tau_0$ is the characteristic dust growth timescale. We look at this term in detail in \ref{ss:setup_model}.

The fourth term of Eq. \eqref{eq:dust_evo} accounts for dust destruction due to exploding SNe (both CC and Type Ia), whose shocks efficiently destroy interstellar dust grains and return the heavy elements locked up in dust grains back to the gas phase (see, e.g. \citealt{McKee89,Jones94}). Dust destruction is also expressed in terms of a grain destruction timescale 
$\tau_{\rm destr}$, which can be written as:
\begin{equation}
    \tau_{\rm destr}=\frac{M_{\rm gas}}{M_{\rm clear}\,SN_{\rm rate}}
\end{equation}
where $SN_{\rm rate}$ is the sum of the CC and Type Ia SN rates, and $M_{\rm clear}$ is the ISM mass completely cleared out of dust per SN. We discuss this last term in \ref{ss:setup_model}.

Finally, the last term of the Eq. \eqref{eq:dust_evo} accounts for the amount of dust lost due to galactic winds. As mentioned in \ref{sss:general_chemical}, we do not assume differential winds. Therefore, we assume that gas and dust are well mixed in galactic outflows.

\subsection{Model setup}
\label{ss:setup_model}

\begin{figure*}
    \centering
    \includegraphics[width=0.99\textwidth]{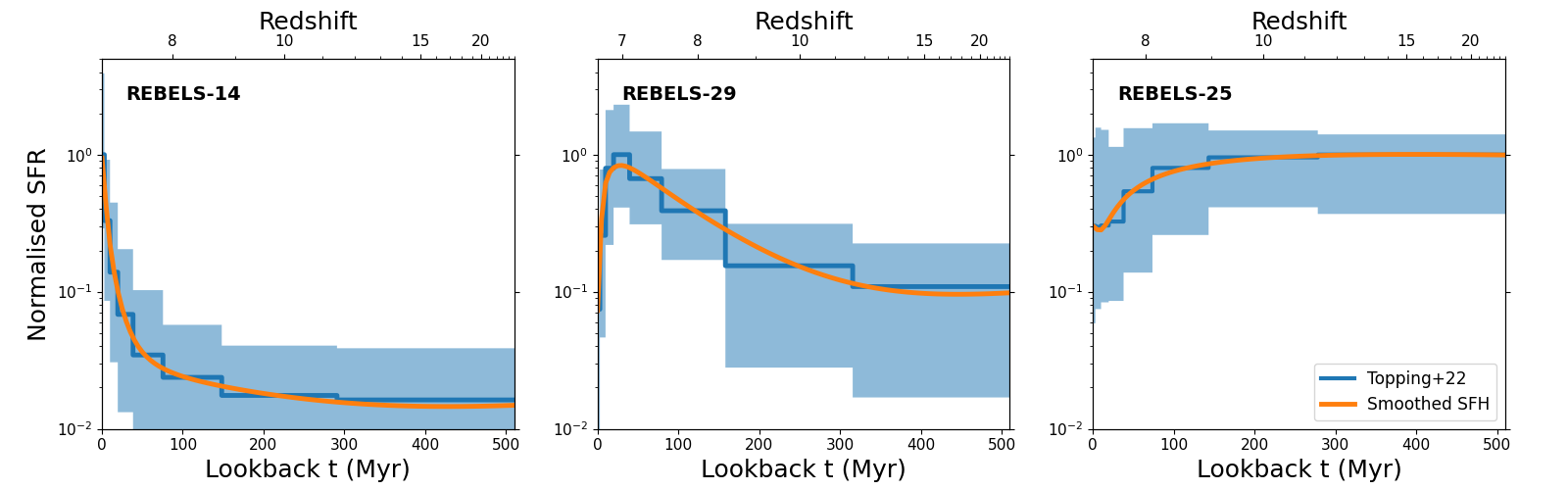}
    \caption{Non-parametric star formation histories (SFHs) for REBELS galaxies from \citet{Topping22}. Left, central and right panels show the SFH of REBELS-14, REBELS-29, and REBELS-25. The shaded areas represent the 1-$\sigma$ confidence interval for the SED-modelled SFHs. 
    The orange lines are the smoothed SFHs used as inputs for our galactic chemical evolution models.}
    \label{fig:SFHs}
\end{figure*}

As mentioned in Section \ref{s:data}, here we adopt the stellar masses from non-parametric SFHs presented in \citet{Topping22}. Therefore, the input SFHs for our models are taken from this set of data. 

In Fig. \ref{fig:SFHs} we show the SFHs (blue lines) from \citet{Topping22} for REBELS-14, 29 and 25, i.e. three galaxies with rather different SFHs and stellar masses (from $M_*\simeq 10^9 {\rm M_\odot}$ to $M_*\simeq 2\times 10^{10} {\rm M_\odot}$) within our sample. 
In this Figure, we also show the SFHs used as input for our chemical evolution models for each galaxy (orange lines). In fact, to avoid unrealistic spikes in the chemical enrichment, we smooth the \citet{Topping22} SFH by adopting non-parametric regression from the \texttt{PyQt-Fit} package\footnote{\url{https://pyqt-fit.readthedocs.io/en/latest/intro.html}}.
The resulting smoothed SFH is then multiplied by a renormalization factor to keep the total stellar mass the same as the one from the original non-parametric SFH.\\

To let the chemical evolution work, we also need ISM gas in our models to fuel star formation. Since variations in the gas mass are expected by varying the galactic metallicity (see Eq. \eqref{eq:Heintz21}), we test multiple setups for the infall mass, i.e. the total amount of gas accreting the galactic system (see, e.g. \citealt{Palla20a}). By having the galactic SFH fixed for each galaxy, the infall mass represents the main parameter with which we can vary the efficiency of star formation, i.e. the amount of gas converted into stars per unit time, which in turn has a direct effect on galactic metallicity, with lower efficiencies leading to lower levels of metal enrichment (e.g. \citealt{Matteucci12,Matteucci21} and references therein). As a further proof of this, despite the importance of galactic outflows, it was demonstrated by several works (e.g. \citealt{Jin11,Gioannini17,DeVis21}) that galaxies with large gas mass fractions require models with low efficiencies of star formation to match their different observables.

For what concerns the gas infall parameters (see Eq. \eqref{eq:infall}), we set the typical infall timescale $\tau_{\rm inf}$ as the evolutionary time at which half of the total mass of the galaxy has been assembled, to be in line with observed scaling relations between SFR and gas mass (see, e.g. \citealt{Kennicutt98}). 
Moreover, we also run additional models in which we allow for multiple gas accretion episodes in case the SFH shows distinct peaks in the SFR, i.e. when the SFR varies by at least a factor of 2 in a time interval less than $\Delta t<0.1$ Gyr. 
When these conditions are met, an additional infall episode modeled as in Eq. \eqref{eq:infall} starts, with the fraction of gas accreted by this episode being equal to the fraction of stellar mass formed in the time between the previous and the subsequent burst of SF. This choice is made in order to mimic a burst of star formation triggered by a large amount of gas that becomes available after, e.g. a galaxy merger. It should be noted that we tested the viability of the above model assumption by varying the specific conditions assumed for the SFR change factor and the time interval $\Delta t$.\\

\begin{figure*}
    \centering
    \includegraphics[width=0.99\textwidth]{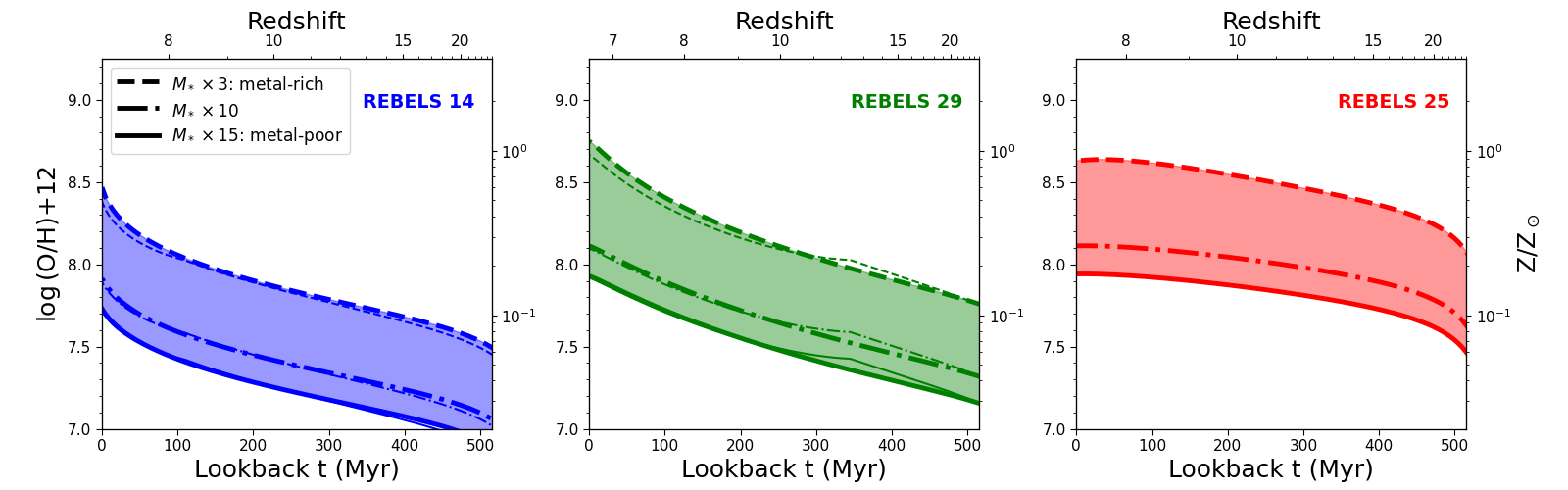}
    \caption{Evolution of the metallicity (both in terms of the oxygen abundance $\log$(O/H)$+12$ and solar scaled metallicity Z/Z$_\odot$) in modelled REBELS galaxies assuming different gas infall masses ($M_{\rm gas,tot}=\times 3,\, \times 10,\, \times 15$) and therefore metallicity evolution scenarios (see also Tab. \ref{tab:param_chem}). Left, central and right panels show the evolution for REBELS-14, REBELS-29 and REBELS-25. The thin lines in the left and central panels represent models in which multiple gas accretion episodes are allowed, as described in \ref{ss:setup_model}. }
    \label{fig:metallicities}
\end{figure*}

In Fig. \ref{fig:metallicities}, we show the ISM metallicity $Z$ evolution for the three REBELS galaxies with SFHs shown in Fig. \ref{fig:SFHs} assuming different gas infall masses, namely 3, 10 and 15 times the stellar mass $M_*$ computed for that REBELS target (see Appendix \ref{app:parameters} for a summary table on the prescriptions adopted).
We stress that such an exploration of the parameter space is necessary due to the lack of metallicity measurements in REBELS galaxies (see also Section \ref{s:intro}). In Fig. \ref{fig:metallicities}, the different model tracks and the colored areas in between want to highlight the large discrepancy in metallicity evolution by having different efficiencies of star formation, which in turn are caused by the different gas accretion histories assumed in the models.
We indeed clearly see that by reducing the gas infall mass we monotonically increase the metallicity at a given evolutionary time. 
However, the amount of gas mass accreted to the galaxy has a minor effect on the shape of metallicity evolution in each of the panels in Fig. \ref{fig:metallicities}. This happens because of the fixed SFHs for the model galaxies (see Fig. \ref{fig:SFHs}). This is especially true for the enrichment of oxygen (adopted as global metallicity indicator due to its mass dominance over other metals), which closely traces the star formation rate in galaxies (see e.g. \citealt{Maiolino19} and references therein), at variance with other elements for which the production is either largely dependent on the initial metallicity of a stellar population (e.g. nitrogen, \citealt{Romano10,Romano17}) or for which the chemical enrichment by a stellar population can be significantly delayed with respect to the time of its birth (e.g. iron, \citealt{Matteucci12} and references therein).

We also note that the models with the largest amount of gas infall $M_{\rm gas,tot}=15\, M_*$) show an average level of chemical enrichment around 1/10 solar ($Z_\odot=0.014$, \citealt{Asplund09}), which is indicative of a metal-poor scenario. On the other hand, models with the lowest amount of gas accreted onto the galaxies ($M_{\rm gas,tot}=3\, M_*$) have metallicities around the solar value (see Fig. \ref{fig:metallicities}, central and right panels). Therefore, we can label such models as indicative of a metal-rich scenario. 
We also check that these model setups lead to $\lq\lq$realistic" efficiencies of star formation in the framework of chemical evolution models. We find values in between $0.1$ Gyr$^{-1}$ and $10$ Gyr$^{-1}$, which are typical in chemical evolution models from local up to high-redshift galaxies (e.g. \citealt{Gioannini17,Gjergo20,Palla20b,Palla20a,Palla22,DeVis21,Galliano21}).
Therefore, we will consider these two scenarios for our model galaxies to test the impact of metallicity variations on galactic dust evolution.

We also show the metallicity evolution for models where multiple gas accretion events are included in case of a star formation burst added to the SFHs from \citet{Topping22}  (see Fig. \ref{fig:metallicities}, thin lines). The difference with the standard models with a single gas accretion episode is small, and therefore the adoption of different setups has negligible impact also on the evolution of the dust quantities.
For this reason, in the rest of the paper we look at the results for the models with single gas infall episodes only.

\subsubsection{Dust evolution scenarios}
\label{sss:dust_setup}

To probe different scenarios of dust evolution in EoR galaxies, we test various prescriptions for stardust production, dust growth, and SN dust destruction.

As described in \ref{sss:general_dust}, we adopt the dust yields from \citet{DellAgli17,Ventura18,Ventura20,Ventura21} for AGB stars, while for CC-SNe we use the dust yields from \citet{Marassi19}. For these latter, we adopt the set ROT FE, which uses the same chemical network and explosion setup as the adopted SN yields from \citet{Limongi18} (set R300). Therefore, our dust yields match perfectly with the adopted chemical yields. 
This is at variance with a large number of previous works on dust evolution, assuming dust and chemical yields from different stellar models or using scaling factors (e.g. \citealt{Calura08,Zhukovska14,Gioannini17,DeVis17,DeVis21}, however see also e.g. \citealt{Valiante14,Mancini15,Ginolfi18,Graziani20} for a consistent use of chemical and dust yields).
However, SN dust yields are affected by severe uncertainties due to the largely unknown effect of the reverse shock (\citealt{Gall18,Micelotta19}), which is not included in the SN dust formation calculations (\citealt{Marassi19}, see however \citealt{Bianchi07} for previous yield grids). 
For this reason, we allow the reduction of the SN dust yields by a factor $K=1,3,10$ to mimic the effect of the SN reverse shock and test its effect in the dust scaling relations for these galaxies.

For what concerns the dust accretion timescale, this is dependent on a characteristic timescale $\tau_0$. Among the several formulations in the literature, we use the one from \citet{Mattsson12}:
\begin{equation}
    \tau_0=\frac{M_{\rm gas}}{\xi\, Z(t)\, \psi(t)},
    \label{eq:Mattsson12}
\end{equation}
where $\xi$ is a free parameter, whose expected value is very uncertain, with results from local galaxies ranging from values of a few hundreds to greater than three thousand (e.g. \citealt{DeLooze20,Galliano21,DeVis21,Calura23}). Therefore, we let this parameter vary in our models by testing values of $\xi=200,1000,3000$. The two extreme values are chosen in order to replicate the conclusions by \citet{DeLooze20} (low $\xi$ values) and \citet{Galliano21} (high $\xi$ values).

Finally, the dust destruction rate in galaxies depends critically on the $M_{\rm clear}$ parameter, which indicates the ISM mass completely cleared out of dust in the ISM. Throughout this paper, we assume the formulation of \citet{Priestley22}:
\begin{equation}
    M_{\rm clear}=\frac{M_0}{1+Z/(0.14\, Z_\odot)},
    \label{eq:Priestley22}
\end{equation}
where $M_{\rm clear}$ and $M_0$ are expressed in ${\rm M_\odot}$. Here, we allow $M_0$ to vary to test different dust destruction efficiencies. In particular, we set $M_0=2\times10^3,\,8\times10^3,\,1.5\times10^4\, {\rm M_\odot}$. The intermediate value resembles the one adopted in \citet{Priestley22}, which is set in order to have $M_{\rm clear}\sim 1000 {\rm M_\odot}$ under local ISM conditions (e.g. \citealt{Jones94,Hu19}). However, there are arguments either in favour of lower values (e.g. \citealt{MartinezGonzales19,Priestley21}) or higher ones (e.g. \citealt{Calura23}) which motivate our choice of varying $M_0$.\\

Thanks to these parametrisations, in this paper we concentrate on three different scenario for dust evolution in REBELS galaxies, which we summarise below (see also Appendix \ref{app:parameters}):
\begin{itemize}
    \item {\it dust-poor scenario}: in this scenario, we reduce the SN dust yields by a factor $K=10$. The dust growth parameter $\xi$ is set to $200$ (lowest value), while $M_{0}$ for dust destruction to $1.5\times 10^4\, {\rm M_\odot}$ (highest value);
    \item {\it intermediate-dust scenario}: here, we reduce the SN dust yields by a factor $K=3$. The dust growth parameter $\xi$ and the dust destruction scaling $M_{0}$ are set to the intermediate values among those tested, i.e. $1000$ and $8\times10^3\,{\rm M_\odot}$, respectively;
    \item {\it dust-rich scenario}: in this case, the SN dust yields are the same as the ones presented in \citet{Marassi19}. The dust growth parameter $\xi$ is set to $3000$ (highest value), while $M_{0}$ for dust destruction to $2\times 10^3\, {\rm M_\odot}$ (lowest value).
\end{itemize}

However, we also test "mixed" dust scenarios, i.e. models where dust process recipes favouring or disfavouring the dust mass buildup are mixed together (see Appendix \ref{app:moremix}). In turn, this allows us to probe in more detail the relative influence that the different dust processes have on shaping the dust scaling relations.

\begin{figure*}
    \centering
    \includegraphics[width=0.99\textwidth]{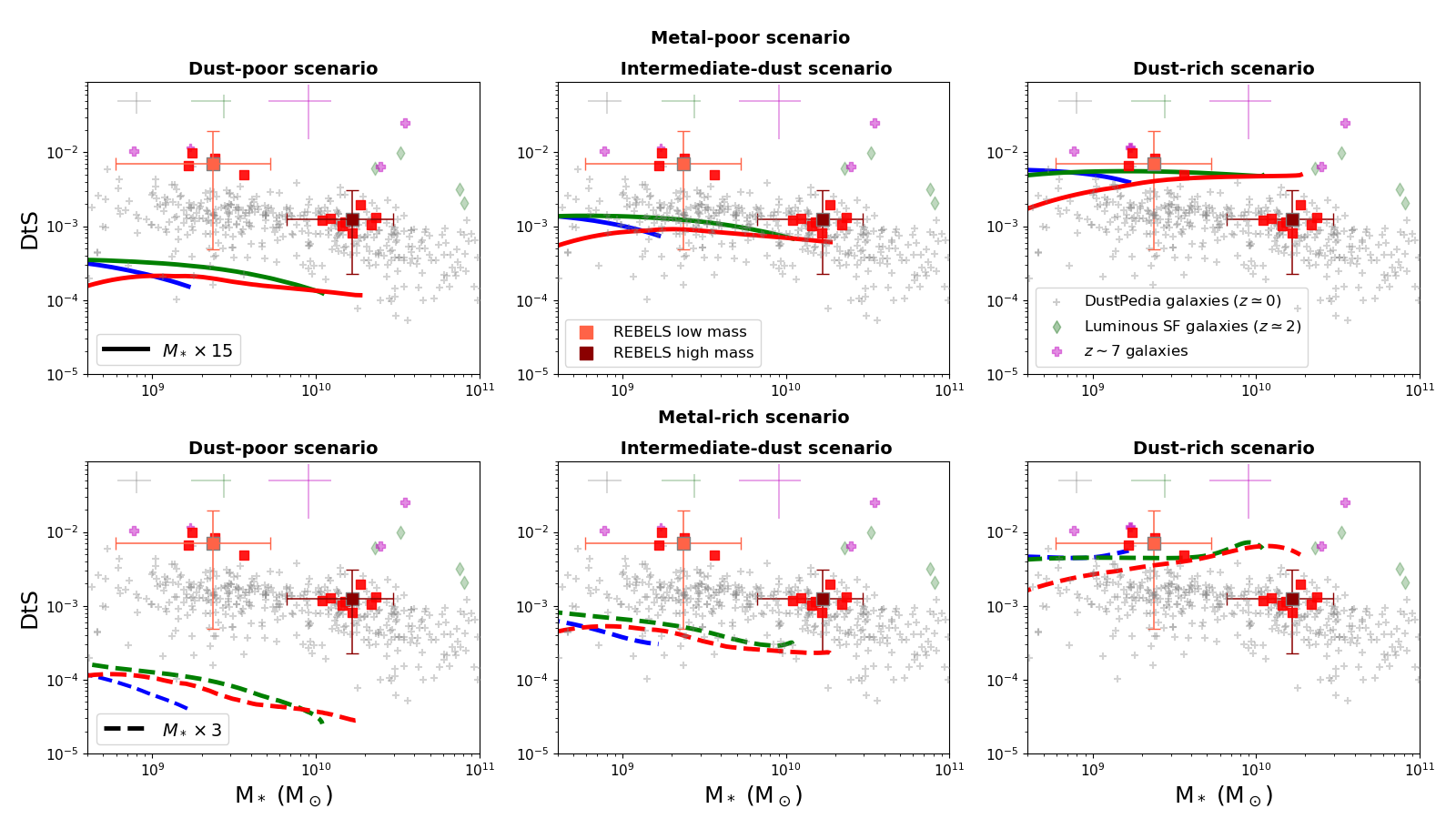}
    \caption{Evolution of the dust-to-stellar (DtS) mass ratio as a function of galactic stellar mass $M_*$ predicted by models for REBELS-14, REBELS-29 and REBELS-25 for different metallicity and dust evolutionary scenarios. Top panels show the results for galaxy evolution models assuming a metal-poor scenario, while lower panels for models assuming a metal-rich scenario (see also Tab. \ref{tab:param_chem}). 
    Left, central and right panels indicate different dust evolutionary scenarios, i.e. dust-poor, intermediate-dust and dust-rich, respectively (see also Tab. \ref{tab:param_dust}). 
    The color code for the galaxies with different star formation histories is the same as in Fig. \ref{fig:metallicities}. 
    Small red squares are REBELS data for individual galaxies, whereas light orange and dark red large squares with errorbars represent low and high stellar mass REBELS galaxy bins with the average error within the bin, respectively. Additional data are from \citet{Casasola20} (DustPedia galaxies), \citet{Shapley20} (luminous SF galaxies), \citet{Watson15,Hashimoto19,Reuter20,Bakx21,Fujimoto22} ($z\sim 7$ galaxies). Average errorbars for these data sets are indicated in the top-left part of each panel.}  
    \label{fig:DtS_Mstar}
\end{figure*}

\section{Current constraints: dust evolution without metallicity information}
\label{s:curr_constr}

In this Section, we show what the current data products can reveal about the dust evolution in the REBELS galaxy sample. Until now, our
understanding of the dust evolution in the EoR has been limited to few observational constraints, i.e. the galactic dust mass and stellar mass (e.g., \citealt{Watson15,Hashimoto19,Reuter20,Algera23}). In the following, we present how these constraints can be interpreted in light of the chemical and dust evolution models presented in Section \ref{s:models}.

In Fig. \ref{fig:DtS_Mstar}, we show a common diagnostic when looking at high-redshift galaxies, i.e. the DtS evolution as a function of the stellar mass $M_*$. 
In particular, we compare the results of the galaxy models shown in Fig. \ref{fig:metallicities} (REBELS-14, 29, 25) with the REBELS data binned according to their stellar mass (REBELS-8, 14, 19, 39 for the low stellar mass bin, REBELS-5, 12, 18, 25, 27, 29, 32, 38, 40 for the high-mass one). 
The choice of comparing the three individual galaxy models with binned data is justified by the fact that REBELS-14 show a SFH that is quite typical of the low stellar mass bin, while REBELS-29 and 25 have two SFHs that are recurrent in the high stellar mass bin. In addition, this choice allows us to better show the typical evolution of an individual galaxy for different dust evolutionary setups.
As mentioned in \ref{ss:setup_model}, in Fig. \ref{fig:DtS_Mstar} we only display the results for models with the largest and smallest amounts of gas mass accretion, i.e. $M_{\rm gas,tot}=15\, M_*$, indicative of a metal-poor chemical evolution scenario (Fig. \ref{fig:DtS_Mstar} upper panels), and $M_{\rm gas,tot}=3\, M_*$, indicative of a metal-rich scenario (Fig. \ref{fig:DtS_Mstar} lower panels).

By looking at the Figure, it can be noted that the uncertainty related to the dust and stellar mass in REBELS galaxies (light orange and dark red errorbars indicate typical uncertainties in the low and high mass bins) prevents to draw definitive conclusions on the efficiency of dust buildup in our target sample. 
Only the dust models assuming relatively slow dust buildup (the dust-poor scenario in \ref{sss:dust_setup}) in Fig. \ref{fig:DtS_Mstar} left panels can be ruled out due to their very low DtS predictions at all stellar masses, in agreement with conclusions drawn in previous work (e.g. \citealt{Dayal22}). 
The other setups (Fig. \ref{fig:DtS_Mstar} central and right panels) show in general values that are compatible with current observations: galaxies in the lowest $M_*$ bin show a preference for very efficient dust buildup, while REBELS galaxies with large stellar masses seem to support a slower dust buildup scenario.
In fact, as can be noted in Tab. \ref{tab:data_summary}, the dust masses show a rather flat trend compared to stellar mass, which instead can vary up to an order of magnitude: therefore, a decreasing DtS ratio with stellar mass is observed for REBELS galaxies. This is in reasonable agreement with the trend observed in local galaxies (from \citealt{Casasola20}, small grey crosses), which shows, however, a shallower relation.

\begin{figure}
    \centering
    \includegraphics[width=0.95\columnwidth]{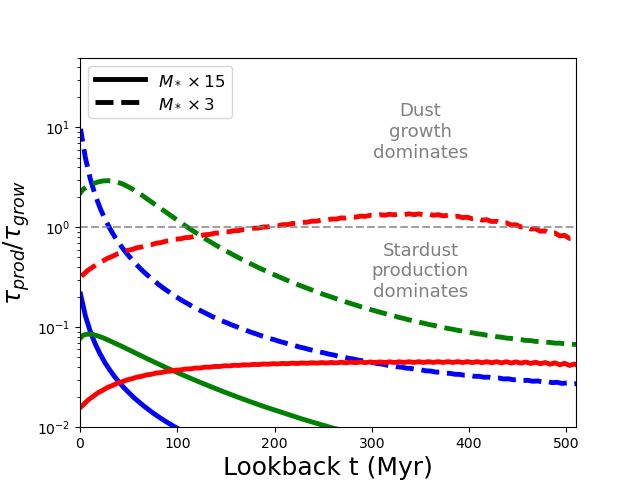}
    \caption{Evolution of the ratio between stardust production and dust growth timescales 
    for gas-rich (i.e. more metal-poor, solid lines) and gas-poor (i.e. more metal-rich, dashed lines) models in the dust-rich scenario (see \ref{sss:dust_setup}). The color code for the galaxies with different star formation history is the same as in Fig. \ref{fig:metallicities}. The thin, grey dashed line indicates a ratio equal to 1, corresponding to the equilibrium between the two dust processes.}
    \label{fig:ratio_timescales}
\end{figure}

The DtS vs. $M_*$ diagnostic diagram (as well as the equivalent $M_{\rm dust}$ vs. $M_*$ relation) also definitely struggles in discriminating the level of metal/chemical enrichment in galaxies. 
Despite a difference in metallicity of almost an order of magnitude between the different scenarios of metal enrichment (see Fig. \ref{fig:metallicities}), Fig. \ref{fig:DtS_Mstar} upper and lower central panels show a rather small difference (less than a factor of 2) in the predicted dust masses.
This happens because of the physical quantities affecting the SN dust destruction rate, which in such models is only secondary to stardust production. The destruction timescale is both proportional to the gas mass and the metallicity $\tau_{\rm destr}\propto M_{\rm gas} (1+Z)$. Since models with larger gas masses have lower metallicity (and vice versa) the two effects tend to cancel out, with the gas mass being only slightly more influential than metallicity.
The difference in the model predictions is even smaller between Fig. \ref{fig:DtS_Mstar} upper and lower right panels, where the DtS overlap for models with different metallicities.
The almost identical predicted dust masses for metal-rich and metal-poor model galaxies is not only a problem from the point of view of galaxy metal enrichment at high-redshift, but also for the understanding of the relative contributions between "dust-cycle" processes in the buildup of the dust mass. To show an example, in Fig. \ref{fig:ratio_timescales} we plot the ratios between stardust production and grain growth timescales\footnote{for dust timescale we mean the ratio between the total dust mass and the respective rate (of stardust production, dust growth or dust destruction). The lower the dust timescale, the larger is the importance of a dust process in regulating the dust budget.} (both dominating over destruction in this case) for the dust-rich models presented in Fig. \ref{fig:DtS_Mstar} right panels. As expected, metal-poor models (solid lines) exhibit a dominance of stardust production over dust growth at all galactic evolutionary times, since the lack of metals in the ISM halts the metal accretion on dust seeds (e.g. \citealt{Mattsson12,Asano13}). On the other hand, metal-rich models (dashed lines) show that the dust growth rate can be comparable or larger than stardust production rate (see, e.g. \citealt{Mancini15,Graziani20,DiCesare23}). The trends also resemble the SFHs shown in Fig. \ref{fig:SFHs}, and therefore highlight the role of the SFR evolution together with the metallic content in shaping this ratio.\\

In conclusion, the current constraints we have at our disposal for high-redshift galaxies are not sufficient to give a compelling view on galactic dust evolution in the early Universe.
Information on the galactic metallicity is fundamental to start overcoming such limitations.
Metallicity itself allows to better constrain the metal enrichment and gas infall history, while comparison of the metallicity and dust constraints tells us how efficiently metals are being locked up in dust grains and can give an indication of what grain formation processes are needed to reproduce the observations. 
In this way, we can constrain the metal and dust evolutionary scenarios in more detail similar to studies in local up to intermediate-redshift ($z<4$) galaxies (e.g. \citealt{RemyRuyer15,DeLooze20}).


\section{Dust vs. metallicity: predictions by models}
\label{s:results1}

In this Section, we investigate several metal and dust enrichment scenarios that will become possible to constrain with future spectroscopic analyses of JWST data of REBELS and other high-z galaxies. 
In particular, we focus on the predictions of various dust scaling relations under the assumption of different dust and metallicity scenarios tested in this work. 
In doing this, we also provide a first important test for recently inferred, metallicity dependent $L_{\rm [C\,II]}-M_{\rm{gas}}$ calibrations for high-redshift galaxies.
All these findings are then framed in the context of the new mass-metallicity relations at $z>4$ obtained by means of JWST large programs in order to provide the first suggestions on the most likely evolutionary pictures.

\subsection{Dust scaling relations}
\label{ss:results_dustrel}

In the following, we look at the three most common dust diagnostic diagrams used for the analysis of dust evolution at lower redshift, i.e., the dust-to-gas (DtG), dust-to-stellar (DtS) and dust-to-metal (DtM) mass ratios as a function of the metallicity (in terms of the oxygen abundance $12+\log({\rm O/H})$). 
The results of these predictions are compared with the dust, stellar, and gas masses presented in Section \ref{s:data} (see Tab. \ref{tab:data_summary}) at different metallicities. 

As done in Fig. \ref{fig:DtS_Mstar}, we bin the data according to the stellar mass of the galaxies as computed by \citet{Topping22}.
It is worth mentioning that this choice of data binning leads to a smaller scatter than the average error for the studied quantities (see Fig. \ref{fig:DtS_Mstar}).
In addition, here we group the galactic evolution models for individual REBELS sources accordingly, to ensure direct comparison between models and data of similar stellar mass and SFH.

\subsubsection{Metal-poor scenario}
\label{sss:metal_poor}

We start by looking at the results from the models with the largest amount of gas accreting onto the galaxies, i.e. $M_{\rm gas,tot}=15\, M_*$, indicative of a metal-poor scenario.

\begin{figure*}
    \centering
    \includegraphics[width=0.99\textwidth]{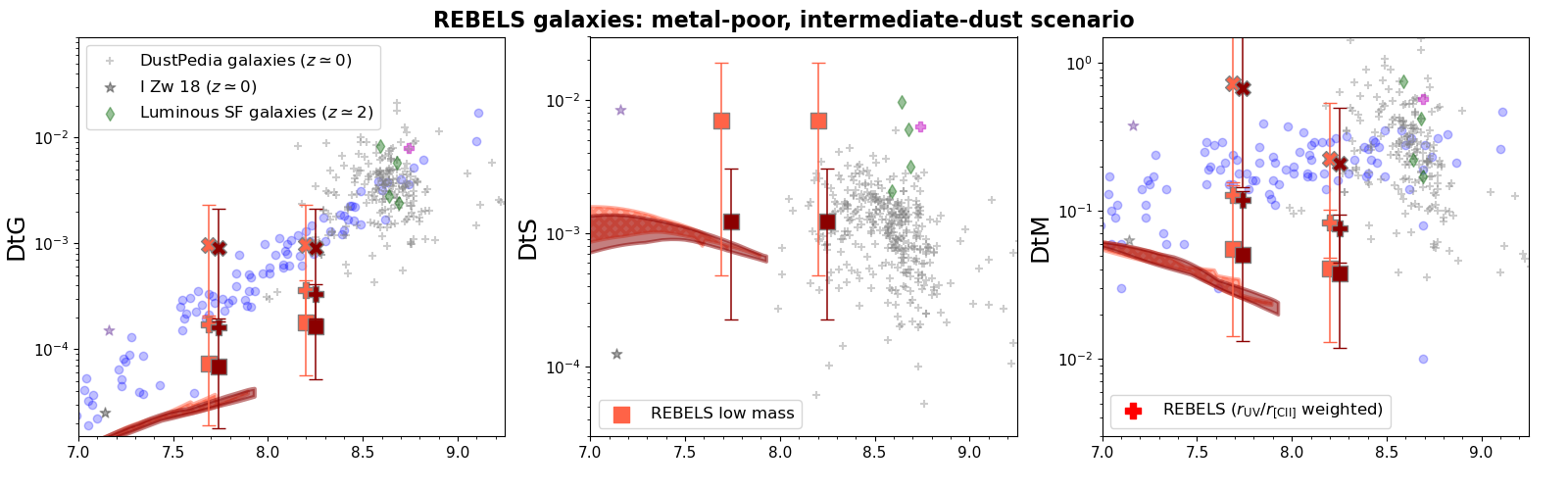}\\
    \includegraphics[width=0.99\textwidth]{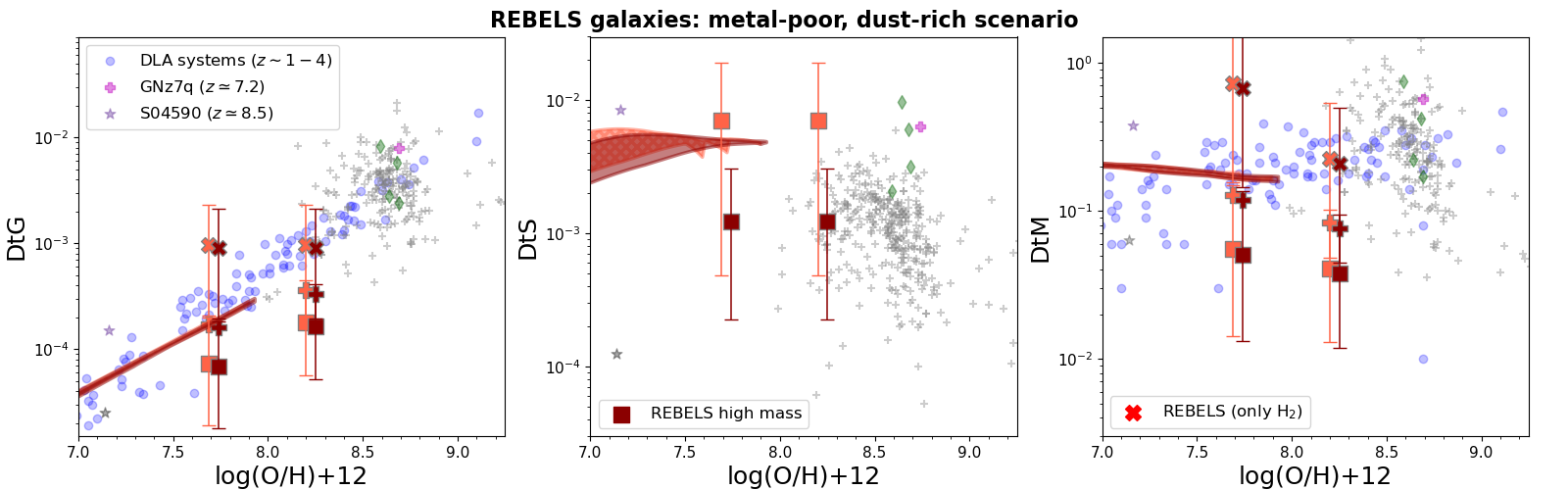}
    \caption{Dust-to-gas (DtG), dust-to-stellar (DtS) and dust-to-metal (DtM) mass ratios predictions from galaxy evolution models assuming a metal-poor scenario. Top panels: predictions for an intermediate-dust scenario (see \ref{sss:dust_setup}). Bottom panels: predictions for a dust-rich scenario (see \ref{sss:dust_setup}).
    Light orange and dark red shaded regions show the theoretical model predictions for REBELS galaxies belonging to the low and high stellar mass bins, respectively. 
    Squares with same color code represent low and high stellar mass REBELS galaxies data bins, respectively. Plus and crosses are the same as squares, but for $r_{\rm UV}/r_{\rm [C\,{\sc{II}}]}$ weighted gas masses and ${\rm H_2}$ gas masses only (see \ref{sss:metal_poor}). High stellar mass bin data are shifted by $0.05 {\rm dex}$ in metallicity for sake of clarity. Additional data are from \citet{Casasola20} (DustPedia galaxies), \citet{Shapley20} (luminous SF galaxies),  \citet{Peroux20} (DLA systems), \citet{Heintz23} (I Zw 18, S04590), \citet{Fujimoto22} (GNz7q). }
    \label{fig:lowmetal_scenario}
\end{figure*}

In Fig. \ref{fig:lowmetal_scenario} we show the predictions of the chemical and dust evolution models in the case dust evolution parameters have intermediate values (intermediate-dust scenario, top panels) or dust production is particularly favoured (dust-rich scenario, bottom panels).
The results of the models are compared with the DtG, DtS and DtM ratios computed from the values in Tab. \ref{tab:data_summary} by assuming a metallicity typical of the output of the models, i.e. 1/10 Z$_\odot$, and a metallicity of 1/3 Z$_\odot$ (both indicated by filled squares).
Due to the uncertainties in the gas mass calibrations in EoR galaxies, we show two additional data points for each metallicity in the left and right panels of Fig \ref{fig:lowmetal_scenario}.  Crosses should be seen as a sort of lower limit for the gas mass by considering only the [C\,{\sc{ii}}]-H$_2$ calibration of \citet{Vizgan22} for the total gas mass. The plus symbols instead indicate an intermediate scenario, where UV/[C\,{\sc{ii}}] weighted gas masses, i.e. gas masses scaled to the UV-to-[C\,{\sc{ii}}] effective radii ratio in REBELS (see Appendix \ref{app:UVCII_weigth}), are considered.

We note that the models that belong to a dust-poor setup are not shown in Fig. \ref{fig:lowmetal_scenario}. This is due to the extremely low values in the dust ratios (see also Fig. \ref{fig:DtS_Mstar}). An analog situation is also found for chemical evolution models evolving to larger metallicities (see \ref{sss:metal_rich}). Therefore, we can exclude that models with such a slow dust buildup can be a reliable solution to explain REBELS targets.

By looking at Fig. \ref{fig:lowmetal_scenario} (upper panels), we note that the simulations for the intermediate-dust scenario are usually within the errorbars of the inferred DtG, DtS and DtM for REBELS targets. However, it is worth noting that the DtS predicted for the low $M_*$ bin is at the lower edge of the large data errorbar. In general, the predictions of the models are below the data bins, therefore suggesting a slightly faster dust buildup to match the observed DtS and the DtG and DtM using HI+H$_2$ gas estimates from \citet{Heintz21} and \citet{Vizgan22}. 

The models assuming a "dust-rich scenario" agree well with the DtG and DtM computed using the weighted UV/[C\,{\sc{ii}}] gas masses (see Fig. \ref{fig:lowmetal_scenario} bottom panels). The evolutionary tracks for these models also agree well with the DtG and DtM vs. metallicity trends observed in Damped Lyman-$\alpha$ systems (DLAs) at redshifts $z\sim1-4$ (\citealt{Peroux20}).
By looking at the DtS, the models for the high-mass REBELS galaxies overestimate the quantities inferred for the correspondent bin. Such a discrepancy cannot be attributed to a significant scatter in the data, since the observed DtS have low rms ($<10^{-3}$) relative to the mean observational uncertainties (see also Fig. \ref{fig:DtS_Mstar}). At the same time, the models for low-mass galaxies are in better agreement with the DtS data with respect to the ones shown in Fig. \ref{fig:lowmetal_scenario} upper panels. 
However, we still notice an underestimation of the dust mass relative to the stellar mass by the models for the low-mass REBELS targets.\\

The adoption of a top-heavy IMF (i.e. favouring the formation of massive stars) may help alleviating the discrepancies between the observed trends for the DtG and the DtS for low-mass REBELS targets. 
In fact, a top-heavy IMF increases the level of chemical enrichment for a given star formation history, due to the  larger number of CC-SNe enriching the ISM with metals. The faster chemical enrichment also corresponds to a faster dust mass buildup and, as a consequence, to larger DtS ratios, but not to larger DtG ratios (see Fig. \ref{fig:topheavyIMF} in Appendix \ref{app:moreplots}).
This happens because the top-heavy IMF enhances dust production from SNe and favours dust growth due to the larger amount of metals available, but at the same time boosts the amount of gas available in the ISM due to the larger mass ejection rates from evolved stellar populations, and in particular from massive stars (see, e.g. \citealt{Palla20b}).

Another explanation for the very high dust-to-stellar mass ratio for the low-mass subsample could also be driven by a significant underestimate of the stellar mass $M_*$ for these targets. In fact, {\it Hubble Space Telescope} (HST) counterparts of REBELS-14 and REBELS-39 (which are part of the low-mass subsample) show very strong emission by rest-frame optical nebular lines (\citealt{Bowler17}), which can be explained only by the presence of a very recent, large burst of star formation. In turn, this can limit the ability of non-parametric SFHs to overcome the so-called outshining problem, i.e. the underestimation of the contribution of older stellar populations due to overwhelming light form recent bursts (e.g. \citealt{Leja19}).
The hypothesis of a stellar mass underestimation is also supported by the very large dynamical mass estimated from the [C\,{\sc{ii}}] line width and size\footnote{for these estimates, the rotation-dominated regime as described in \citet{DeCarli18} is adopted.} (see Schouws et al., in prep; \citealt{Topping22}) of REBELS-39 ($>10^{11} {\rm M_\odot}$). The derived stellar mass for this galaxy contributes only to $\sim 1\%$ of the total dynamical mass within the [C\,{\sc{ii}}]-emitting region, therefore leaving room for the presence of older stellar populations, significantly enhancing the total stellar mass.

\subsubsection{Metal-rich scenario}
\label{sss:metal_rich}

\begin{figure*}
    \centering
    \includegraphics[width=0.95\textwidth]{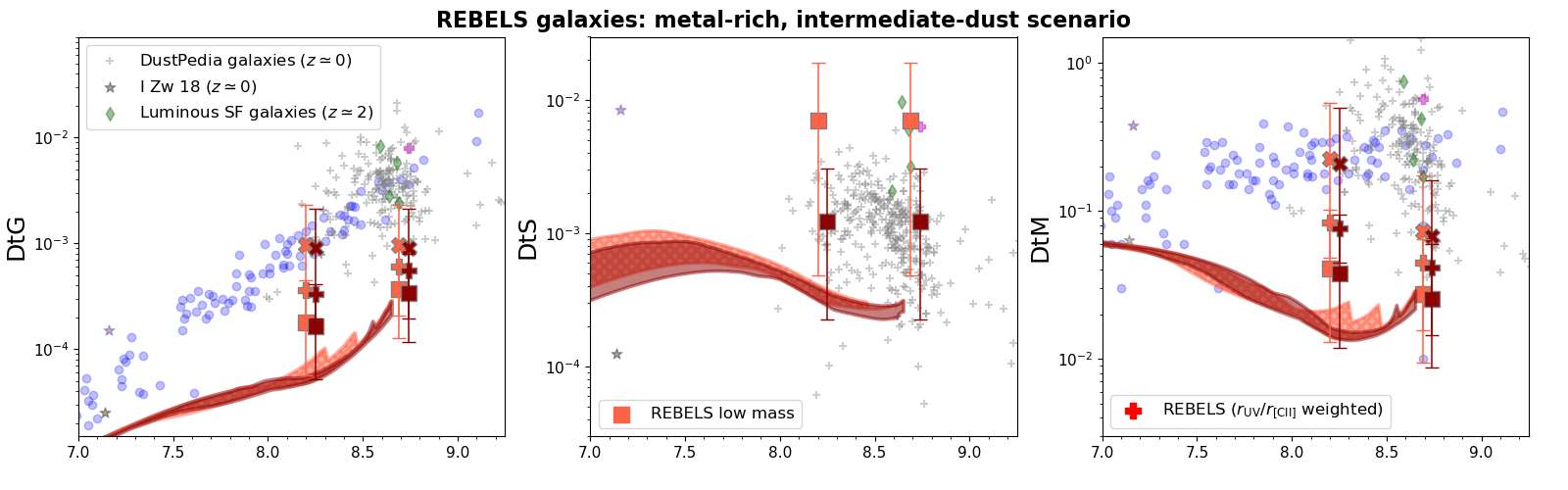}\\
    \includegraphics[width=0.95\textwidth]{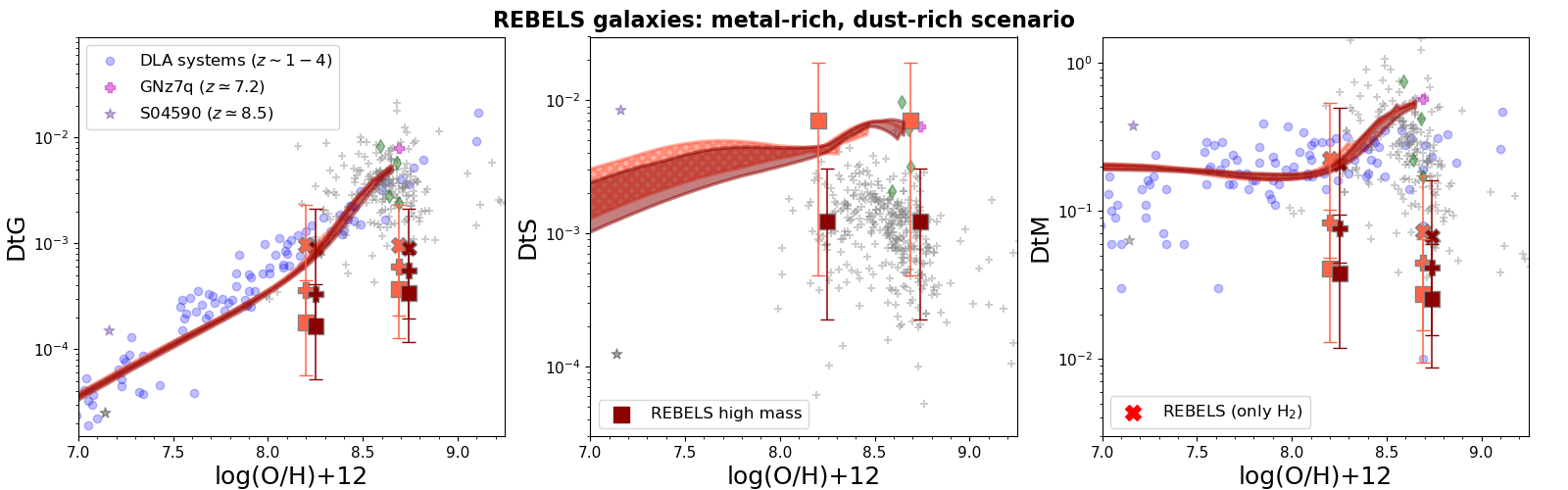}
    \caption{Dust-to-gas (DtG), dust-to-stellar (DtS) and dust-to-metal (DtM) ratios predictions from galaxy evolution models assuming a metal-rich scenario. Top panels: predictions for an intermediate-dust scenario (see \ref{sss:dust_setup}). Bottom panels: predictions for a dust-rich scenario (see \ref{sss:dust_setup}).
    Models and data legend are as in Fig. \ref{fig:lowmetal_scenario}.}
    \label{fig:highmetal_scenario}
\end{figure*}

Moving to the models suited to reproduce a metal-rich scenario for REBELS targets (see Fig. \ref{fig:highmetal_scenario}), we show the predictions of the models in the case dust evolution parameters have intermediate values (intermediate-dust scenario, top panels) or dust production is particularly favoured (dust-rich scenario, bottom panels). 
The predictions of the models are compared with the DtG, DtS and DtM computed by assuming metallicities of 1/3 Z$_\odot$ and Z$_\odot$, with the latter being a typical output of the models.

The models assuming the intermediate-dust scenario do reproduce the DtG and DtM ratios estimated by considering HI+H$_2$ mass calibrations. However, these models underestimate the DtS ratio. 
The reproduction of the estimated DtG and the underprediction of the DtS by the models may be driven by an overestimation of the gas mass through the [C\,{\sc{ii}}] calibrations: at variance with what happens for low-mass REBELS galaxies in the low-metallicity case, here the problem stands for both low- and high-mass targets, with the latter showing large stellar masses ($\log(M_*/{\rm M_\odot})\gtrsim 10$) for galaxies at $z\gtrsim 6$, which are less likely to be underestimated, despite the uncertainties.
As a further proof, the comparison with the local DustPedia sample (pale grey crosses, \citealt{Casasola17,Casasola20}), redshift $z\simeq2$ strongly star forming objects (green diamonds, \citealt{Shapley20}) and the galaxy GNz7q at $z\simeq7.2$ (\citealt{Fujimoto22}) show comparable DtS ratios to the REBELS ones, but almost an order of magnitude larger DtG ratios. We will come back to this point in \ref{sss:discussion}.

In Fig. \ref{fig:highmetal_scenario} lower panel, we see instead what happens when assuming a dust-rich scenario for the models reaching $\sim$ solar metallicity. The galactic evolution models do reproduce the observed DtS, especially for the low stellar mass bin. However, as also seen for models in the upper panel, we do not find unambiguous agreement between data and models in the different dust scaling relations. Here, we do not find any agreement with the data for DtG and DtM, even considering only the H$_2$ estimated mass contributing to the  galactic gas budget. 
The models show in fact a steep rise in the predicted DtG and DtM between $\log({\rm O/H})+12=8.2\, {\rm dex}$ and $8.5\, {\rm dex}$ that allows reproducing the constraints for the bulk of local galaxies, $z\simeq2$ dusty starbursts and the galaxy GNz7q at $z\simeq7.2$, but at the same time prevents any agreement with the data presented in Section \ref{s:data}, further suggesting an overestimation of the gas mass for the REBELS targets.
The jump to large values in the DtG and DtM is due to the transition between two regimes. In the first, the dust mass buildup is primarily set by stardust production, while in the second one dust growth also plays a fundamental role, leading the dust amount to saturate since no gas-phase metals available to form dust are left (e.g. \citealt{Hirashita15,Aoyama17,Aoyama18,Aoyama20}). This transition corresponds to the so-called "critical metallicity" regime (\citealt{Asano13}, see also \citealt{Gioannini17,Graziani20,DiCesare23}), which happens at metallicities $\gtrsim 0.1 Z_\odot$, with the exact value depending on different galactic evolution parameters, e.g. the galactic timescale/efficiency of star formation and the typical dust growth timescale (see, e.g. \citealt{Hirashita15,Aoyama17,Hirashita19}). 
It is worth noting that the transition between the regimes is also present when looking at models with mixed dust prescriptions (see Appendix \ref{app:moremix}): while for metal-poor models SN dust production is the main driver of the resulting dust mass, in the metal-rich case it is the balance between dust growth and destruction that is discriminant for the level of dust enrichment.

\subsection{Discussion}
\label{sss:discussion}

By looking at Fig. \ref{fig:lowmetal_scenario} and \ref{fig:highmetal_scenario}, a striking feature in the dust evolution tracks is the similarity, at similar levels of efficiency of star formation and therefore metallicity, between models with rather different SFHs (see Fig. \ref{fig:SFHs}). 
Such an outcome highlights how the dust evolution modelled in high-redshift galaxies is mostly influenced by parameters setting the efficiency of the different dust processes, i.e. production, growth and destruction (see also Appendix \ref{app:moremix}). \citet{Popping22} showed that the assumption of different dust parameters and/or parametrisation is the most critical ingredient influencing the predictions of dust evolution at different redshifts in the framework of several SAMs and hydro simulations (e.g. \citealt{Popping17,Li19,Triani20,Graziani20}), however not considering the effect of the different modeling of the rest of the baryonic physics in the simulations. Here, we further support this conclusion by adopting a simpler framework on a galactic scale (instead of cosmic), by showing the very weak dependence on the particular SFH shapes for the inferred dust scaling relations.\\

\begin{figure*}
    \centering
    \includegraphics[width=0.69\textwidth]{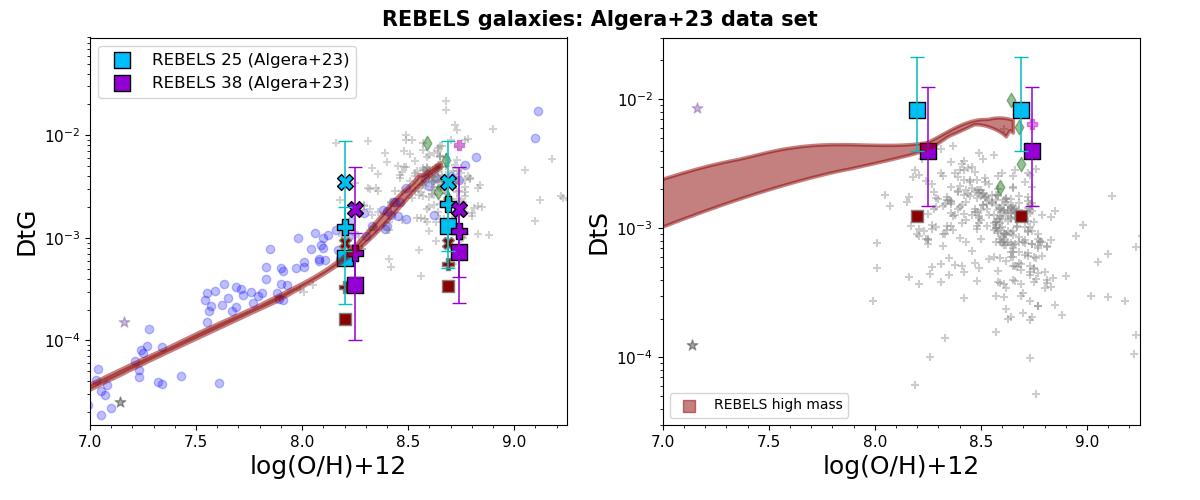}
    \caption{Dust-to-gas (DtG) and dust-to-stellar (DtS) mass ratios predictions from galaxy evolution models for the high stellar mass subsample assuming a metal-rich, dust-rich scenario (dark red shaded regions). Purple and cyan symbols refer to the dust mass estimates by \citet{Algera23} for REBELS-25 and REBELS-38, respectively. REBELS-38 data are shifted by $0.05 {\rm dex}$ in metallicity for the sake of clarity. 
    Dark red symbols and additional data are as in Fig. \ref{fig:lowmetal_scenario}.}
    \label{fig:dust_Algera}
\end{figure*}

The comparison of our model tracks for different metal enrichment scenarios with the REBELS data also highlights other important features.
On the one hand, our metal-poor models generally reproduce the constraints from REBELS galaxies without assuming extreme prescriptions for dust evolution (see Fig. \ref{fig:lowmetal_scenario}), in agreement with \citet{DiCesare23}; Schneider et al. (in prep.). Only for the REBELS galaxies with the lowest stellar masses we discuss the possibility of a top-heavy IMF to reconcile the DtG and DtS measurements. These results are partly in tension with what was found in \citet{Dayal22}, where some of the REBELS galaxies where only reproduced by a "maximal dust model" for which extreme dust recipes (e.g. extremely fast dust growth timescale, no dust destruction by SNe) were adopted. However, it is worth noting that the stellar masses adopted here (from \citealt{Topping22}) are different from those adopted in \citet{Dayal22} (from Stefanon et al., in prep.). In turn, the different stellar masses lead to different dust masses in the methodology employed by \citet{Sommovigo22a,Sommovigo22b}. Therefore, it is likely that the different dataset of stellar masses might lie at the heart of this need for extreme dust prescriptions and therefore of the discrepancy with our results.

\begin{figure*}
    \centering
    \includegraphics[width=0.99\textwidth]{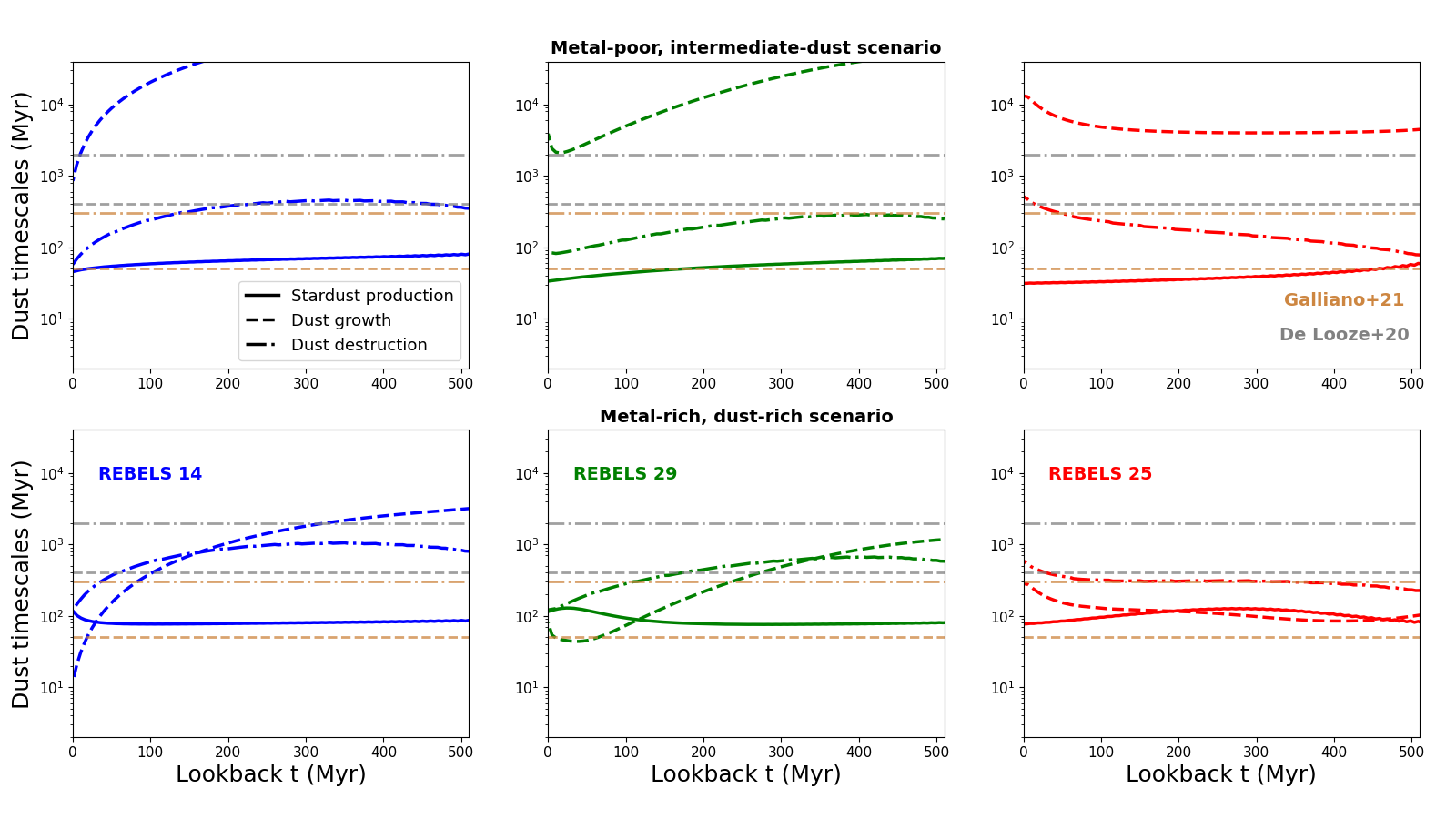}
    \caption{Evolution of dust formation and destruction timescales predicted by models for REBELS-14, REBELS-29 and REBELS-25 for different metallicity and dust evolutionary scenarios. Top panels show the results for galaxy evolution models assuming a metal-poor, intermediate-dust scenario, while lower panels for models assuming a metal-rich, dust-rich scenario (see also Tabs. \ref{tab:param_chem} and \ref{tab:param_dust}).
    Solid lines represent stardust production, dashed lines dust growth, and dash-dotted lines dust destruction. For the sake of comparison, we overlay the average dust formation and destruction timescales obtained for local galaxies by \citet{DeLooze20} and \citet{Galliano21}.}
    \label{fig:dust_timescales}
\end{figure*}

On the other hand, the models tracks mimicking a metal-rich scenario (see Fig. \ref{fig:highmetal_scenario}) are unable to simultaneously reproduce the data for the DtG, DtM and DtS ratios independently of the assumed dust evolution scenario.  
Regarding this latter point, by adopting the most recent dust mass estimates by \citet{Algera23} for REBELS-25 ($\log(M_{\rm dust}/{\rm M_\odot})=8.13^{+0.38}_{-0.27}$) and REBELS-38 ($\log(M_{\rm dust}/{\rm M_\odot})=7.90^{+0.39}_{-0.35}$)\footnote{we adopt the fiducial estimates by \citet{Algera23} for optically thin dust with dust emissivity index $\beta=2.0$. For optically thick dust ($\lambda_{\rm thick} = 100 \mu m$) $M_{\rm dust}$ estimates are reduced by $\sim 0.3 {\rm dex}$.} based on additional sampling of the FIR continuum (see Section \ref{s:data}), the model difficulties in simultaneously reproducing the DtG and DtS are partially alleviated. This is because of the lower $T_{\rm dust}$ in \citet{Algera23}, leading to larger dust masses.
The comparison between the models and this other dataset can be observed in Fig. \ref{fig:dust_Algera}, where the DtG and DtS derived using newer dust mass estimates are shown. 
A relatively good agreement between data and model DtGs for the metal-rich case can be recovered for REBELS-25 by assuming molecular gas mass alone, whereas by adding HI masses the models still largely overestimate the dust-to-gas ratios. 
However, the larger dust masses estimated by \citet{Algera23} give a good agreement with the predicted dust-to-stellar mass evolution tracks for the dust-rich scenario. This same conclusion is reached when the comparison is made with \texttt{dustygadget}, supporting the fact that at least some of these galaxies may have lower dust temperature and higher dust masses compared to what is estimated on the basis of single band observations (\citealt{Algera23}). 

This agreement also reinforces our previous statement of no need for extreme assumptions for the buildup of dust in UV-bright, dusty galaxies as the ones in REBELS. The prescriptions and dust evolution parameters used to reproduce the dust content in these high-redshift galaxies do not seem to differ much from the framework used for low-redshift galaxies (e.g. \citealt{Asano13,Gioannini17,Galliano18,Calura23}). This can be clearly seen in Fig. \ref{fig:dust_timescales}, where we show the timescales of stardust production, dust grain growth and dust destruction in the three REBELS galaxies with different SFHs shown in Fig. \ref{fig:SFHs}, i.e. REBELS-14, 29 and 25, compared with dust timescales found in local studies (\citealt{DeLooze20,Galliano21}). In particular, we show the evolution of the different dust timescales in the two antipodal scenarios shown in Figs. \ref{fig:lowmetal_scenario} and \ref{fig:highmetal_scenario}, i.e. the metal-poor, intermediate-dust (top panels) and the metal-rich, dust-rich one (bottom panels). 
Despite of the large differences, i.e. of the order of one magnitude, between the timescale tracks for dust grain growth and SN dust destruction in upper and lower panels, we observe that in both metal and dust scenarios all dust timescales are well above few tenths of Myr at all evolutionary times. In particular, a striking feature for all panels is that the timescales found for individual REBELS galaxies are always comparable to the ones found in local studies, either in the case we have a higher dust destruction per SN event (upper panels) or we favour stardust production and dust grain growth (lower panels).

\subsubsection{A problem of gas mass at high metallicity?}
\label{sss:gasmass_problem}

The marginal global agreement between models and data within the metal-rich scenario raises the question about the origin of this discrepancy: even allowing for additional combinations in the dust evolution parameters (see Appendix \ref{app:moremix}), none of the models is able to simultaneously recover the different dust scaling relations obtained by using the data presented in Section \ref{s:data}.

\begin{figure}
    \centering
    \includegraphics[width=0.95\columnwidth]{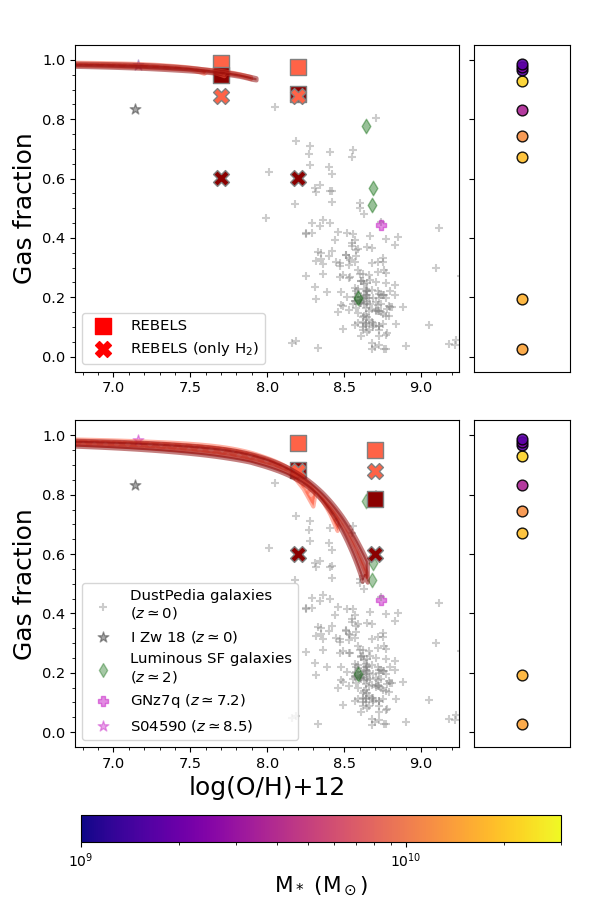}
    \caption{Gas fraction evolution with metallicity as predicted by galaxy evolution models compared with gas fraction derived through different [C\,{\sc{ii}}] calibrations for the gas mass. Data legend is as in Figs. \ref{fig:lowmetal_scenario}. Top panel: prediction for the metal-poor scenario. Bottom panel: prediction for the metal-rich scenario. For both panels, the gas fraction estimated for individual REBELS galaxies using their dynamical mass is shown in the right-side panels according to their stellar mass (see bottom colorbar). 
    }
    \label{fig:gas_fraction}
\end{figure}

To further explore this, we look at the gas fraction $M_{\rm gas,[C\,{\sc{II}}]}/(M_{\rm{gas,}{\rm [C\,{\sc{II}}]}}+M_{*})$ inferred from observations, and compare these with the predictions of our chemical evolution models (see Fig. \ref{fig:gas_fraction} top and bottom panels for the metal-poor and metal-rich scenario, respectively).

The gas fractions obtained using [C\,{\sc{ii}}] calibrated HI+H$_2$ masses yield very large values, i.e. of the order of 0.9. 
While this is consistent with model predictions at low metallicities (Fig. \ref{fig:gas_fraction} upper panel), large gas fractions do not match with what is obtained by galaxy evolution models at high metallicities (Fig. \ref{fig:gas_fraction} lower panel).
The motivation behind this discrepancy can be searched in some basic consideration of the chemical enrichment process that is irrespective of the mechanism of the stellar mass buildup in galaxies, from gas accretion from the cosmic web to galaxy mergers. In fact, to allow for a significant ISM metal enrichment, a large number of subsequent stellar generations are needed. At the same time, the short cosmic timescales for these high-redshift galaxies ($\lesssim 0.85$ Gyr at $z\simeq6.5$) only allow for intermediate-massive stars ($m \gtrsim 3 {\rm M_\odot}$) to contribute to the chemical enrichment process. Due to the negative power-law nature of the stellar IMF at such masses, the large majority of the gas mass accreting the galactic system remains trapped in long-lived stars, therefore lowering significantly the available gas fraction. 
This leads to the suggestion that the gas masses estimated through [C\,{\sc{ii}}] may significantly overestimate the gas content at high metallicities and, as a consequence, be at the basis of the inconsistency between models and data in the different dust scaling relations. 
As a further proof of this, Fig. \ref{fig:gas_fraction} shows that if we only assume the [C\,{\sc{ii}}]-calibrated H$_2$ mass to contribute to the galactic gas budget we better match typical values observed in other intermediate to high-redshift sources with high metallicities (\citealt{Shapley20,Fujimoto22}).
Moreover, the modeled gas mass fractions reproduce the values in the REBELS high-mass bin when considering H$_{\rm 2}$ masses only. Considering the acceptable match between data and models for the DtG and DtS in Fig. \ref{fig:dust_Algera} when adopting only H$_{\rm 2}$ calibration for the gas mass, this clearly highlights the need for lower gas masses derived through [C\,{\sc{ii}}] calibration at high-metallicities.

Therefore, calibrations from additional FIR lines (e.g. [OI]63$\mu$m, see \citealt{Wilson23}) may be needed to depict a clearer view on the atomic (but also total) gas contribution in these galactic sources, which is also crucial to understand their dust buildup.

\subsection{Findings in the context of high-redhsift MZR relations}
\label{sss:MZR_JWST}

Several JWST programs have already started probing fundamental galactic scaling relations, such as the mass-metallicity relation (MZR) at redshifts $z\gtrsim 4$. 
In particular, \citet{Nakajima23} studied the evolution of the MZR at $z = 4-10$ derived from $\sim$ 130 galaxies observed with JWST/NIRSpec taken from three major public spectroscopy programs (ERO, GLASS, and CEERS), using a similar stellar mass determination method to the one employed in this work (see also \citealt{Harikane23}).
They found that galaxies show a small evolution relative to the MZR at $z\sim3$ (\citealt{Sanders21}), deriving this linear relation:
\begin{equation}
    \log({\rm O/H})+12= (8.24 \pm 0.05) + (0.25\pm 0.03) \, \log(\frac{M_*}{10^{10}{\rm M_\odot}}),
    \label{eq:MZR_Naka}
\end{equation}
in the stellar mass regime $\log(M_*/{\rm M_\odot}) \simeq 7.5 - 9.5$. 
\citet{Curti23} found a shallower MZR, especially for redshift $z>6$ galaxies. However, the sample collected by \citet{Curti23} considered targets at the low-mass end of the MZR ($M_*/{\rm M_\odot} \sim 10^{6.5}-10^{8.5}$, from the JADES program) that can significantly influence the inferred slope. Moreover, as also mentioned by the authors, the $z>6$ redshift bin in \citet{Curti23} is sparsely populated and probably subject to strong selection biases.\\

By extrapolating the relation by \citet{Nakajima23} up to $\sim 10^{10} {\rm M_\odot}$, i.e. slightly above the upper mass limit of validity of Eq. \eqref{eq:MZR_Naka}, we find that our galaxies should have a metallicity between $\log({\rm O/H})+12\simeq 8 \, {\rm dex}$ (for the low-mass end of REBELS) and $\log({\rm O/H})+12\simeq 8.35 \, {\rm dex}$ (for the high-mass end).
This means that low-mass REBELS galaxies should have a global metallicity of around $0.2 {\rm Z_\odot}$, i.e. between the low and intermediate metallicity data shown in Fig. \ref{fig:lowmetal_scenario}.  
Therefore, low-mass REBELS galaxies should be in agreement with the metal-poor, dust-rich scenario outlined in \ref{sss:metal_poor}, where most of the dust is the result of stardust production by SNe, with other dust processes as destruction and especially dust growth playing a minor role in shaping the dust budget (see also Appendix \ref{app:moremix}).
Moreover, the comparison of the different dust scaling relations may suggest the presence of an IMF favoring the formation of massive stars or a slight revision of the computed stellar masses.

The most massive REBELS galaxies instead should have a metallicity of $\sim 0.5\, {\rm Z_\odot}$, that would be compatible with the model predictions presented in \ref{sss:metal_rich} for high-metallicity galaxies. This would indicate either i) rapid dust buildup in case gas masses are overestimated by [C\,{\sc{ii}}]-$M_{\rm gas}$ calibrations, or ii) slower dust formation in case stellar masses are underestimated. 
The former scenario seems favoured according to the set of stellar masses adopted, which shows larger values than those determined through other SED fitting methods (\citealt{Topping22}).
In addition, the larger dust masses by \citet{Algera23} for two high-mass REBELS galaxies would also make the scenario i) more appealing.
Therefore, different indications point towards a metal-rich, dust-rich scenario, where the large dust-to-gas, stellar and metal ratios are the result of a concurrent production of dust by stars and ISM growth processes. 
In such a case, the positive difference between the dust growth and the dust destruction rate is comparable to the stardust production rate (see also Fig. \ref{fig:dust_timescales} to look at this in terms of dust timescales).\\

With this section, we have tried to place the dust evolution of REBELS galaxies in context of the first JWST results about the MZR relation for galaxies across the first Gyr of cosmic time. 
Drawing a detailed analysis about the relation between REBELS targets and the high-redshift galaxy scaling relations (which are prone to be revised in the next JWST cycles) is beyond the scope of this work. 
Conversely, we want to provide insights on the scenarios that JWST spectroscopy may open in the analysis of UV-bright, dusty galaxies in the EoR. In this way, we want to stress the urgent need for JWST follow-up for a statistically consistent number of FIR detected sources to characterize in much greater detail the dust evolution in the high-redshift Universe.

\section{Conclusions}
\label{s:conclusion}

In this paper, we explored
the possible scenarios of metal and dust evolution that apply to the ALMA REBELS (\citealt{Bowens22}) galaxy sample during the Epoch of Reionization (EoR). 
We took advantage of the sources with both [C\,{\sc{ii}}]$158\mu$m line and FIR continuum detections, which makes this sample of galaxies the largest with simultaneous stellar and dust mass determinations in the EoR (\citealt{Sommovigo22a,Inami22}), and extracted the gas masses from recent, metallicity dependent calibrations based on the [C\,{\sc{ii}}] line luminosity (\citealt{Heintz21,Vizgan22}). 
We then modeled the chemical and dust evolution in these galaxies by means of detailed models adopting the non-parametric star formation histories (SFHs) derived by \citet{Topping22} for each galaxy of the REBELS sample. In particular, we allow the gas flows and dust evolutionary history to vary in the model galaxies in order to simulate the different scenarios of metal and dust enrichment that may be found for such objects.   

In this way, we wanted to provide clues on the dominant dust production mechanisms in EoR galaxies in different scenarios of metal enrichment, which can be probed by means of rest-frame optical ISM line spectroscopy from JWST.
Moreover, the analysis performed throughout the paper also provides a test for [C\,{\sc{ii}}]-gas mass calibrations (\citealt{Zanella18,Heintz21,Vizgan22}), which play a crucial role in the study of galaxies at redshift $z\gtrsim 6$.\\

Our results can be summarized as follows:
\begin{enumerate}
    \item at similar levels of efficiency in star formation and therefore metallicity, the predicted dust evolutionary tracks for different SFHs are very similar when assuming the same parametrizations for the different dust processes, i.e. stardust production, dust growth and dust destruction. This highlights that the modelling of dust evolution in high-redshift galaxies is primarily influenced by the assumptions and parameters setting the efficiency of the different dust processes in the ISM, in agreement with \citet{Popping22};
    \item as expected from previous dust modelling at $z\gtrsim 5$ (\citealt{Dayal22,DiCesare23}), models showing a slow dust buildup (dust-poor scenario) are far from reproducing the dust constraints from REBELS galaxies and therefore should be excluded when studying "normal" star-forming galaxies in the EoR. 
    At the same time, there is no need for extreme assumptions about dust formation and evolution for these sources, with dust timescales comparable to what claimed by several studies focused on local galaxies (e.g. \citealt{DeLooze20,Galliano21}). 
    However, current observational constraints, i.e. the galactic dust and stellar masses, are still insufficient to discriminate between different galactic evolution scenarios that in turn influence the relative importance of the different dust  processes in reproducing the observed dust masses.
    Information on ISM metallicity will therefore have a crucial importance to start placing tighter constraints on the contribution of dust evolution processes;
    \item in case JWST observations will reveal low metal enrichment ($Z\simeq0.1\,Z_\odot$), we demonstrate that dust evolution is mainly driven by stardust production, whose rate dominates over ISM dust processes (growth, destruction). Either models where dust parameters are set to have a fast dust buildup (dust-rich scenario) or a quieter evolution (intermediate-dust scenario) are compatible with the constraints from dust-to-gas (DtG), dust-to-stellar (DtS) and dust-to-metal (DtM) mass ratios.
    However, models for REBELS targets with the lowest stellar masses ($M_*\sim 10^9 {\rm M_\odot}$) generally underpredict the observed DtS.
    A possible way to reconcile models with observations for these sources is to adopt a top-heavy IMF (i.e. favouring the production of massive stars, see e.g. \citealt{Palla20b}). However, indications from HST (\citealt{Bowler17}) and dynamical mass estimates from the [C\,{\sc{ii}}] line (\citealt{Topping22}) may also indicate an underestimation of the stellar mass for these targets;
    \item in the case of sources evolving up to solar metallicity, the level of dust enrichment in galaxies is mostly driven by the balance between ISM dust growth and destruction processes.
    The comparison with REBELS galaxy measurements reveal that different galaxy models struggle to reproduce simultaneously the estimated DtG/DtM and DtS.
    In particular, models with a intermediate-dust scenario do reproduce the estimated DtG/DtM ratios while they underestimate the observed DtS. At the same time, models with a fast dust buildup do reproduce the observed DtS, while they significantly overestimate the DtG/DtM. 
    This may be due to an overestimation of the gas mass by means of [C\,{\sc{ii}}] line calibrations (\citealt{Heintz21,Vizgan22}) at high metallicities: gas fractions using these calibrations yield very large values which are not compatible with the presence of a chemically evolved galaxy basing on general consideration of galactic chemical enrichment. To alleviate the disagreement, lower gas masses relative to the reference dataset should be considered;
    \item looking at the first JWST probes of the mass-metallicity relation (MZR) at high-redshift (e.g. \citealt{Nakajima23}), REBELS galaxies should exhibit metallicities between $\sim 0.2$ and $\sim 0.5 Z_\odot$. 
    The galaxies with the lowest stellar masses should be representative of the metal-poor case mentioned in point iii) of this Section. For these galaxies, our modeling suggests a dust-rich scenario. 
    Galaxies in the upper mass end should instead represent a case compatible to the metal-rich one. For such galaxies, either a dust-rich scenario in case gas masses are overestimated or an intermediate-dust scenario in case stellar masses are underestimated may be suggested, with the former more likely according to point (iv) of this Section.
\end{enumerate}

Present and upcoming JWST/NIRSpec programs at intermediate- to high-resolution ($\mathcal{R}\sim1000-2700$) are allowing to  resolve several nebular emission lines in galaxy spectra in $z>6$ galaxies, such as [OII]$\lambda\lambda3727, 3729$ and [OIII]$\lambda\lambda4960, 5008$ doublets, [NII]$\lambda6585$ and H$\alpha$ lines  (for $z<6.7$ galaxies) H$\beta$, H$\gamma$ and possibly even the auroral [OIII]$\lambda$4363 transition. 
This opens up the possiblity to determine the level of chemical enrichment in REBELS sources as we get access to the most common strong line metallicity indicators (e.g. O3, O32, R2, R23), even when direct-$T_e$ measurements will not be possible.
Considering uncertainties of the order of 0.2-0.3 dex within $\log$(O/H)$+12$ determinations (e.g. \citealt{Nakajima23,Bunker23,Curti23}), such metallicity measurements will be sufficient to start discriminating between the different scenarios illustrated in this paper. 

Therefore, the results presented here may be used as a viable tool to interpret and constrain the metal and dust evolution in EoR galaxies. At the same time, these results highlight even more the great potentials of the ALMA-JWST synergy to unveil some of the most debated issues in contemporary astronomy.

\section*{Acknowledgements}
The authors thank Dr. Shohei Aoyama for the reading of the manuscript and useful suggestions to improve the paper.
MP, IDL SvdG acknowledge funding support from ERC starting grant 851622 DustOrigin. MP also thank Laura Sommovigo for providing data adopted in this paper. MR acknowledges financial support by the research projects AYA2017-84897-P and PID2020-113689GB-I00, financed by MCIN/AEI/10.13039/501100011033.
PD acknowledges support from the NWO grant 016.VIDI.189.162 (``ODIN") and from the European Commission's and University of Groningen's CO-FUND Rosalind Franklin program. 
HSBA and HI acknowledge support from the NAOJ ALMA Scientific Research Grant Code 2021-19A. 
MA acknowledges support from FONDECYT grant 1211951 and ANID BASAL project FB210003. 
RB acknowledges support from an STFC Ernest Rutherford Fellowship [grant number ST/T003596/1].
JH acknowledges support from the ERC Consolidator Grant 101088676 (VOYAJ) and the VIDI research programme with project number 639.042.611, which is (partly) financed by the Netherlands Organisation for Scientific Research (NWO).

\section*{Data Availability}

The data underlying this paper will be shared on reasonable request to the corresponding author.



\bibliographystyle{mnras}
\bibliography{paper} 

\begin{thebibliography}{}
\makeatletter
\relax
\def\mn@urlcharsother{\let\do\@makeother \do\$\do\&\do\#\do\^\do\_\do\%\do\~}
\def\mn@doi{\begingroup\mn@urlcharsother \@ifnextchar [ {\mn@doi@}
  {\mn@doi@[]}}
\def\mn@doi@[#1]#2{\def\@tempa{#1}\ifx\@tempa\@empty \href
  {http://dx.doi.org/#2} {doi:#2}\else \href {http://dx.doi.org/#2} {#1}\fi
  \endgroup}
\def\mn@eprint#1#2{\mn@eprint@#1:#2::\@nil}
\def\mn@eprint@arXiv#1{\href {http://arxiv.org/abs/#1} {{\tt arXiv:#1}}}
\def\mn@eprint@dblp#1{\href {http://dblp.uni-trier.de/rec/bibtex/#1.xml}
  {dblp:#1}}
\def\mn@eprint@#1:#2:#3:#4\@nil{\def\@tempa {#1}\def\@tempb {#2}\def\@tempc
  {#3}\ifx \@tempc \@empty \let \@tempc \@tempb \let \@tempb \@tempa \fi \ifx
  \@tempb \@empty \def\@tempb {arXiv}\fi \@ifundefined
  {mn@eprint@\@tempb}{\@tempb:\@tempc}{\expandafter \expandafter \csname
  mn@eprint@\@tempb\endcsname \expandafter{\@tempc}}}

\bibitem[\protect\citeauthoryear{{Ag{\'u}ndez} \& {Wakelam}}{{Ag{\'u}ndez} \&
  {Wakelam}}{2013}]{Agundez13}
{Ag{\'u}ndez} M.,  {Wakelam} V.,  2013, \mn@doi [Chemical Reviews]
  {10.1021/cr4001176}, \href
  {https://ui.adsabs.harvard.edu/abs/2013ChRv..113.8710A} {113, 8710}

\bibitem[\protect\citeauthoryear{{Algera} et~al.,}{{Algera}
  et~al.}{2023}]{Algera23}
{Algera} H. S.~B.,  et~al., 2023, \mn@doi [\mnras] {10.1093/mnras/stad3111},
  \href {https://ui.adsabs.harvard.edu/abs/2023MNRAS.tmp.2998A} {}

\bibitem[\protect\citeauthoryear{{Aoyama}, {Hou}, {Shimizu}, {Hirashita},
  {Todoroki}, {Choi}  \& {Nagamine}}{{Aoyama} et~al.}{2017}]{Aoyama17}
{Aoyama} S.,  {Hou} K.-C.,  {Shimizu} I.,  {Hirashita} H.,  {Todoroki} K.,
  {Choi} J.-H.,   {Nagamine} K.,  2017, \mn@doi [\mnras]
  {10.1093/mnras/stw3061}, \href
  {https://ui.adsabs.harvard.edu/abs/2017MNRAS.466..105A} {466, 105}

\bibitem[\protect\citeauthoryear{{Aoyama}, {Hou}, {Hirashita}, {Nagamine}  \&
  {Shimizu}}{{Aoyama} et~al.}{2018}]{Aoyama18}
{Aoyama} S.,  {Hou} K.-C.,  {Hirashita} H.,  {Nagamine} K.,   {Shimizu} I.,
  2018, \mn@doi [\mnras] {10.1093/mnras/sty1431}, \href
  {https://ui.adsabs.harvard.edu/abs/2018MNRAS.478.4905A} {478, 4905}

\bibitem[\protect\citeauthoryear{{Aoyama}, {Hirashita}  \& {Nagamine}}{{Aoyama}
  et~al.}{2020}]{Aoyama20}
{Aoyama} S.,  {Hirashita} H.,   {Nagamine} K.,  2020, \mn@doi [\mnras]
  {10.1093/mnras/stz3253}, \href
  {https://ui.adsabs.harvard.edu/abs/2020MNRAS.491.3844A} {491, 3844}

\bibitem[\protect\citeauthoryear{{Aravena} et~al.,}{{Aravena}
  et~al.}{2023}]{Aravena23}
{Aravena} M.,  et~al., 2023, \mn@doi [arXiv e-prints]
  {10.48550/arXiv.2309.15948}, \href
  {https://ui.adsabs.harvard.edu/abs/2023arXiv230915948A} {p. arXiv:2309.15948}

\bibitem[\protect\citeauthoryear{{Asano}, {Takeuchi}, {Hirashita}  \&
  {Inoue}}{{Asano} et~al.}{2013}]{Asano13}
{Asano} R.~S.,  {Takeuchi} T.~T.,  {Hirashita} H.,   {Inoue} A.~K.,  2013,
  \mn@doi [Earth, Planets and Space] {10.5047/eps.2012.04.014}, \href
  {https://ui.adsabs.harvard.edu/abs/2013EP&S...65..213A} {65, 213}

\bibitem[\protect\citeauthoryear{{Asplund}, {Grevesse}, {Sauval}  \&
  {Scott}}{{Asplund} et~al.}{2009}]{Asplund09}
{Asplund} M.,  {Grevesse} N.,  {Sauval} A.~J.,   {Scott} P.,  2009, \mn@doi
  [\araa] {10.1146/annurev.astro.46.060407.145222}, \href
  {https://ui.adsabs.harvard.edu/abs/2009ARA&A..47..481A} {47, 481}

\bibitem[\protect\citeauthoryear{{Atek} et~al.,}{{Atek} et~al.}{2023}]{Atek22}
{Atek} H.,  et~al., 2023, \mn@doi [\mnras] {10.1093/mnras/stac3144}, \href
  {https://ui.adsabs.harvard.edu/abs/2023MNRAS.519.1201A} {519, 1201}

\bibitem[\protect\citeauthoryear{{Bakx} et~al.,}{{Bakx} et~al.}{2020}]{Bakx20}
{Bakx} T. J.~L.~C.,  et~al., 2020, \mn@doi [\mnras] {10.1093/mnras/staa509},
  \href {https://ui.adsabs.harvard.edu/abs/2020MNRAS.493.4294B} {493, 4294}

\bibitem[\protect\citeauthoryear{{Bakx} et~al.,}{{Bakx} et~al.}{2021}]{Bakx21}
{Bakx} T. J.~L.~C.,  et~al., 2021, \mn@doi [\mnras] {10.1093/mnrasl/slab104},
  \href {https://ui.adsabs.harvard.edu/abs/2021MNRAS.508L..58B} {508, L58}

\bibitem[\protect\citeauthoryear{{Bianchi} \& {Schneider}}{{Bianchi} \&
  {Schneider}}{2007}]{Bianchi07}
{Bianchi} S.,  {Schneider} R.,  2007, \mn@doi [\mnras]
  {10.1111/j.1365-2966.2007.11829.x}, \href
  {https://ui.adsabs.harvard.edu/abs/2007MNRAS.378..973B} {378, 973}

\bibitem[\protect\citeauthoryear{{Bocchio}, {Marassi}, {Schneider}, {Bianchi},
  {Limongi}  \& {Chieffi}}{{Bocchio} et~al.}{2016}]{Bocchio16}
{Bocchio} M.,  {Marassi} S.,  {Schneider} R.,  {Bianchi} S.,  {Limongi} M.,
  {Chieffi} A.,  2016, \mn@doi [\aap] {10.1051/0004-6361/201527432}, \href
  {https://ui.adsabs.harvard.edu/abs/2016A&A...587A.157B} {587, A157}

\bibitem[\protect\citeauthoryear{{Bouwens} et~al.,}{{Bouwens}
  et~al.}{2010}]{Bowens10}
{Bouwens} R.~J.,  et~al., 2010, \mn@doi [\apjl] {10.1088/2041-8205/709/2/L133},
  \href {https://ui.adsabs.harvard.edu/abs/2010ApJ...709L.133B} {709, L133}

\bibitem[\protect\citeauthoryear{{Bouwens} et~al.,}{{Bouwens}
  et~al.}{2022}]{Bowens22}
{Bouwens} R.~J.,  et~al., 2022, \mn@doi [\apj] {10.3847/1538-4357/ac5a4a},
  \href {https://ui.adsabs.harvard.edu/abs/2022ApJ...931..160B} {931, 160}

\bibitem[\protect\citeauthoryear{{Bowler}, {Dunlop}, {McLure}  \&
  {McLeod}}{{Bowler} et~al.}{2017}]{Bowler17}
{Bowler} R.~A.~A.,  {Dunlop} J.~S.,  {McLure} R.~J.,   {McLeod} D.~J.,  2017,
  \mn@doi [\mnras] {10.1093/mnras/stw3296}, \href
  {https://ui.adsabs.harvard.edu/abs/2017MNRAS.466.3612B} {466, 3612}

\bibitem[\protect\citeauthoryear{{Bunker} et~al.,}{{Bunker}
  et~al.}{2023}]{Bunker23}
{Bunker} A.~J.,  et~al., 2023, \mn@doi [\aap] {10.1051/0004-6361/202346159},
  \href {https://ui.adsabs.harvard.edu/abs/2023A&A...677A..88B} {677, A88}

\bibitem[\protect\citeauthoryear{{Calura}, {Pipino}  \& {Matteucci}}{{Calura}
  et~al.}{2008}]{Calura08}
{Calura} F.,  {Pipino} A.,   {Matteucci} F.,  2008, \mn@doi [\aap]
  {10.1051/0004-6361:20078090}, \href
  {https://ui.adsabs.harvard.edu/abs/2008A&A...479..669C} {479, 669}

\bibitem[\protect\citeauthoryear{{Calura} et~al.,}{{Calura}
  et~al.}{2023}]{Calura23}
{Calura} F.,  et~al., 2023, \mn@doi [\mnras] {10.1093/mnras/stad1316}, \href
  {https://ui.adsabs.harvard.edu/abs/2023MNRAS.tmp.1344C} {}

\bibitem[\protect\citeauthoryear{{Calzetti}, {Armus}, {Bohlin}, {Kinney},
  {Koornneef}  \& {Storchi-Bergmann}}{{Calzetti} et~al.}{2000}]{Calzetti00}
{Calzetti} D.,  {Armus} L.,  {Bohlin} R.~C.,  {Kinney} A.~L.,  {Koornneef} J.,
   {Storchi-Bergmann} T.,  2000, \mn@doi [\apj] {10.1086/308692}, \href
  {https://ui.adsabs.harvard.edu/abs/2000ApJ...533..682C} {533, 682}

\bibitem[\protect\citeauthoryear{{Casasola} et~al.,}{{Casasola}
  et~al.}{2017}]{Casasola17}
{Casasola} V.,  et~al., 2017, \mn@doi [\aap] {10.1051/0004-6361/201731020},
  \href {https://ui.adsabs.harvard.edu/abs/2017A&A...605A..18C} {605, A18}

\bibitem[\protect\citeauthoryear{{Casasola} et~al.,}{{Casasola}
  et~al.}{2020}]{Casasola20}
{Casasola} V.,  et~al., 2020, \mn@doi [\aap] {10.1051/0004-6361/201936665},
  \href {https://ui.adsabs.harvard.edu/abs/2020A&A...633A.100C} {633, A100}

\bibitem[\protect\citeauthoryear{{Casey} et~al.,}{{Casey}
  et~al.}{2014}]{Casey14}
{Casey} C.~M.,  et~al., 2014, \mn@doi [\apj] {10.1088/0004-637X/796/2/95},
  \href {https://ui.adsabs.harvard.edu/abs/2014ApJ...796...95C} {796, 95}

\bibitem[\protect\citeauthoryear{{Castellano} et~al.,}{{Castellano}
  et~al.}{2023}]{Castellano22}
{Castellano} M.,  et~al., 2023, \mn@doi [\apjl] {10.3847/2041-8213/accea5},
  \href {https://ui.adsabs.harvard.edu/abs/2023ApJ...948L..14C} {948, L14}

\bibitem[\protect\citeauthoryear{{Chabrier}}{{Chabrier}}{2003}]{Chabrier03}
{Chabrier} G.,  2003, \mn@doi [\pasp] {10.1086/376392}, \href
  {https://ui.adsabs.harvard.edu/abs/2003PASP..115..763C} {115, 763}

\bibitem[\protect\citeauthoryear{{C{\^o}t{\'e}}, {Ritter}, {O'Shea}, {Herwig},
  {Pignatari}, {Jones}  \& {Fryer}}{{C{\^o}t{\'e}} et~al.}{2016}]{Cote16}
{C{\^o}t{\'e}} B.,  {Ritter} C.,  {O'Shea} B.~W.,  {Herwig} F.,  {Pignatari}
  M.,  {Jones} S.,   {Fryer} C.~L.,  2016, \mn@doi [\apj]
  {10.3847/0004-637X/824/2/82}, \href
  {https://ui.adsabs.harvard.edu/abs/2016ApJ...824...82C} {824, 82}

\bibitem[\protect\citeauthoryear{{Curti}, {Mannucci}, {Cresci}  \&
  {Maiolino}}{{Curti} et~al.}{2020}]{Curti20}
{Curti} M.,  {Mannucci} F.,  {Cresci} G.,   {Maiolino} R.,  2020, \mn@doi
  [\mnras] {10.1093/mnras/stz2910}, \href
  {https://ui.adsabs.harvard.edu/abs/2020MNRAS.491..944C} {491, 944}

\bibitem[\protect\citeauthoryear{{Curti} et~al.,}{{Curti}
  et~al.}{2023}]{Curti23}
{Curti} M.,  et~al., 2023, \mn@doi [arXiv e-prints]
  {10.48550/arXiv.2304.08516}, \href
  {https://ui.adsabs.harvard.edu/abs/2023arXiv230408516C} {p. arXiv:2304.08516}

\bibitem[\protect\citeauthoryear{{Dayal}, {Ferrara}  \& {Saro}}{{Dayal}
  et~al.}{2010}]{Dayal10}
{Dayal} P.,  {Ferrara} A.,   {Saro} A.,  2010, \mn@doi [\mnras]
  {10.1111/j.1365-2966.2009.15995.x}, \href
  {https://ui.adsabs.harvard.edu/abs/2010MNRAS.402.1449D} {402, 1449}

\bibitem[\protect\citeauthoryear{{Dayal} et~al.,}{{Dayal}
  et~al.}{2022}]{Dayal22}
{Dayal} P.,  et~al., 2022, \mn@doi [\mnras] {10.1093/mnras/stac537}, \href
  {https://ui.adsabs.harvard.edu/abs/2022MNRAS.512..989D} {512, 989}

\bibitem[\protect\citeauthoryear{{De Looze} et~al.,}{{De Looze}
  et~al.}{2014}]{DeLooze14}
{De Looze} I.,  et~al., 2014, \mn@doi [\aap] {10.1051/0004-6361/201322489},
  \href {https://ui.adsabs.harvard.edu/abs/2014A&A...568A..62D} {568, A62}

\bibitem[\protect\citeauthoryear{{De Looze} et~al.,}{{De Looze}
  et~al.}{2020}]{DeLooze20}
{De Looze} I.,  et~al., 2020, \mn@doi [\mnras] {10.1093/mnras/staa1496}, \href
  {https://ui.adsabs.harvard.edu/abs/2020MNRAS.496.3668D} {496, 3668}

\bibitem[\protect\citeauthoryear{{De Vis} et~al.,}{{De Vis}
  et~al.}{2017}]{DeVis17}
{De Vis} P.,  et~al., 2017, \mn@doi [\mnras] {10.1093/mnras/stw2501}, \href
  {https://ui.adsabs.harvard.edu/abs/2017MNRAS.464.4680D} {464, 4680}

\bibitem[\protect\citeauthoryear{{De Vis}, {Maddox}, {Gomez}, {Jones}  \&
  {Dunne}}{{De Vis} et~al.}{2021}]{DeVis21}
{De Vis} P.,  {Maddox} S.~J.,  {Gomez} H.~L.,  {Jones} A.~P.,   {Dunne} L.,
  2021, \mn@doi [\mnras] {10.1093/mnras/stab1604}, \href
  {https://ui.adsabs.harvard.edu/abs/2021MNRAS.505.3228D} {505, 3228}

\bibitem[\protect\citeauthoryear{{Decarli} et~al.,}{{Decarli}
  et~al.}{2018}]{DeCarli18}
{Decarli} R.,  et~al., 2018, \mn@doi [\apj] {10.3847/1538-4357/aaa5aa}, \href
  {https://ui.adsabs.harvard.edu/abs/2018ApJ...854...97D} {854, 97}

\bibitem[\protect\citeauthoryear{{Dell'Agli}, {Garc{\'\i}a-Hern{\'a}ndez},
  {Schneider}, {Ventura}, {La Franca}, {Valiante}, {Marini}  \& {Di
  Criscienzo}}{{Dell'Agli} et~al.}{2017}]{DellAgli17}
{Dell'Agli} F.,  {Garc{\'\i}a-Hern{\'a}ndez} D.~A.,  {Schneider} R.,  {Ventura}
  P.,  {La Franca} F.,  {Valiante} R.,  {Marini} E.,   {Di Criscienzo} M.,
  2017, \mn@doi [\mnras] {10.1093/mnras/stx387}, \href
  {https://ui.adsabs.harvard.edu/abs/2017MNRAS.467.4431D} {467, 4431}

\bibitem[\protect\citeauthoryear{{Di Cesare}, {Graziani}, {Schneider},
  {Ginolfi}, {Venditti}, {Santini}  \& {Hunt}}{{Di Cesare}
  et~al.}{2023}]{DiCesare23}
{Di Cesare} C.,  {Graziani} L.,  {Schneider} R.,  {Ginolfi} M.,  {Venditti} A.,
   {Santini} P.,   {Hunt} L.~K.,  2023, \mn@doi [\mnras]
  {10.1093/mnras/stac3702}, \href
  {https://ui.adsabs.harvard.edu/abs/2023MNRAS.519.4632D} {519, 4632}

\bibitem[\protect\citeauthoryear{{Draine}}{{Draine}}{1989}]{Draine89}
{Draine} B.~T.,  1989, in {B{\"o}hm-Vitense} E.,  ed., Infrared Spectroscopy in
  Astronomy. p.~93

\bibitem[\protect\citeauthoryear{{Draine}}{{Draine}}{2003}]{Draine03}
{Draine} B.~T.,  2003, \mn@doi [\apj] {10.1086/379118}, \href
  {https://ui.adsabs.harvard.edu/abs/2003ApJ...598.1017D} {598, 1017}

\bibitem[\protect\citeauthoryear{{Draine} \& {Hensley}}{{Draine} \&
  {Hensley}}{2012}]{Draine12}
{Draine} B.~T.,  {Hensley} B.,  2012, \mn@doi [\apj]
  {10.1088/0004-637X/757/1/103}, \href
  {https://ui.adsabs.harvard.edu/abs/2012ApJ...757..103D} {757, 103}

\bibitem[\protect\citeauthoryear{{Dwek}}{{Dwek}}{1998}]{Dwek98}
{Dwek} E.,  1998, \mn@doi [\apj] {10.1086/305829}, \href
  {https://ui.adsabs.harvard.edu/abs/1998ApJ...501..643D} {501, 643}

\bibitem[\protect\citeauthoryear{{Dwek}}{{Dwek}}{2004}]{Dwek04}
{Dwek} E.,  2004, \mn@doi [\apj] {10.1086/382653}, \href
  {https://ui.adsabs.harvard.edu/abs/2004ApJ...607..848D} {607, 848}

\bibitem[\protect\citeauthoryear{{Ferrara}, {Viti}  \& {Ceccarelli}}{{Ferrara}
  et~al.}{2016}]{Ferrara16}
{Ferrara} A.,  {Viti} S.,   {Ceccarelli} C.,  2016, \mn@doi [\mnras]
  {10.1093/mnrasl/slw165}, \href
  {https://ui.adsabs.harvard.edu/abs/2016MNRAS.463L.112F} {463, L112}

\bibitem[\protect\citeauthoryear{{Ferrara} et~al.,}{{Ferrara}
  et~al.}{2022}]{Ferrara22}
{Ferrara} A.,  et~al., 2022, \mn@doi [\mnras] {10.1093/mnras/stac460}, \href
  {https://ui.adsabs.harvard.edu/abs/2022MNRAS.512...58F} {512, 58}

\bibitem[\protect\citeauthoryear{{Finkelstein} et~al.,}{{Finkelstein}
  et~al.}{2015}]{Finkelstein15}
{Finkelstein} S.~L.,  et~al., 2015, \mn@doi [\apj]
  {10.1088/0004-637X/810/1/71}, \href
  {https://ui.adsabs.harvard.edu/abs/2015ApJ...810...71F} {810, 71}

\bibitem[\protect\citeauthoryear{{Fudamoto} et~al.,}{{Fudamoto}
  et~al.}{2021}]{Fudamoto21}
{Fudamoto} Y.,  et~al., 2021, \mn@doi [\nat] {10.1038/s41586-021-03846-z},
  \href {https://ui.adsabs.harvard.edu/abs/2021Natur.597..489F} {597, 489}

\bibitem[\protect\citeauthoryear{{Fudamoto} et~al.,}{{Fudamoto}
  et~al.}{2022}]{Fudamoto22}
{Fudamoto} Y.,  et~al., 2022, \mn@doi [\apj] {10.3847/1538-4357/ac7a47}, \href
  {https://ui.adsabs.harvard.edu/abs/2022ApJ...934..144F} {934, 144}

\bibitem[\protect\citeauthoryear{{Fujimoto} et~al.,}{{Fujimoto}
  et~al.}{2022}]{Fujimoto22}
{Fujimoto} S.,  et~al., 2022, \mn@doi [\nat] {10.1038/s41586-022-04454-1},
  \href {https://ui.adsabs.harvard.edu/abs/2022Natur.604..261F} {604, 261}

\bibitem[\protect\citeauthoryear{{Gall} \& {Hjorth}}{{Gall} \&
  {Hjorth}}{2018}]{Gall18}
{Gall} C.,  {Hjorth} J.,  2018, \mn@doi [\apj] {10.3847/1538-4357/aae520},
  \href {https://ui.adsabs.harvard.edu/abs/2018ApJ...868...62G} {868, 62}

\bibitem[\protect\citeauthoryear{{Galliano}, {Galametz}  \& {Jones}}{{Galliano}
  et~al.}{2018}]{Galliano18}
{Galliano} F.,  {Galametz} M.,   {Jones} A.~P.,  2018, \mn@doi [\araa]
  {10.1146/annurev-astro-081817-051900}, \href
  {https://ui.adsabs.harvard.edu/abs/2018ARA&A..56..673G} {56, 673}

\bibitem[\protect\citeauthoryear{{Galliano} et~al.,}{{Galliano}
  et~al.}{2021}]{Galliano21}
{Galliano} F.,  et~al., 2021, \mn@doi [\aap] {10.1051/0004-6361/202039701},
  \href {https://ui.adsabs.harvard.edu/abs/2021A&A...649A..18G} {649, A18}

\bibitem[\protect\citeauthoryear{{Ginolfi}, {Graziani}, {Schneider}, {Marassi},
  {Valiante}, {Dell'Agli}, {Ventura}  \& {Hunt}}{{Ginolfi}
  et~al.}{2018}]{Ginolfi18}
{Ginolfi} M.,  {Graziani} L.,  {Schneider} R.,  {Marassi} S.,  {Valiante} R.,
  {Dell'Agli} F.,  {Ventura} P.,   {Hunt} L.~K.,  2018, \mn@doi [\mnras]
  {10.1093/mnras/stx2572}, \href
  {https://ui.adsabs.harvard.edu/abs/2018MNRAS.473.4538G} {473, 4538}

\bibitem[\protect\citeauthoryear{{Ginolfi} et~al.,}{{Ginolfi}
  et~al.}{2020}]{Ginolfi20}
{Ginolfi} M.,  et~al., 2020, \mn@doi [\aap] {10.1051/0004-6361/201936872},
  \href {https://ui.adsabs.harvard.edu/abs/2020A&A...633A..90G} {633, A90}

\bibitem[\protect\citeauthoryear{{Gioannini}, {Matteucci}  \&
  {Calura}}{{Gioannini} et~al.}{2017}]{Gioannini17}
{Gioannini} L.,  {Matteucci} F.,   {Calura} F.,  2017, \mn@doi [\mnras]
  {10.1093/mnras/stx1914}, \href
  {https://ui.adsabs.harvard.edu/abs/2017MNRAS.471.4615G} {471, 4615}

\bibitem[\protect\citeauthoryear{{Gjergo}, {Palla}, {Matteucci}, {Lacchin},
  {Biviano}  \& {Fan}}{{Gjergo} et~al.}{2020}]{Gjergo20}
{Gjergo} E.,  {Palla} M.,  {Matteucci} F.,  {Lacchin} E.,  {Biviano} A.,
  {Fan} X.,  2020, \mn@doi [\mnras] {10.1093/mnras/staa431}, \href
  {https://ui.adsabs.harvard.edu/abs/2020MNRAS.493.2782G} {493, 2782}

\bibitem[\protect\citeauthoryear{{Graziani}, {Schneider}, {Ginolfi}, {Hunt},
  {Maio}, {Glatzle}  \& {Ciardi}}{{Graziani} et~al.}{2020}]{Graziani20}
{Graziani} L.,  {Schneider} R.,  {Ginolfi} M.,  {Hunt} L.~K.,  {Maio} U.,
  {Glatzle} M.,   {Ciardi} B.,  2020, \mn@doi [\mnras] {10.1093/mnras/staa796},
  \href {https://ui.adsabs.harvard.edu/abs/2020MNRAS.494.1071G} {494, 1071}

\bibitem[\protect\citeauthoryear{{Harikane} et~al.,}{{Harikane}
  et~al.}{2022}]{Harikane22}
{Harikane} Y.,  et~al., 2022, \mn@doi [\apj] {10.3847/1538-4357/ac53a9}, \href
  {https://ui.adsabs.harvard.edu/abs/2022ApJ...929....1H} {929, 1}

\bibitem[\protect\citeauthoryear{{Harikane} et~al.,}{{Harikane}
  et~al.}{2023}]{Harikane23}
{Harikane} Y.,  et~al., 2023, \mn@doi [\apjs] {10.3847/1538-4365/acaaa9}, \href
  {https://ui.adsabs.harvard.edu/abs/2023ApJS..265....5H} {265, 5}

\bibitem[\protect\citeauthoryear{{Hashimoto} et~al.,}{{Hashimoto}
  et~al.}{2019}]{Hashimoto19}
{Hashimoto} T.,  et~al., 2019, \mn@doi [\pasj] {10.1093/pasj/psz049}, \href
  {https://ui.adsabs.harvard.edu/abs/2019PASJ...71...71H} {71, 71}

\bibitem[\protect\citeauthoryear{{Heintz}, {Watson}, {Oesch}, {Narayanan}  \&
  {Madden}}{{Heintz} et~al.}{2021}]{Heintz21}
{Heintz} K.~E.,  {Watson} D.,  {Oesch} P.~A.,  {Narayanan} D.,   {Madden}
  S.~C.,  2021, \mn@doi [\apj] {10.3847/1538-4357/ac2231}, \href
  {https://ui.adsabs.harvard.edu/abs/2021ApJ...922..147H} {922, 147}

\bibitem[\protect\citeauthoryear{{Heintz} et~al.,}{{Heintz}
  et~al.}{2022}]{Heintz22}
{Heintz} K.~E.,  et~al., 2022, \mn@doi [\apjl] {10.3847/2041-8213/ac8057},
  \href {https://ui.adsabs.harvard.edu/abs/2022ApJ...934L..27H} {934, L27}

\bibitem[\protect\citeauthoryear{{Heintz} et~al.,}{{Heintz}
  et~al.}{2023}]{Heintz23}
{Heintz} K.~E.,  et~al., 2023, \mn@doi [\apjl] {10.3847/2041-8213/acb2cf},
  \href {https://ui.adsabs.harvard.edu/abs/2023ApJ...944L..30H} {944, L30}

\bibitem[\protect\citeauthoryear{{Hirashita}}{{Hirashita}}{2015}]{Hirashita15}
{Hirashita} H.,  2015, \mn@doi [\mnras] {10.1093/mnras/stu2617}, \href
  {https://ui.adsabs.harvard.edu/abs/2015MNRAS.447.2937H} {447, 2937}

\bibitem[\protect\citeauthoryear{{Hirashita} \& {Kuo}}{{Hirashita} \&
  {Kuo}}{2011}]{Hirashita11}
{Hirashita} H.,  {Kuo} T.-M.,  2011, \mn@doi [\mnras]
  {10.1111/j.1365-2966.2011.19131.x}, \href
  {https://ui.adsabs.harvard.edu/abs/2011MNRAS.416.1340H} {416, 1340}

\bibitem[\protect\citeauthoryear{{Hirashita} \& {Murga}}{{Hirashita} \&
  {Murga}}{2020}]{Hirashita19}
{Hirashita} H.,  {Murga} M.~S.,  2020, \mn@doi [\mnras]
  {10.1093/mnras/stz3640}, \href
  {https://ui.adsabs.harvard.edu/abs/2020MNRAS.492.3779H} {492, 3779}

\bibitem[\protect\citeauthoryear{{Hodge} \& {da Cunha}}{{Hodge} \& {da
  Cunha}}{2020}]{Hodge20}
{Hodge} J.~A.,  {da Cunha} E.,  2020, \mn@doi [Royal Society Open Science]
  {10.1098/rsos.200556}, \href
  {https://ui.adsabs.harvard.edu/abs/2020RSOS....700556H} {7, 200556}

\bibitem[\protect\citeauthoryear{{Hu}, {Zhukovska}, {Somerville}  \&
  {Naab}}{{Hu} et~al.}{2019}]{Hu19}
{Hu} C.-Y.,  {Zhukovska} S.,  {Somerville} R.~S.,   {Naab} T.,  2019, \mn@doi
  [\mnras] {10.1093/mnras/stz1481}, \href
  {https://ui.adsabs.harvard.edu/abs/2019MNRAS.487.3252H} {487, 3252}

\bibitem[\protect\citeauthoryear{{Hygate} et~al.,}{{Hygate}
  et~al.}{2023}]{Hygate23}
{Hygate} A.~P.~S.,  et~al., 2023, \mn@doi [\mnras] {10.1093/mnras/stad1212},
  \href {https://ui.adsabs.harvard.edu/abs/2023MNRAS.524.1775H} {524, 1775}

\bibitem[\protect\citeauthoryear{{Inami} et~al.,}{{Inami}
  et~al.}{2022}]{Inami22}
{Inami} H.,  et~al., 2022, \mn@doi [\mnras] {10.1093/mnras/stac1779}, \href
  {https://ui.adsabs.harvard.edu/abs/2022MNRAS.515.3126I} {515, 3126}

\bibitem[\protect\citeauthoryear{{Iwamoto}, {Brachwitz}, {Nomoto}, {Kishimoto},
  {Umeda}, {Hix}  \& {Thielemann}}{{Iwamoto} et~al.}{1999}]{Iwamoto99}
{Iwamoto} K.,  {Brachwitz} F.,  {Nomoto} K.,  {Kishimoto} N.,  {Umeda} H.,
  {Hix} W.~R.,   {Thielemann} F.-K.,  1999, \mn@doi [\apjs] {10.1086/313278},
  \href {https://ui.adsabs.harvard.edu/abs/1999ApJS..125..439I} {125, 439}

\bibitem[\protect\citeauthoryear{{Jakobsen} et~al.,}{{Jakobsen}
  et~al.}{2022}]{Jakobsen22}
{Jakobsen} P.,  et~al., 2022, \mn@doi [\aap] {10.1051/0004-6361/202142663},
  \href {https://ui.adsabs.harvard.edu/abs/2022A&A...661A..80J} {661, A80}

\bibitem[\protect\citeauthoryear{{Johnson}, {Leja}, {Conroy}  \&
  {Speagle}}{{Johnson} et~al.}{2021}]{Johnson21}
{Johnson} B.~D.,  {Leja} J.,  {Conroy} C.,   {Speagle} J.~S.,  2021, \mn@doi
  [\apjs] {10.3847/1538-4365/abef67}, \href
  {https://ui.adsabs.harvard.edu/abs/2021ApJS..254...22J} {254, 22}

\bibitem[\protect\citeauthoryear{{Jones}, {Tielens}, {Hollenbach}  \&
  {McKee}}{{Jones} et~al.}{1994}]{Jones94}
{Jones} A.~P.,  {Tielens} A.~G.~G.~M.,  {Hollenbach} D.~J.,   {McKee} C.~F.,
  1994, \mn@doi [\apj] {10.1086/174689}, \href
  {https://ui.adsabs.harvard.edu/abs/1994ApJ...433..797J} {433, 797}

\bibitem[\protect\citeauthoryear{{Kennicutt}}{{Kennicutt}}{1998}]{Kennicutt98}
{Kennicutt} Robert~C. J.,  1998, \mn@doi [\apj] {10.1086/305588}, \href
  {https://ui.adsabs.harvard.edu/abs/1998ApJ...498..541K} {498, 541}

\bibitem[\protect\citeauthoryear{{Kewley} \& {Ellison}}{{Kewley} \&
  {Ellison}}{2008}]{Kewley08}
{Kewley} L.~J.,  {Ellison} S.~L.,  2008, \mn@doi [\apj] {10.1086/587500}, \href
  {https://ui.adsabs.harvard.edu/abs/2008ApJ...681.1183K} {681, 1183}

\bibitem[\protect\citeauthoryear{{Kewley}, {Nicholls}  \&
  {Sutherland}}{{Kewley} et~al.}{2019}]{Kewley19}
{Kewley} L.~J.,  {Nicholls} D.~C.,   {Sutherland} R.~S.,  2019, \mn@doi [\araa]
  {10.1146/annurev-astro-081817-051832}, \href
  {https://ui.adsabs.harvard.edu/abs/2019ARA&A..57..511K} {57, 511}

\bibitem[\protect\citeauthoryear{{Kirchschlager}, {Schmidt}, {Barlow},
  {Fogerty}, {Bevan}  \& {Priestley}}{{Kirchschlager}
  et~al.}{2019}]{Kirchschlager19}
{Kirchschlager} F.,  {Schmidt} F.~D.,  {Barlow} M.~J.,  {Fogerty} E.~L.,
  {Bevan} A.,   {Priestley} F.~D.,  2019, \mn@doi [\mnras]
  {10.1093/mnras/stz2399}, \href
  {https://ui.adsabs.harvard.edu/abs/2019MNRAS.489.4465K} {489, 4465}

\bibitem[\protect\citeauthoryear{{Kirchschlager}, {Schmidt}, {Barlow}, {De
  Looze}  \& {Sartorio}}{{Kirchschlager} et~al.}{2023}]{Kirchschlager23}
{Kirchschlager} F.,  {Schmidt} F.~D.,  {Barlow} M.~J.,  {De Looze} I.,
  {Sartorio} N.~S.,  2023, \mn@doi [\mnras] {10.1093/mnras/stad290}, \href
  {https://ui.adsabs.harvard.edu/abs/2023MNRAS.520.5042K} {520, 5042}

\bibitem[\protect\citeauthoryear{{Kobayashi} \& {Ferrara}}{{Kobayashi} \&
  {Ferrara}}{2023}]{Koba23}
{Kobayashi} C.,  {Ferrara} A.,  2023, \mn@doi [arXiv e-prints]
  {10.48550/arXiv.2308.15583}, \href
  {https://ui.adsabs.harvard.edu/abs/2023arXiv230815583K} {p. arXiv:2308.15583}

\bibitem[\protect\citeauthoryear{{Konstantopoulou} et~al.,}{{Konstantopoulou}
  et~al.}{2022}]{Konstantopoulou22}
{Konstantopoulou} C.,  et~al., 2022, \mn@doi [\aap]
  {10.1051/0004-6361/202243994}, \href
  {https://ui.adsabs.harvard.edu/abs/2022A&A...666A..12K} {666, A12}

\bibitem[\protect\citeauthoryear{{Langeroodi} et~al.,}{{Langeroodi}
  et~al.}{2022}]{Langeroodi22}
{Langeroodi} D.,  et~al., 2022, \mn@doi [arXiv e-prints]
  {10.48550/arXiv.2212.02491}, \href
  {https://ui.adsabs.harvard.edu/abs/2022arXiv221202491L} {p. arXiv:2212.02491}

\bibitem[\protect\citeauthoryear{{Laporte} et~al.,}{{Laporte}
  et~al.}{2017}]{Laporte17}
{Laporte} N.,  et~al., 2017, \mn@doi [\apjl] {10.3847/2041-8213/aa62aa}, \href
  {https://ui.adsabs.harvard.edu/abs/2017ApJ...837L..21L} {837, L21}

\bibitem[\protect\citeauthoryear{{Leja}, {Carnall}, {Johnson}, {Conroy}  \&
  {Speagle}}{{Leja} et~al.}{2019}]{Leja19}
{Leja} J.,  {Carnall} A.~C.,  {Johnson} B.~D.,  {Conroy} C.,   {Speagle} J.~S.,
   2019, \mn@doi [\apj] {10.3847/1538-4357/ab133c}, \href
  {https://ui.adsabs.harvard.edu/abs/2019ApJ...876....3L} {876, 3}

\bibitem[\protect\citeauthoryear{{Li}, {Narayanan}  \& {Dav{\'e}}}{{Li}
  et~al.}{2019}]{Li19}
{Li} Q.,  {Narayanan} D.,   {Dav{\'e}} R.,  2019, \mn@doi [\mnras]
  {10.1093/mnras/stz2684}, \href
  {https://ui.adsabs.harvard.edu/abs/2019MNRAS.490.1425L} {490, 1425}

\bibitem[\protect\citeauthoryear{{Limongi} \& {Chieffi}}{{Limongi} \&
  {Chieffi}}{2018}]{Limongi18}
{Limongi} M.,  {Chieffi} A.,  2018, \mn@doi [\apjs] {10.3847/1538-4365/aacb24},
  \href {https://ui.adsabs.harvard.edu/abs/2018ApJS..237...13L} {237, 13}

\bibitem[\protect\citeauthoryear{{Lodders}}{{Lodders}}{2010}]{Lodders10}
{Lodders} K.,  2010, in Principles and Perspectives in Cosmochemistry. p.~379
  (\mn@eprint {arXiv} {1010.2746}), \mn@doi{10.1007/978-3-642-10352-0_8}

\bibitem[\protect\citeauthoryear{{Maiolino} \& {Mannucci}}{{Maiolino} \&
  {Mannucci}}{2019}]{Maiolino19}
{Maiolino} R.,  {Mannucci} F.,  2019, \mn@doi [\aapr]
  {10.1007/s00159-018-0112-2}, \href
  {https://ui.adsabs.harvard.edu/abs/2019A&ARv..27....3M} {27, 3}

\bibitem[\protect\citeauthoryear{{Mancini}, {Schneider}, {Graziani},
  {Valiante}, {Dayal}, {Maio}, {Ciardi}  \& {Hunt}}{{Mancini}
  et~al.}{2015}]{Mancini15}
{Mancini} M.,  {Schneider} R.,  {Graziani} L.,  {Valiante} R.,  {Dayal} P.,
  {Maio} U.,  {Ciardi} B.,   {Hunt} L.~K.,  2015, \mn@doi [\mnras]
  {10.1093/mnrasl/slv070}, \href
  {https://ui.adsabs.harvard.edu/abs/2015MNRAS.451L..70M} {451, L70}

\bibitem[\protect\citeauthoryear{{Maoz} \& {Graur}}{{Maoz} \&
  {Graur}}{2017}]{Maoz17}
{Maoz} D.,  {Graur} O.,  2017, \mn@doi [\apj] {10.3847/1538-4357/aa8b6e}, \href
  {https://ui.adsabs.harvard.edu/abs/2017ApJ...848...25M} {848, 25}

\bibitem[\protect\citeauthoryear{{Marassi}, {Schneider}, {Limongi}, {Chieffi},
  {Graziani}  \& {Bianchi}}{{Marassi} et~al.}{2019}]{Marassi19}
{Marassi} S.,  {Schneider} R.,  {Limongi} M.,  {Chieffi} A.,  {Graziani} L.,
  {Bianchi} S.,  2019, \mn@doi [\mnras] {10.1093/mnras/sty3323}, \href
  {https://ui.adsabs.harvard.edu/abs/2019MNRAS.484.2587M} {484, 2587}

\bibitem[\protect\citeauthoryear{{Mart{\'\i}nez-Gonz{\'a}lez}, {W{\"u}nsch},
  {Silich}, {Tenorio-Tagle}, {Palou{\v{s}}}  \&
  {Ferrara}}{{Mart{\'\i}nez-Gonz{\'a}lez} et~al.}{2019}]{MartinezGonzales19}
{Mart{\'\i}nez-Gonz{\'a}lez} S.,  {W{\"u}nsch} R.,  {Silich} S.,
  {Tenorio-Tagle} G.,  {Palou{\v{s}}} J.,   {Ferrara} A.,  2019, \mn@doi [\apj]
  {10.3847/1538-4357/ab571b}, \href
  {https://ui.adsabs.harvard.edu/abs/2019ApJ...887..198M} {887, 198}

\bibitem[\protect\citeauthoryear{{Matteucci}}{{Matteucci}}{2012}]{Matteucci12}
{Matteucci} F.,  2012, {Chemical Evolution of Galaxies},
  \mn@doi{10.1007/978-3-642-22491-1.
}

\bibitem[\protect\citeauthoryear{{Matteucci}}{{Matteucci}}{2021}]{Matteucci21}
{Matteucci} F.,  2021, \mn@doi [\aapr] {10.1007/s00159-021-00133-8}, \href
  {https://ui.adsabs.harvard.edu/abs/2021A&ARv..29....5M} {29, 5}

\bibitem[\protect\citeauthoryear{{Mattsson}, {Andersen}  \&
  {Munkhammar}}{{Mattsson} et~al.}{2012}]{Mattsson12}
{Mattsson} L.,  {Andersen} A.~C.,   {Munkhammar} J.~D.,  2012, \mn@doi [\mnras]
  {10.1111/j.1365-2966.2012.20575.x}, \href
  {https://ui.adsabs.harvard.edu/abs/2012MNRAS.423...26M} {423, 26}

\bibitem[\protect\citeauthoryear{{McKee}}{{McKee}}{1989}]{McKee89}
{McKee} C.,  1989, in {Allamandola} L.~J.,  {Tielens} A.~G.~G.~M.,  eds,  Vol.
  135, Interstellar Dust. p.~431

\bibitem[\protect\citeauthoryear{{McKinnon}, {Vogelsberger}, {Torrey},
  {Marinacci}  \& {Kannan}}{{McKinnon} et~al.}{2018}]{Mckinnon18}
{McKinnon} R.,  {Vogelsberger} M.,  {Torrey} P.,  {Marinacci} F.,   {Kannan}
  R.,  2018, \mn@doi [\mnras] {10.1093/mnras/sty1248}, \href
  {https://ui.adsabs.harvard.edu/abs/2018MNRAS.478.2851M} {478, 2851}

\bibitem[\protect\citeauthoryear{{McLure} et~al.,}{{McLure}
  et~al.}{2013}]{McLure13}
{McLure} R.~J.,  et~al., 2013, \mn@doi [\mnras] {10.1093/mnras/stt627}, \href
  {https://ui.adsabs.harvard.edu/abs/2013MNRAS.432.2696M} {432, 2696}

\bibitem[\protect\citeauthoryear{{Micelotta}, {Matsuura}  \&
  {Sarangi}}{{Micelotta} et~al.}{2019}]{Micelotta19}
{Micelotta} E.~R.,  {Matsuura} M.,   {Sarangi} A.,  2019, in {Bykov} A.,
  {Roger} C.,  {Raymond} J.,  {Thielemann} F.-K.,  {Falanga} M.,   {von
  Steiger} R.,  eds, , Vol.~68, Supernovae. Series: Space Sciences Series of
  ISSI.
pp 361--418, \mn@doi{10.1007/978-94-024-1581-0_14}

\bibitem[\protect\citeauthoryear{{Naab} \& {Ostriker}}{{Naab} \&
  {Ostriker}}{2017}]{Naab17}
{Naab} T.,  {Ostriker} J.~P.,  2017, \mn@doi [\araa]
  {10.1146/annurev-astro-081913-040019}, \href
  {https://ui.adsabs.harvard.edu/abs/2017ARA&A..55...59N} {55, 59}

\bibitem[\protect\citeauthoryear{{Naidu} et~al.,}{{Naidu}
  et~al.}{2022}]{Naidu22}
{Naidu} R.~P.,  et~al., 2022, \mn@doi [\apjl] {10.3847/2041-8213/ac9b22}, \href
  {https://ui.adsabs.harvard.edu/abs/2022ApJ...940L..14N} {940, L14}

\bibitem[\protect\citeauthoryear{{Nakajima}, {Ouchi}, {Isobe}, {Harikane},
  {Zhang}, {Ono}, {Umeda}  \& {Oguri}}{{Nakajima} et~al.}{2023}]{Nakajima23}
{Nakajima} K.,  {Ouchi} M.,  {Isobe} Y.,  {Harikane} Y.,  {Zhang} Y.,  {Ono}
  Y.,  {Umeda} H.,   {Oguri} M.,  2023, \mn@doi [arXiv e-prints]
  {10.48550/arXiv.2301.12825}, \href
  {https://ui.adsabs.harvard.edu/abs/2023arXiv230112825N} {p. arXiv:2301.12825}

\bibitem[\protect\citeauthoryear{{Nanni}, {Burgarella}, {Theul{\'e}},
  {C{\^o}t{\'e}}  \& {Hirashita}}{{Nanni} et~al.}{2020}]{Nanni20}
{Nanni} A.,  {Burgarella} D.,  {Theul{\'e}} P.,  {C{\^o}t{\'e}} B.,
  {Hirashita} H.,  2020, \mn@doi [\aap] {10.1051/0004-6361/202037833}, \href
  {https://ui.adsabs.harvard.edu/abs/2020A&A...641A.168N} {641, A168}

\bibitem[\protect\citeauthoryear{{Oesch} et~al.,}{{Oesch}
  et~al.}{2016}]{Oesch16}
{Oesch} P.~A.,  et~al., 2016, \mn@doi [\apj] {10.3847/0004-637X/819/2/129},
  \href {https://ui.adsabs.harvard.edu/abs/2016ApJ...819..129O} {819, 129}

\bibitem[\protect\citeauthoryear{{Palla}}{{Palla}}{2021}]{Palla21}
{Palla} M.,  2021, \mn@doi [\mnras] {10.1093/mnras/stab293}, \href
  {https://ui.adsabs.harvard.edu/abs/2021MNRAS.503.3216P} {503, 3216}

\bibitem[\protect\citeauthoryear{{Palla}, {Calura}, {Matteucci}, {Fan},
  {Vincenzo}  \& {Lacchin}}{{Palla} et~al.}{2020a}]{Palla20b}
{Palla} M.,  {Calura} F.,  {Matteucci} F.,  {Fan} X.~L.,  {Vincenzo} F.,
  {Lacchin} E.,  2020a, \mn@doi [\mnras] {10.1093/mnras/staa848}, \href
  {https://ui.adsabs.harvard.edu/abs/2020MNRAS.494.2355P} {494, 2355}

\bibitem[\protect\citeauthoryear{{Palla}, {Matteucci}, {Calura}  \&
  {Longo}}{{Palla} et~al.}{2020b}]{Palla20a}
{Palla} M.,  {Matteucci} F.,  {Calura} F.,   {Longo} F.,  2020b, \mn@doi [\apj]
  {10.3847/1538-4357/ab6080}, \href
  {https://ui.adsabs.harvard.edu/abs/2020ApJ...889....4P} {889, 4}

\bibitem[\protect\citeauthoryear{{Palla}, {Santos-Peral}, {Recio-Blanco}  \&
  {Matteucci}}{{Palla} et~al.}{2022}]{Palla22}
{Palla} M.,  {Santos-Peral} P.,  {Recio-Blanco} A.,   {Matteucci} F.,  2022,
  \mn@doi [\aap] {10.1051/0004-6361/202142645}, \href
  {https://ui.adsabs.harvard.edu/abs/2022A&A...663A.125P} {663, A125}

\bibitem[\protect\citeauthoryear{{P{\'e}roux} \& {Howk}}{{P{\'e}roux} \&
  {Howk}}{2020}]{Peroux20}
{P{\'e}roux} C.,  {Howk} J.~C.,  2020, \mn@doi [\araa]
  {10.1146/annurev-astro-021820-120014}, \href
  {https://ui.adsabs.harvard.edu/abs/2020ARA&A..58..363P} {58, 363}

\bibitem[\protect\citeauthoryear{{Pizzati}, {Ferrara}, {Pallottini},
  {Gallerani}, {Vallini}, {Decataldo}  \& {Fujimoto}}{{Pizzati}
  et~al.}{2020}]{Pizzati20}
{Pizzati} E.,  {Ferrara} A.,  {Pallottini} A.,  {Gallerani} S.,  {Vallini} L.,
  {Decataldo} D.,   {Fujimoto} S.,  2020, \mn@doi [\mnras]
  {10.1093/mnras/staa1163}, \href
  {https://ui.adsabs.harvard.edu/abs/2020MNRAS.495..160P} {495, 160}

\bibitem[\protect\citeauthoryear{{Popping} \& {P{\'e}roux}}{{Popping} \&
  {P{\'e}roux}}{2022}]{Popping22}
{Popping} G.,  {P{\'e}roux} C.,  2022, \mn@doi [\mnras]
  {10.1093/mnras/stac695}, \href
  {https://ui.adsabs.harvard.edu/abs/2022MNRAS.513.1531P} {513, 1531}

\bibitem[\protect\citeauthoryear{{Popping}, {Somerville}  \&
  {Galametz}}{{Popping} et~al.}{2017}]{Popping17}
{Popping} G.,  {Somerville} R.~S.,   {Galametz} M.,  2017, \mn@doi [\mnras]
  {10.1093/mnras/stx1545}, \href
  {https://ui.adsabs.harvard.edu/abs/2017MNRAS.471.3152P} {471, 3152}

\bibitem[\protect\citeauthoryear{{Priestley}, {Chawner}, {Matsuura}, {De
  Looze}, {Barlow}  \& {Gomez}}{{Priestley} et~al.}{2021}]{Priestley21}
{Priestley} F.~D.,  {Chawner} H.,  {Matsuura} M.,  {De Looze} I.,  {Barlow}
  M.~J.,   {Gomez} H.~L.,  2021, \mn@doi [\mnras] {10.1093/mnras/staa3445},
  \href {https://ui.adsabs.harvard.edu/abs/2021MNRAS.500.2543P} {500, 2543}

\bibitem[\protect\citeauthoryear{{Priestley}, {De Looze}  \&
  {Barlow}}{{Priestley} et~al.}{2022}]{Priestley22}
{Priestley} F.~D.,  {De Looze} I.,   {Barlow} M.~J.,  2022, \mn@doi [\mnras]
  {10.1093/mnrasl/slab114}, \href
  {https://ui.adsabs.harvard.edu/abs/2022MNRAS.509L...6P} {509, L6}

\bibitem[\protect\citeauthoryear{{R{\'e}my-Ruyer} et~al.,}{{R{\'e}my-Ruyer}
  et~al.}{2014}]{RemyRuyer14}
{R{\'e}my-Ruyer} A.,  et~al., 2014, \mn@doi [\aap]
  {10.1051/0004-6361/201322803}, \href
  {https://ui.adsabs.harvard.edu/abs/2014A&A...563A..31R} {563, A31}

\bibitem[\protect\citeauthoryear{{R{\'e}my-Ruyer} et~al.,}{{R{\'e}my-Ruyer}
  et~al.}{2015}]{RemyRuyer15}
{R{\'e}my-Ruyer} A.,  et~al., 2015, \mn@doi [\aap]
  {10.1051/0004-6361/201526067}, \href
  {https://ui.adsabs.harvard.edu/abs/2015A&A...582A.121R} {582, A121}

\bibitem[\protect\citeauthoryear{{Reuter} et~al.,}{{Reuter}
  et~al.}{2020}]{Reuter20}
{Reuter} C.,  et~al., 2020, \mn@doi [\apj] {10.3847/1538-4357/abb599}, \href
  {https://ui.adsabs.harvard.edu/abs/2020ApJ...902...78R} {902, 78}

\bibitem[\protect\citeauthoryear{{Riechers} et~al.,}{{Riechers}
  et~al.}{2013}]{Riechers13}
{Riechers} D.~A.,  et~al., 2013, \mn@doi [\nat] {10.1038/nature12050}, \href
  {https://ui.adsabs.harvard.edu/abs/2013Natur.496..329R} {496, 329}

\bibitem[\protect\citeauthoryear{{Romano}, {Karakas}, {Tosi}  \&
  {Matteucci}}{{Romano} et~al.}{2010}]{Romano10}
{Romano} D.,  {Karakas} A.~I.,  {Tosi} M.,   {Matteucci} F.,  2010, \mn@doi
  [\aap] {10.1051/0004-6361/201014483}, \href
  {https://ui.adsabs.harvard.edu/abs/2010A&A...522A..32R} {522, A32}

\bibitem[\protect\citeauthoryear{{Romano}, {Matteucci}, {Zhang}, {Papadopoulos}
   \& {Ivison}}{{Romano} et~al.}{2017}]{Romano17}
{Romano} D.,  {Matteucci} F.,  {Zhang} Z.~Y.,  {Papadopoulos} P.~P.,   {Ivison}
  R.~J.,  2017, \mn@doi [\mnras] {10.1093/mnras/stx1197}, \href
  {https://ui.adsabs.harvard.edu/abs/2017MNRAS.470..401R} {470, 401}

\bibitem[\protect\citeauthoryear{{Sanders} et~al.,}{{Sanders}
  et~al.}{2021}]{Sanders21}
{Sanders} R.~L.,  et~al., 2021, \mn@doi [\apj] {10.3847/1538-4357/abf4c1},
  \href {https://ui.adsabs.harvard.edu/abs/2021ApJ...914...19S} {914, 19}

\bibitem[\protect\citeauthoryear{{Savage} \& {Sembach}}{{Savage} \&
  {Sembach}}{1996}]{Savage96}
{Savage} B.~D.,  {Sembach} K.~R.,  1996, \mn@doi [\araa]
  {10.1146/annurev.astro.34.1.279}, \href
  {https://ui.adsabs.harvard.edu/abs/1996ARA&A..34..279S} {34, 279}

\bibitem[\protect\citeauthoryear{{Schaerer} et~al.,}{{Schaerer}
  et~al.}{2020}]{Schaerer20}
{Schaerer} D.,  et~al., 2020, \mn@doi [\aap] {10.1051/0004-6361/202037617},
  \href {https://ui.adsabs.harvard.edu/abs/2020A&A...643A...3S} {643, A3}

\bibitem[\protect\citeauthoryear{{Schneider} \& {Maiolino}}{{Schneider} \&
  {Maiolino}}{2023}]{Schneider23}
{Schneider} R.,  {Maiolino} R.,  2023, \mn@doi [arXiv e-prints]
  {10.48550/arXiv.2310.00053}, \href
  {https://ui.adsabs.harvard.edu/abs/2023arXiv231000053S} {p. arXiv:2310.00053}

\bibitem[\protect\citeauthoryear{{Schneider}, {Hunt}  \&
  {Valiante}}{{Schneider} et~al.}{2016}]{Schneider16}
{Schneider} R.,  {Hunt} L.,   {Valiante} R.,  2016, \mn@doi [\mnras]
  {10.1093/mnras/stw114}, \href
  {https://ui.adsabs.harvard.edu/abs/2016MNRAS.457.1842S} {457, 1842}

\bibitem[\protect\citeauthoryear{{Schouws} et~al.,}{{Schouws}
  et~al.}{2022}]{Schouws22}
{Schouws} S.,  et~al., 2022, \mn@doi [\apj] {10.3847/1538-4357/ac4605}, \href
  {https://ui.adsabs.harvard.edu/abs/2022ApJ...928...31S} {928, 31}

\bibitem[\protect\citeauthoryear{{Shapley}, {Cullen}, {Dunlop}, {McLure},
  {Kriek}, {Reddy}  \& {Sanders}}{{Shapley} et~al.}{2020}]{Shapley20}
{Shapley} A.~E.,  {Cullen} F.,  {Dunlop} J.~S.,  {McLure} R.~J.,  {Kriek} M.,
  {Reddy} N.~A.,   {Sanders} R.~L.,  2020, \mn@doi [\apjl]
  {10.3847/2041-8213/abc006}, \href
  {https://ui.adsabs.harvard.edu/abs/2020ApJ...903L..16S} {903, L16}

\bibitem[\protect\citeauthoryear{{Slavin}, {Dwek}, {Mac Low}  \&
  {Hill}}{{Slavin} et~al.}{2020}]{Slavin20}
{Slavin} J.~D.,  {Dwek} E.,  {Mac Low} M.-M.,   {Hill} A.~S.,  2020, \mn@doi
  [\apj] {10.3847/1538-4357/abb5a4}, \href
  {https://ui.adsabs.harvard.edu/abs/2020ApJ...902..135S} {902, 135}

\bibitem[\protect\citeauthoryear{{Sommovigo}, {Ferrara}, {Carniani}, {Zanella},
  {Pallottini}, {Gallerani}  \& {Vallini}}{{Sommovigo}
  et~al.}{2021}]{Sommovigo21}
{Sommovigo} L.,  {Ferrara} A.,  {Carniani} S.,  {Zanella} A.,  {Pallottini} A.,
   {Gallerani} S.,   {Vallini} L.,  2021, \mn@doi [\mnras]
  {10.1093/mnras/stab720}, \href
  {https://ui.adsabs.harvard.edu/abs/2021MNRAS.503.4878S} {503, 4878}

\bibitem[\protect\citeauthoryear{{Sommovigo} et~al.,}{{Sommovigo}
  et~al.}{2022a}]{Sommovigo22a}
{Sommovigo} L.,  et~al., 2022a, \mn@doi [\mnras] {10.1093/mnras/stac302}, \href
  {https://ui.adsabs.harvard.edu/abs/2022MNRAS.513.3122S} {513, 3122}

\bibitem[\protect\citeauthoryear{{Sommovigo} et~al.,}{{Sommovigo}
  et~al.}{2022b}]{Sommovigo22b}
{Sommovigo} L.,  et~al., 2022b, \mn@doi [\mnras] {10.1093/mnras/stac2997},
  \href {https://ui.adsabs.harvard.edu/abs/2022MNRAS.517.5930S} {517, 5930}

\bibitem[\protect\citeauthoryear{{Stefanon} et~al.,}{{Stefanon}
  et~al.}{2019}]{Stefanon19}
{Stefanon} M.,  et~al., 2019, \mn@doi [\apj] {10.3847/1538-4357/ab3792}, \href
  {https://ui.adsabs.harvard.edu/abs/2019ApJ...883...99S} {883, 99}

\bibitem[\protect\citeauthoryear{{Tamura} et~al.,}{{Tamura}
  et~al.}{2019}]{Tamura18}
{Tamura} Y.,  et~al., 2019, \mn@doi [\apj] {10.3847/1538-4357/ab0374}, \href
  {https://ui.adsabs.harvard.edu/abs/2019ApJ...874...27T} {874, 27}

\bibitem[\protect\citeauthoryear{{Topping} et~al.,}{{Topping}
  et~al.}{2022}]{Topping22}
{Topping} M.~W.,  et~al., 2022, \mn@doi [\mnras] {10.1093/mnras/stac2291},
  \href {https://ui.adsabs.harvard.edu/abs/2022MNRAS.516..975T} {516, 975}

\bibitem[\protect\citeauthoryear{{Totani}, {Morokuma}, {Oda}, {Doi}  \&
  {Yasuda}}{{Totani} et~al.}{2008}]{Totani08}
{Totani} T.,  {Morokuma} T.,  {Oda} T.,  {Doi} M.,   {Yasuda} N.,  2008,
  \mn@doi [\pasj] {10.1093/pasj/60.6.1327}, \href
  {https://ui.adsabs.harvard.edu/abs/2008PASJ...60.1327T} {60, 1327}

\bibitem[\protect\citeauthoryear{{Triani}, {Sinha}, {Croton}, {Pacifici}  \&
  {Dwek}}{{Triani} et~al.}{2020}]{Triani20}
{Triani} D.~P.,  {Sinha} M.,  {Croton} D.~J.,  {Pacifici} C.,   {Dwek} E.,
  2020, \mn@doi [\mnras] {10.1093/mnras/staa446}, \href
  {https://ui.adsabs.harvard.edu/abs/2020MNRAS.493.2490T} {493, 2490}

\bibitem[\protect\citeauthoryear{{Valiante}, {Schneider}, {Bianchi}  \&
  {Andersen}}{{Valiante} et~al.}{2009}]{Valiante09}
{Valiante} R.,  {Schneider} R.,  {Bianchi} S.,   {Andersen} A.~C.,  2009,
  \mn@doi [\mnras] {10.1111/j.1365-2966.2009.15076.x}, \href
  {https://ui.adsabs.harvard.edu/abs/2009MNRAS.397.1661V} {397, 1661}

\bibitem[\protect\citeauthoryear{{Valiante}, {Schneider}, {Salvadori}  \&
  {Gallerani}}{{Valiante} et~al.}{2014}]{Valiante14}
{Valiante} R.,  {Schneider} R.,  {Salvadori} S.,   {Gallerani} S.,  2014,
  \mn@doi [\mnras] {10.1093/mnras/stu1613}, \href
  {https://ui.adsabs.harvard.edu/abs/2014MNRAS.444.2442V} {444, 2442}

\bibitem[\protect\citeauthoryear{{Valiante}, {Gioannini}, {Schneider},
  {Matteucci}, {Dell'Agli}  \& {Di Criscienzo}}{{Valiante}
  et~al.}{2017}]{Valiante17}
{Valiante} R.,  {Gioannini} L.,  {Schneider} R.,  {Matteucci} F.,  {Dell'Agli}
  F.,   {Di Criscienzo} M.,  2017, \memsai, \href
  {https://ui.adsabs.harvard.edu/abs/2017MmSAI..88..420V} {88, 420}

\bibitem[\protect\citeauthoryear{{Venemans} et~al.,}{{Venemans}
  et~al.}{2018}]{Venemans18}
{Venemans} B.~P.,  et~al., 2018, \mn@doi [\apj] {10.3847/1538-4357/aadf35},
  \href {https://ui.adsabs.harvard.edu/abs/2018ApJ...866..159V} {866, 159}

\bibitem[\protect\citeauthoryear{{Ventura}, {Di Criscienzo}, {Carini}  \&
  {D'Antona}}{{Ventura} et~al.}{2013}]{Ventura13}
{Ventura} P.,  {Di Criscienzo} M.,  {Carini} R.,   {D'Antona} F.,  2013,
  \mn@doi [\mnras] {10.1093/mnras/stt444}, \href
  {https://ui.adsabs.harvard.edu/abs/2013MNRAS.431.3642V} {431, 3642}

\bibitem[\protect\citeauthoryear{{Ventura}, {Karakas}, {Dell'Agli},
  {Garc{\'\i}a-Hern{\'a}ndez}  \& {Guzman-Ramirez}}{{Ventura}
  et~al.}{2018}]{Ventura18}
{Ventura} P.,  {Karakas} A.,  {Dell'Agli} F.,  {Garc{\'\i}a-Hern{\'a}ndez}
  D.~A.,   {Guzman-Ramirez} L.,  2018, \mn@doi [\mnras]
  {10.1093/mnras/stx3338}, \href
  {https://ui.adsabs.harvard.edu/abs/2018MNRAS.475.2282V} {475, 2282}

\bibitem[\protect\citeauthoryear{{Ventura}, {Dell'Agli}, {Lugaro}, {Romano},
  {Tailo}  \& {Yag{\"u}e}}{{Ventura} et~al.}{2020}]{Ventura20}
{Ventura} P.,  {Dell'Agli} F.,  {Lugaro} M.,  {Romano} D.,  {Tailo} M.,
  {Yag{\"u}e} A.,  2020, \mn@doi [\aap] {10.1051/0004-6361/202038289}, \href
  {https://ui.adsabs.harvard.edu/abs/2020A&A...641A.103V} {641, A103}

\bibitem[\protect\citeauthoryear{{Ventura} et~al.,}{{Ventura}
  et~al.}{2021}]{Ventura21}
{Ventura} P.,  et~al., 2021, \mn@doi [\aap] {10.1051/0004-6361/202141017},
  \href {https://ui.adsabs.harvard.edu/abs/2021A&A...655A...6V} {655, A6}

\bibitem[\protect\citeauthoryear{{Vincenzo}, {Matteucci}  \&
  {Spitoni}}{{Vincenzo} et~al.}{2017}]{Vincenzo17}
{Vincenzo} F.,  {Matteucci} F.,   {Spitoni} E.,  2017, \mn@doi [\mnras]
  {10.1093/mnras/stw3369}, \href
  {https://ui.adsabs.harvard.edu/abs/2017MNRAS.466.2939V} {466, 2939}

\bibitem[\protect\citeauthoryear{{Vizgan}, {Heintz}, {Greve}, {Narayanan},
  {Dav{\'e}}, {Olsen}, {Popping}  \& {Watson}}{{Vizgan}
  et~al.}{2022}]{Vizgan22}
{Vizgan} D.,  {Heintz} K.~E.,  {Greve} T.~R.,  {Narayanan} D.,  {Dav{\'e}} R.,
  {Olsen} K.~P.,  {Popping} G.,   {Watson} D.,  2022, \mn@doi [\apjl]
  {10.3847/2041-8213/ac982c}, \href
  {https://ui.adsabs.harvard.edu/abs/2022ApJ...939L...1V} {939, L1}

\bibitem[\protect\citeauthoryear{{Vladilo}, {Gioannini}, {Matteucci}  \&
  {Palla}}{{Vladilo} et~al.}{2018}]{Vladilo18}
{Vladilo} G.,  {Gioannini} L.,  {Matteucci} F.,   {Palla} M.,  2018, \mn@doi
  [\apj] {10.3847/1538-4357/aae8dc}, \href
  {https://ui.adsabs.harvard.edu/abs/2018ApJ...868..127V} {868, 127}

\bibitem[\protect\citeauthoryear{{Watson}, {Christensen}, {Knudsen}, {Richard},
  {Gallazzi}  \& {Micha{\l}owski}}{{Watson} et~al.}{2015}]{Watson15}
{Watson} D.,  {Christensen} L.,  {Knudsen} K.~K.,  {Richard} J.,  {Gallazzi}
  A.,   {Micha{\l}owski} M.~J.,  2015, \mn@doi [\nat] {10.1038/nature14164},
  \href {https://ui.adsabs.harvard.edu/abs/2015Natur.519..327W} {519, 327}

\bibitem[\protect\citeauthoryear{{Wilson} et~al.,}{{Wilson}
  et~al.}{2023}]{Wilson23}
{Wilson} S.~N.,  et~al., 2023, \mn@doi [arXiv e-prints]
  {10.48550/arXiv.2305.05213}, \href
  {https://ui.adsabs.harvard.edu/abs/2023arXiv230505213W} {p. arXiv:2305.05213}

\bibitem[\protect\citeauthoryear{{Yates}, {Henriques}, {Thomas}, {Kauffmann},
  {Johansson}  \& {White}}{{Yates} et~al.}{2013}]{Yates13}
{Yates} R.~M.,  {Henriques} B.,  {Thomas} P.~A.,  {Kauffmann} G.,  {Johansson}
  J.,   {White} S. D.~M.,  2013, \mn@doi [\mnras] {10.1093/mnras/stt1542},
  \href {https://ui.adsabs.harvard.edu/abs/2013MNRAS.435.3500Y} {435, 3500}

\bibitem[\protect\citeauthoryear{{Yin}, {Matteucci}  \& {Vladilo}}{{Yin}
  et~al.}{2011}]{Jin11}
{Yin} J.,  {Matteucci} F.,   {Vladilo} G.,  2011, \mn@doi [\aap]
  {10.1051/0004-6361/201015022}, \href
  {https://ui.adsabs.harvard.edu/abs/2011A&A...531A.136Y} {531, A136}

\bibitem[\protect\citeauthoryear{{Zanella} et~al.,}{{Zanella}
  et~al.}{2018}]{Zanella18}
{Zanella} A.,  et~al., 2018, \mn@doi [\mnras] {10.1093/mnras/sty2394}, \href
  {https://ui.adsabs.harvard.edu/abs/2018MNRAS.481.1976Z} {481, 1976}

\bibitem[\protect\citeauthoryear{{Zhukovska}}{{Zhukovska}}{2014}]{Zhukovska14}
{Zhukovska} S.,  2014, \mn@doi [\aap] {10.1051/0004-6361/201322989}, \href
  {https://ui.adsabs.harvard.edu/abs/2014A&A...562A..76Z} {562, A76}

\bibitem[\protect\citeauthoryear{{de Bennassuti}, {Schneider}, {Valiante}  \&
  {Salvadori}}{{de Bennassuti} et~al.}{2014}]{deBenassutti14}
{de Bennassuti} M.,  {Schneider} R.,  {Valiante} R.,   {Salvadori} S.,  2014,
  in Journal of Physics Conference Series. p. 012010,
  \mn@doi{10.1088/1742-6596/566/1/012010}

\makeatother
\end{thebibliography}




\appendix

\section{Maximum dust-to-metal (DtM) ratio}
\label{app:DtMmax}

In the paper, we adopt a maximum value for the dust-to-metal (DtM) ratio of 0.6. In this appendix, we explain why this value should be considered basing on simple consideration on elemental abundances and known dust compounds in galaxies.\\

If we consider a solar abundance pattern (\citealt{Asplund09}), we get that the number abundance of Mg, Si, Fe atoms is similar, i.e. $\log({\rm Mg/H})+12\simeq 7.61$, $\log({\rm Si/H})+12\simeq 7.51$, $\log({\rm Fe/H})+12\simeq 7.5$. Other fundamental dust constituents, such as O and C, instead have larger abundances, i.e. $\log({\rm O/H})+12=8.69$, $\log({\rm C/H})+12=8.43$.

We then assume maximum dust depletion for Mg, Si and Fe.
Most of Mg, Si and Fe are found in silicates such as olivines (Mg$_{2x}$ Fe$_{2(1-x)}$ Si O$_4$) and pyroxenes (Mg$_x$ Fe$_{(1-x)}$ Si O$_3$). However, Fe  may also be incorporated in another form of dust, such as metallic Fe particles (\citealt{Dwek04,Draine12,Vladilo18}). In any case, the molecules stoichiometry imposes an intrinsic limit to the level of O depleted by dust in the ISM.

To simplify the calculations, let us assume that silicates compounds are either only pyroxenes or olivines.
In the case we have 100\% of silicates are pyroxenes, then we need 3 oxygen atoms for a Mg+Si or Fe+Si couple. However, Fe is also depleted in other dust forms, so to maximise the level of dust depletion in the ISM we consider only the enstatite compound (MgSiO$_{3}$). For this molecule, for every Si atom (which is less abundant in the ISM) 3 O atoms are needed, whereas around 15 O atoms per Si are available in the ISM. This means that the number of O atoms trapped in silicates are around 20\% of the total.
In the case instead we consider silicates as olivines, for a forsterite compound (Mg$_2$SiO$_4$) we need 2 O atoms per Mg one (with the latter fixing maximum number of molecules that can form). For every Mg atom, more than 12 O atoms are available: therefore, the number of O atoms trapped in forsterite are around 15\% of the total. However, the 20\% of Si remains spare. The latter can form ferrosilite (FeSiO$_3$) or fayalite (Fe$_2$SiO$_4$) molecules. To maximise the O depletion, let us assume that fayalite is formed. In this case around 10\% of oxygen atoms remain trapped in this molecule. Summing it with the fraction of O in forsterite, we get a O depletion level around 25\%.

C dust is instead mostly found in the ISM in the form of graphite and amorphous carbon. However, a significant fraction of C is locked in the very stable CO molecule (around 40\% locally, \citealt{Lodders10,Agundez13}). 
If we consider a lower limit of 20\% of C atoms locked CO, then the maximum level of C depletion can arrive to 80\%.

Up to now we have worked using number ratios between elements. However, the DtM ratio is a mass ratio: therefore, we need to convert the number ratios into mass ratios. By doing and this and then by summing up all the elemental contributions to ISM dust, we obtain that around 50\% of all ISM metals can be depleted into dust.\\

Our calculations approximate dust as formed by a limited number of compounds. Other dust molecules (e.g. silicon carbide, SiC) are also present, however being much less abundant. At the same time, it was shown that interstellar ices can accrete silicate grains and form mantles around them. Nonetheless, several works indicate that these mantles are unlikely to survive to the harsh radiation field from new stellar generations, especially at high redshift (\citealt{Ferrara16,Gall18}). This indicates very short lifetimes and therefore a very small contribution to the total dust mass budget in galactic evolution. 
For these reasons, it is safe to assume for each model timestep a slightly larger limit for the total dust mass, which translate into a maximum DtM of 0.6.
It is also worth noting that such limit was also found in ISM dust depletion studies (e.g. \citealt{Konstantopoulou22}) in several environments, from the MW galaxy up to $z\gtrsim2$ Damped Lyman-$\alpha$ systems.\\

On another note, the interstellar abundance pattern considered for the above calculations is the solar one.
Nonetheless the values displayed can be easy generalised in case of a subsolar chemical enrichment, as the one tested in this paper. In fact, while the magnesium and carbon to oxygen ratios are similar to those of the solar system, Si and mainly Fe are underproduced. This means less seeds for oxygen atoms depleted into dust and therefore slightly smaller maximal DtM.

\section{Galactic evolution parameters}
\label{app:parameters}

\defcitealias{Limongi18}{LC18}
\defcitealias{Iwamoto99}{I99}
\defcitealias{Ventura21}{Vmult}
\defcitealias{Marassi19}{M19}

Throughout Section \ref{s:models} of the paper, we listed and justified the prescriptions and assumptions adopted for model parameters, both at the level of chemical evolution and dust enrichment.

In the following, we summarise our choices to allow a more immediate view and comparison between the model results displayed from Fig. \ref{fig:metallicities} onward.
In particular, in Tab. \ref{tab:param_chem} we list the parameter setup for the parameters responsible for the galactic chemical evolution, while in Tab. \ref{tab:param_dust} we list the parameter choices for different dust evolutionary scenario.

\begin{table*}
    \centering
    \caption{Prescriptions adopted throughout this work for the model galaxies representing different metallicity scenarios. The acronyms in the stellar yields column stand for: \citet{Limongi18} (\citetalias{Limongi18}), \citet{Ventura13,Ventura18,Ventura20,Ventura21} (\citetalias{Ventura21}), \citet{Iwamoto99} (\citetalias{Iwamoto99}).}
    \begin{tabular}{c | c c c c c c c }
        \hline
        Model scenario & \makecell{Redshift $z$ at\\$t=0$}  & IMF & Stellar yields & SFH & $M_{\rm gas,tot}$  & $\tau_{inf}$ & \makecell{Wind factor\\ $\omega$}   \\
        \hline\\[-1.95ex]
        \makecell{\bf Metal-poor\\($M_* \times 15$)} & 25 & \citet{Chabrier03} & \makecell{CC-SNe: \citetalias{Limongi18} (R300)\\ AGBs: \citetalias{Ventura21}\\ SNe Ia: \citetalias{Iwamoto99}} & \makecell{\citet{Topping22}\\ (galaxy dependent)}& $15\, M_*$ & \makecell{$t$ at which\\$M_*(t)=M_*/2$\\(galaxy dependent)} & $\simeq 3$  \\[0.5cm]
        \makecell{\bf Intermediate-\\ \bf metallicity \\($M_* \times 10$)} & 25 & \citet{Chabrier03}& \makecell{CC-SNe: \citetalias{Limongi18} (R300)\\ AGBs: \citetalias{Ventura21}\\ SNe Ia: \citetalias{Iwamoto99}}& \makecell{\citet{Topping22}\\ (galaxy dependent)}& $10\, M_*$ & \makecell{$t$ at which\\$M_*(t)=M_*/2$\\(galaxy dependent)}& $\simeq 3$ \\[0.5cm]
        \makecell{\bf Metal-rich\\($M_* \times 15$)} & 25 & \citet{Chabrier03} & \makecell{CC-SNe: \citetalias{Limongi18} (R300)\\ AGBs: \citetalias{Ventura21}\\ SNe Ia: \citetalias{Iwamoto99}}& \makecell{\citet{Topping22}\\ (galaxy dependent)}& $3\, M_*$ & \makecell{$t$ at which\\$M_*(t)=M_*/2$\\(galaxy dependent)}& $\simeq 3$ \\[0.4cm]
        \hline
    \end{tabular}
    \label{tab:param_chem}
\end{table*}

\begin{table*}
    \centering
    \caption{Prescriptions adopted throughout this work for the model galaxies representing different dust scenarios. The acronyms in the dust yields column stand for: \citet{Marassi19} (\citetalias{Marassi19}), \citet{DellAgli17,Ventura18,Ventura20,Ventura21} (\citetalias{Ventura21}).}    
    \begin{tabular}{c | c c c}
        \hline
        Model scenario & Dust yields & \makecell{Characteristic growth timescale:\\$\tau_0\,({\rm Gyr})$}  & \makecell{Mass cleared out of dust:\\$M_{\rm clear}\,({\rm M_\odot})$} \\
        \hline\\[-1.95ex]
        {\bf Dust-poor} & \makecell{CC-SNe: \citetalias{Marassi19} (ROT FE)$\times (1/10)$\\AGBs: \citetalias{Ventura21}} & $\mathlarger{\frac{M_{gas}}{200 \, Z \, \psi}}$ & $\mathlarger{\frac{1.5 \cdot 10^4}{1+Z(0.14\,Z_\odot)}}$\\[0.35cm]
        \makecell{\bf Intermediate-\\ \bf dust} & \makecell{CC-SNe: \citetalias{Marassi19} (ROT FE)$\times (1/3)$\\AGBs: \citetalias{Ventura21}} & $\mathlarger{\frac{M_{gas}}{1000 \, Z \, \psi}}$ & $\mathlarger{\frac{8 \cdot 10^3}{1+Z(0.14\,Z_\odot)}}$\\[0.35cm]
        {\bf Dust-rich} & \makecell{CC-SNe: \citetalias{Marassi19} (ROT FE)\\AGBs: \citetalias{Ventura21}} & $\mathlarger{\frac{M_{gas}}{3000 \, Z \, \psi}}$ & $\mathlarger{\frac{2 \cdot 10^3}{1+Z(0.14\,Z_\odot)}}$\\[0.25cm]
        \hline 
    \end{tabular}
    \label{tab:param_dust}
\end{table*}

\section{UV/[C\,{\sc{ii}}] weighted gas masses}
\label{app:UVCII_weigth}

In addition to HI+H$_2$ and H$_2$ masses calibrated through the [C\,{\sc{ii}}]158$\mu$m line luminosity (\citealt{Heintz21,Vizgan22}), during this paper we also consider an intermediate value for the gas mass.

In the latter, the gas masses are obtained by multiplying the HI masses from \citet{Heintz21} by the ratio between the UV and [C\,{\sc{ii}}] effective radii $r_{\rm UV}/r_{\rm [C\,{\sc{II}}]}$ from the REBELS stacks (\citealt{Fudamoto22}), in this way:
\begin{equation}
    M_{\rm gas,{\rm[C\,{\sc{II}}]}}=\frac{r_{\rm UV}}{r_{\rm [C\,{\sc{II}}]}}M_{\rm HI,[C\,{\sc{II}}]} + M_{\rm H_2,[C\,{\sc{II}}]},
    \label{eq:weight_UV_CII}
\end{equation}
where $M_{\rm HI,[C\,{\sc{II}}]}$ is the HI mass inferred from the calibration in \citet{Heintz21} relation and $M_{\rm H_2,[C\,{\sc{II}}]}$ is the H$_2$ mass predicted from \citet{Vizgan22} relation. 
The motivation for this may be found in \citet{Fudamoto22}, who showed that the stacked [C\,{\sc{ii}}] emission in REBELS sources is $\simeq2.5$ times more extended than the UV one: this may be indicative of an extended HI gas emission belonging to outflows, as also pointed out in slightly lower redshift ($z\sim5$) observations (e.g. \citealt{Ginolfi20}) and theoretical predictions (e.g. \citealt{Pizzati20}). 
Moreover, it is worth mentioning that the UV size obtained from stacking shows a value similar to the tentative measurement for the dust continuum effective radius (\citealt{Fudamoto22}). In turn, this allows one to identify the UV size as a good proxy for the gas component not involved in galactic outflows, as gas and dust should be mostly co-located.

\section{Additional model runs}
\label{app:moreplots}

\subsection{Top-heavy IMF}
\label{app:moretopheavy}

\begin{figure*}
    \centering
    \includegraphics[width=0.69\textwidth]{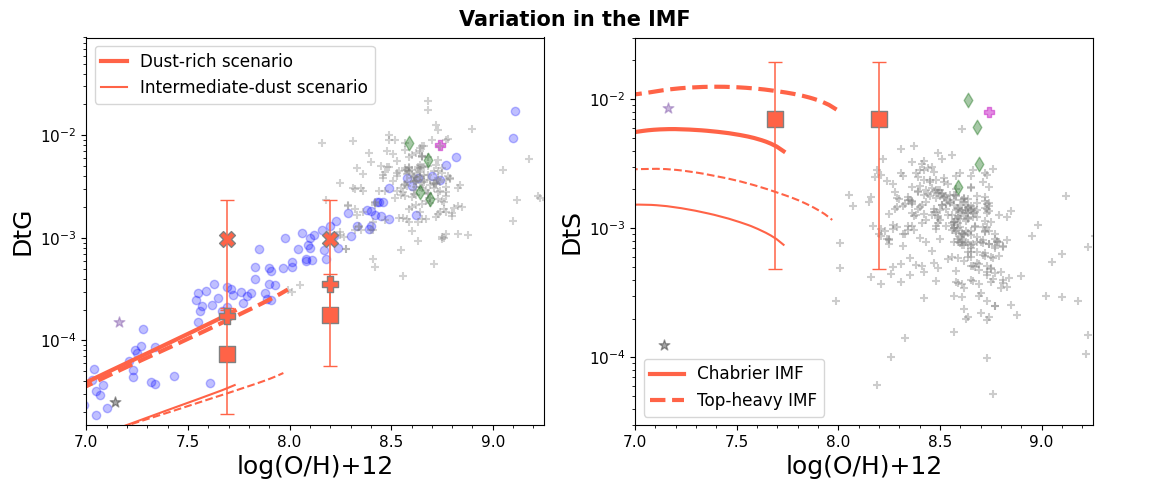}
    \caption{Effects of IMF variation on the dust-to-gas (DtG) and dust-to-stellar (DtS) mass ratios predictions from galaxy evolution models for REBELS-14 (representative of the low-mass REBELS sample) assuming a metal-poor scenario. Solid lines show results for a \citet{Chabrier03} IMF, while dashed lines show results for a top-heavy IMF. Thick lines stand for a dust-rich scenario (see \ref{sss:dust_setup}), while thin lines for an intermediate-dust scenario (see \ref{sss:dust_setup}). Data legend is as in Fig. \ref{fig:lowmetal_scenario}.}
    \label{fig:topheavyIMF}
\end{figure*}

In Section \ref{s:results1} of the paper we comment about the effects of a top-heavy IMF (i.e. that favours the formation of massive stars) on the evolution of dust quantities as a possible solution for the underprediction of the DtS of low-mass galaxies in the metal-poor scenario (see \ref{sss:metal_poor}).

In Fig. \ref{fig:topheavyIMF}, we show the results of the additional model runs with such a top-heavy IMF.
For the latter, we allow the high-mass end index of the standard \citet{Chabrier03} IMF to vary from $x=-1.3$ to $x=-0.85$. 


\subsection{"Mixed" dust prescriptions}
\label{app:moremix}

In the paper, we focus on the outcome of different dust evolution scenarios. These scenarios adopt for the different dust processes, i.e. stardust production, dust growth, dust destruction, prescriptions that either favour or disfavour the dust mass buildup.
However, we also test "mixed" setup, i.e. models where dust processes recipes favouring or disfavouring the dust mass buildup are mixed together. This allows us to have a clearer view on the relative influence that the different dust processes have in shaping the dust scaling relations. 

\begin{figure*}
    \centering
    \includegraphics[width=0.69\textwidth]{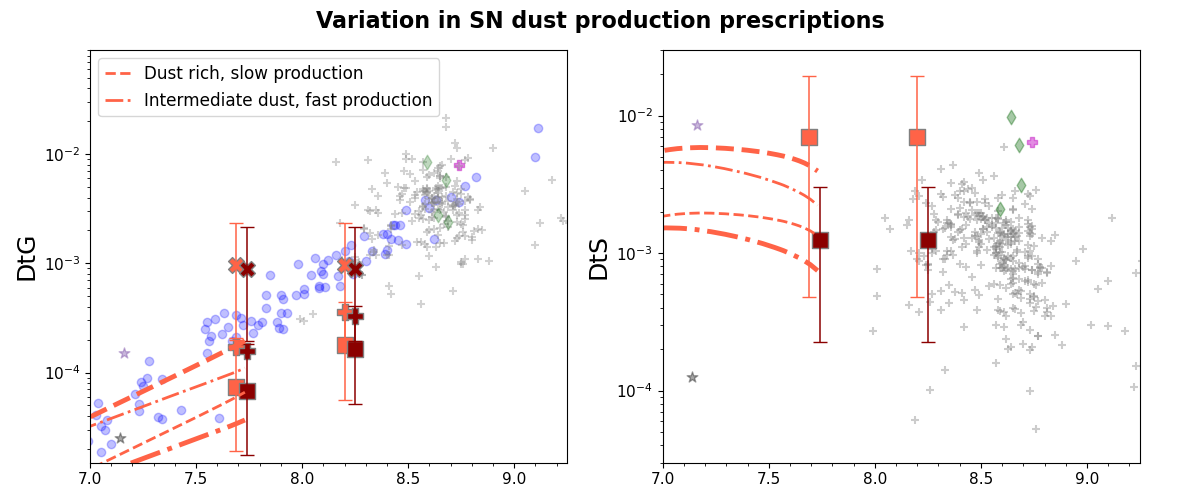}\\
     \includegraphics[width=0.69\textwidth]{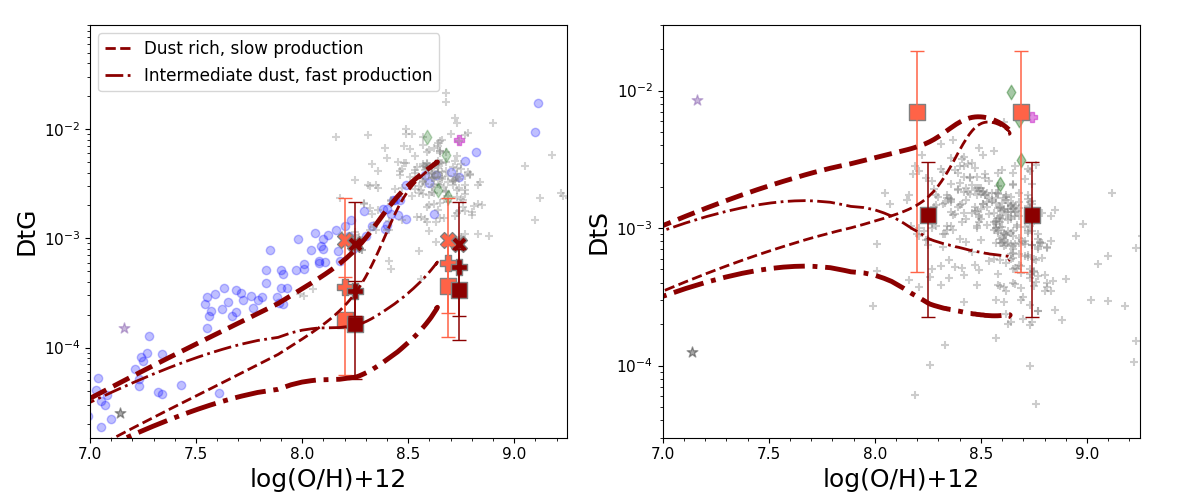}
    \caption{Effect of SN dust production prescriptions on the dust-to-gas (DtG) and dust-to-stellar (DtS) mass ratios predictions from galaxy evolution models for REBELS-14 assuming a metal-poor scenario (upper panels) and REBELS-25 assuming a metal-rich scenario (lower panels). Thick dashed and dash-dotted lines show results for the dust-rich and intermediate-dust scenario (see \ref{sss:dust_setup}) analysed in Section \ref{s:results1}. The thin dashed and dash-dotted lines represent a dust-rich scenario with smaller SN dust production and an intermediate-dust scenario with larger SN dust production, respectively. Data legend is as in Fig. \ref{fig:lowmetal_scenario}.}
    \label{fig:mixprod}
\end{figure*}

\begin{figure*}
    \centering
    \includegraphics[width=0.69\textwidth]{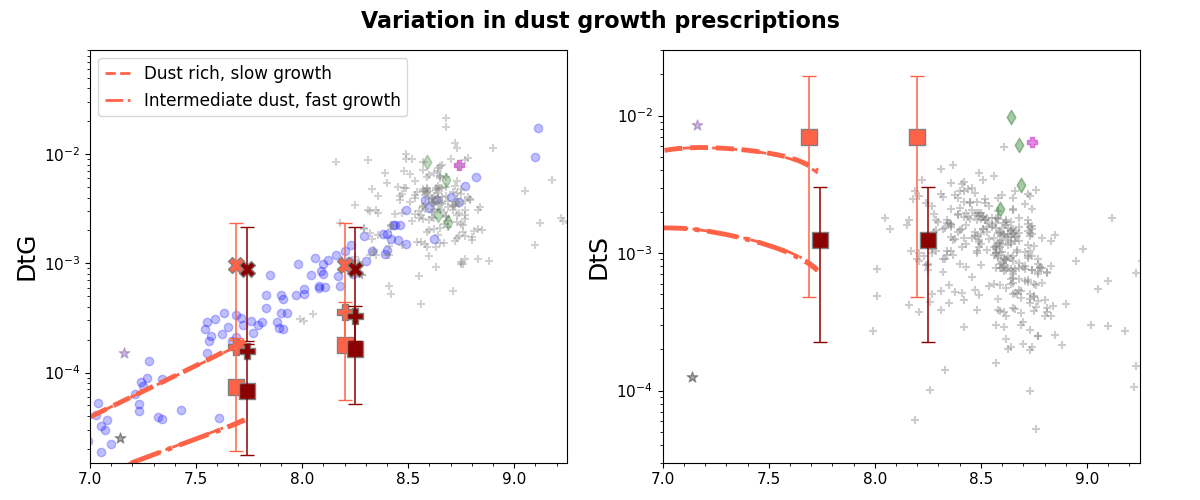}\\
     \includegraphics[width=0.69\textwidth]{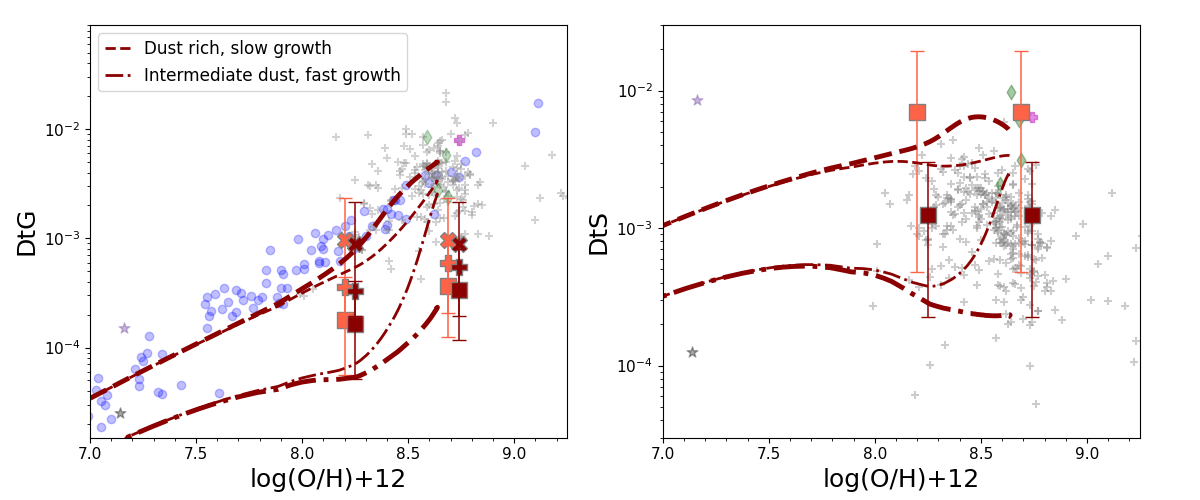}
    \caption{Effect of dust growth prescriptions on the dust-to-gas (DtG) and dust-to-stellar (DtS) mass ratios predictions from galaxy evolution models for REBELS-14 assuming a metal-poor scenario (upper panels) and REBELS-25 assuming a metal-rich scenario (lower panels). Thick dashed and dash-dotted lines show results for the dust-rich and intermediate-dust scenario (see \ref{sss:dust_setup}) analysed in Section \ref{s:results1}. The thin dashed and dash-dotted lines represent a dust-rich scenario with larger dust growth and an intermediate-dust scenario with smaller dust growth, respectively. Data legend is as in Fig. \ref{fig:lowmetal_scenario}.}
    \label{fig:mixgrow}
\end{figure*}

\begin{figure*}
    \centering
    \includegraphics[width=0.69\textwidth]{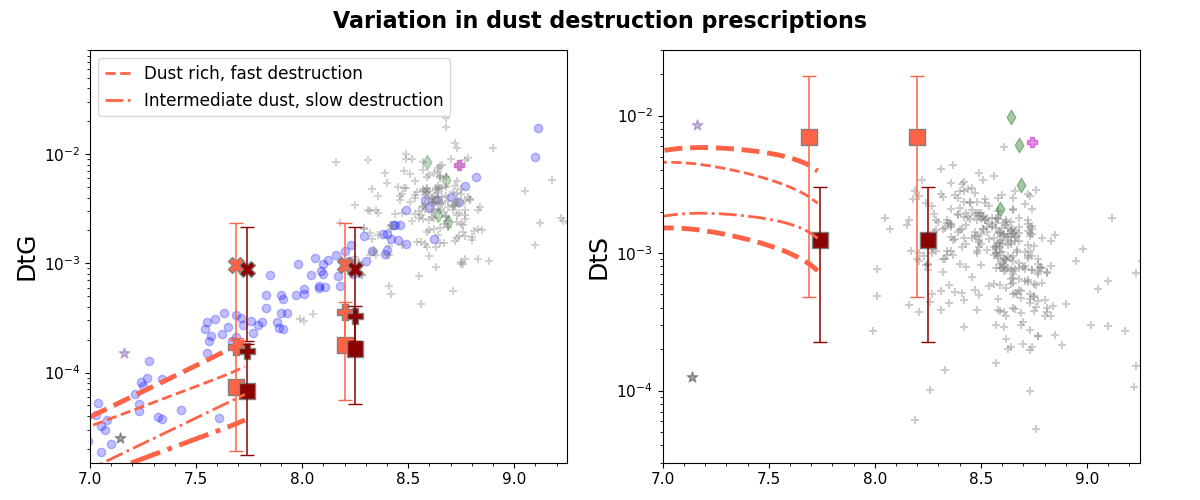}\\
     \includegraphics[width=0.69\textwidth]{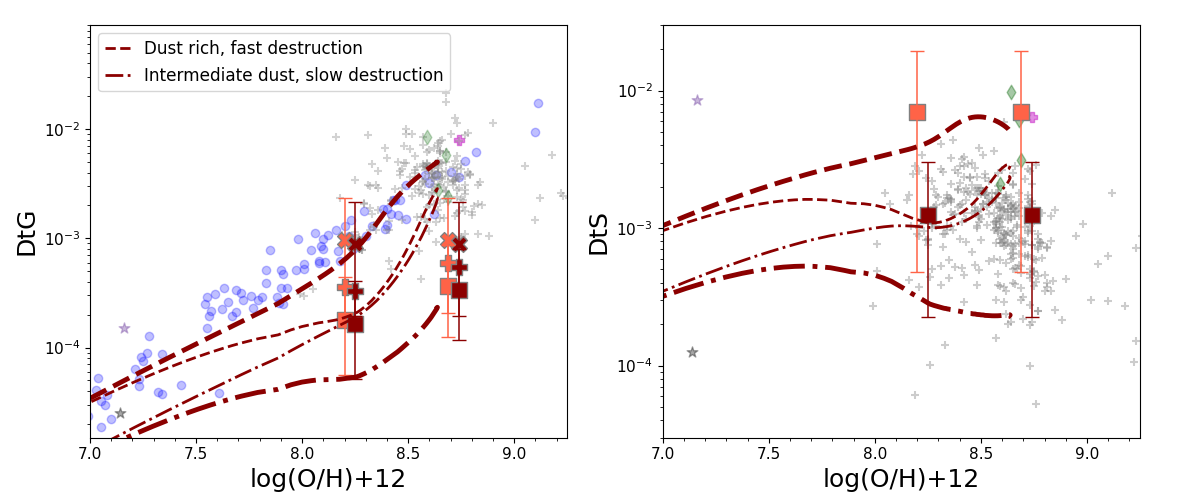}
    \caption{Effect of SN dust destruction prescriptions on the dust-to-gas (DtG) and dust-to-stellar (DtS) mass ratios predictions from galaxy evolution models for REBELS-14 assuming a metal-poor scenario (upper panels) and REBELS-25 assuming a metal-rich scenario (lower panels). Thick dashed and dash-dotted lines show results for the dust-rich and intermediate-dust scenario (see \ref{sss:dust_setup}) analysed in Section \ref{s:results1}. The thin dashed and dash-dotted lines represent a dust-rich scenario with larger SN dust destruction and an intermediate-dust scenario with smaller SN dust destruction, respectively. Data legend is as in Fig. \ref{fig:lowmetal_scenario}.}
    \label{fig:mixdestr}
\end{figure*}

In Fig. \ref{fig:mixprod}, \ref{fig:mixgrow}, \ref{fig:mixdestr} we show the results for these model runs with mixed dust prescriptions in both metal-poor (upper panels) and metal-rich (lower panels) scenarios. 
To have a clearer view on the importance of individual dust processes in shaping the dust evolution, we allow the prescriptions to vary one at a time relative the setups shown in Fig. \ref{fig:lowmetal_scenario}, \ref{fig:highmetal_scenario}. In particular, in Fig. \ref{fig:mixprod} we display the effect of changing SN dust yields, in Fig. \ref{fig:mixgrow} we show the effect of different dust growth timescales and in Fig. \ref{fig:mixdestr} we plot the consequence of changing SN dust destruction efficiencies.

\bsp	
\label{lastpage}
\end{document}